\documentstyle[11pt]{article}

\def\tabaddress#1{{\small\it\begin{tabular}[t]{c}#1
\\[1.2ex]\end{tabular}}}
\def\UPCMAT{Departamento de Matem\'atica Aplicada y Telem\'atica\\
   Edificio C-3, Campus Norte UPC\\
   C/ Jordi Girona 1. E-08034 BARCELONA. SPAIN}

\font\fr=eufm10 scaled \magstep 1 
\font\es=msbm10                   
\newtheorem{teor}{Theorem}
\newtheorem{prop}{Proposition}

\newtheorem{definition}{Definition}
\newtheorem{lem}{Lemma}
\def\beq{\begin{equation}}
\def\eeq{\end{equation}}
\def\bea{\begin{eqnarray}}
\def\eea{\end{eqnarray}}
\def\beann{\begin{eqnarray*}}
\def\eeann{\end{eqnarray*}}
\def\ben{\begin{enumerate}}
\def\een{\end{enumerate}}
\def\bit{\begin{itemize}}
\def\eit{\end{itemize}}
\def\dst{\(\displaystyle}
\def\derpar#1#2{\frac{\partial{#1}}{\partial{#2}}}
\def\bm#1{\hbox{\boldmath$#1$}}
\def\trans#1{\presup{t}#1}
\def\map#1{\mathrel{\mathop{\to}\limits^{#1}}}
\def\mapping#1{\mathrel{\mathop{\longrightarrow}\limits^{#1}}}
\def\feble#1{\mathrel{\mathop =\limits_{#1}}}
\def\coor#1#2#3{{#1}^{#2}, \ldots, {#1}^{#3}}
\def\moment#1#2#3{{#1}_{#2}, \ldots, {#1}_{#3}}
\def\blank{\leavevmode\hbox{\tt\char`\ }}
\def\qed{\ifvmode\removelastskip\fi
{\unskip\nobreak\hfil\penalty50\hbox{}\nobreak\hfil \hbox{\vrule
height1.2ex width1.2ex}\parfillskip=0pt \finalhyphendemerits=0
\par\smallskip}}
\def\vf{\mbox{\fr X}}
\def\df{{\mit\Omega}}
\def\Lag{{\cal L}}
\def\lag{\pounds}
\def\Diff{{\rm Diff}}
\def\d{{\rm d}}
\def\Nat{\mbox{\es N}}
\def\Zahl{\mbox{\es Z}}
\def\Real{\mbox{\es R}}
\def\Complex{\mbox{\es C}}
\def\Ker{\mathop{\rm Ker}\nolimits}
\def\Img{\mathop{\rm Im}\nolimits}
\def\inn{\mathop{i}\nolimits}
\def\Tan{{\rm T}}
\def\Ver{\mathop{V}\nolimits}
\def\Lie{\mathop{\rm L}\nolimits}
\def\ls{(J^1E,\Omega_\Lag)}
\def\hsjpi{(J^1\pi^*,\Omega_{h_\mu})}
\def\hspi{(\Pi,\Omega_{h_\delta})}
\def\hsljpi{(J^1\pi^*,\Omega_{{\rm h}_\mu})}
\def\hslpi{(\Pi,\Omega_{{\rm h}_\delta})}
\def\hspio{(\Pi,P,\Omega_{\hat{\rm h}_\delta}^0)}
\def\hsjpio{(J^1\pi^*,{\cal P},\Omega_{\tilde{\rm h}_\mu}^0)}
\def\del{{\cal E}^{\nabla}_{\Lag}}
\def\fel{{\rm E}^{\nabla}_{\Lag}}
\def\mcb#1#2{\Lambda^{#1}{\Tan^*#2}}
\def\Cinfty{{\rm C}^\infty}
\def\proof{( {\sl Proof} )\quad}

\title{GEOMETRY OF MULTISYMPLECTIC HAMILTONIAN FIRST-ORDER
       FIELD THEORIES}
\author{\sc Arturo Echeverr\'ia-Enr\'iquez,
   \\
{\sc Miguel C. Mu\~noz-Lecanda\thanks{{\bf e}-{\it mail}:
 MATMCML@MAT.UPC.ES}},
   \\
{\sc Narciso Rom\'an-Roy\thanks{{\bf e}-{\it mail}:
 MATNRR@MAT.UPC.ES}},
   \\
   \tabaddress{\UPCMAT}}

\begin{document}
\maketitle
\thispagestyle{empty}
\setcounter{page}{0}

 \begin{abstract}
 In the jet bundle description of Field Theories (multisymplectic
 models, in particular), there are several choices for the
 multimomentum bundle where the covariant Hamiltonian formalism
 takes place. As a consequence, several proposals for this
 formalism can be stated, and, on each one of them, the
 differentiable structures needed for setting the formalism are
 obtained in different ways. In this work we make an accurate study
 of some of these Hamiltonian formalisms, showing their equivalence.
 In particular, the geometrical structures (canonical
 or not) needed for the Hamiltonian formalism, are introduced and
 compared, and the derivation of Hamiltonian
 field equations from the corresponding variational principle is
 shown in detail. Furthermore, the Hamiltonian formalism of
 systems described by Lagrangians is performed, both for
 the hyper-regular and almost-regular cases.
 Finally, the role of connections in the
 construction of Hamiltonian Field theories is clarified.
 \end{abstract}


\bigskip
\bigskip

\vbox{\raggedleft AMS s.\,c.\,(2000):
 53C05, 53C80, 55R10, 55R99, 58A20, 70S05.\\
PACS (1999): 02.40.Hw, 02.40.Vh, 11.10.Ef, 45.10.Na }\null

 \clearpage

 \tableofcontents

 \section{Introduction}

 The application of techniques of differential geometry to the
 study of physical theories has been revealed as a very suitable
 method for better understanding many features of these theories.
 In particular, the geometric description of classical Field
 Theories is an area of increasing interest.

 The standard geometrical techniques used
 for the covariant  Lagrangian description
 of first-order Field Theories,
 involve first order jet bundles
 $J^1E\stackrel{\pi^1}{\to}E\stackrel{\pi}{\to}M$ and their
 canonical structures (see, for instance,
 \cite{EMR-96}, and references quoted therein).
 Nevertheless, for the covariant Hamiltonian
 formalism of these theories the situation is rather different, and
 there are different kinds of geometrical descriptions for
 this formalism. For instance, we can find models such as those described
 in \cite{Aw-92}, \cite{No-93} and \cite{Pu-88},
 which use {\sl $k$-symplectic forms}, or
 in \cite{LMS-88a}, \cite{LMS-88b}, \cite{LMO-98} and \cite{LMS-99},
 where the essential geometric structure are the
 {\sl $k$-cosymplectic forms}, or also as in \cite{Gu-87},
 \cite{Ka-97a} and \cite{Ka-98},
 where use is made of {\sl polysymplectic forms}
 (in fact, $k$-symplectic, $k$-cosymplectic and polysymplectic structures are
 essentially equivalent objects).
 In this work, we consider only the  {\sl multisymplectic} models
 \cite{CIL-96b}, \cite{CIL-98}, \cite{Hr-99a},
 \cite{Hr-99b}, \cite{IEMR-98}, \cite{Ma-88a}, \cite{Ma-88b},
 and depending on the choice of the
 {\sl multimomentum phase space} there are different ones. In fact:
 \ben
 \item
 There are some models where the multimomentum phase space
 is taken to be
 ${\cal M}\pi\equiv\Lambda_1^m\Tan^*E$, the bundle of $m$-forms on
 $E$ ($m$ being the dimension of $M$) vanishing by the action of
 two $\pi$-vertical vector fields.
 This choice is made in works such as  \cite{Go-91b}, \cite{Go-91c}
 and \cite{GIMMSY-mm}, as a refinement
 of the techniques previously given in \cite{Ki-73},
 \cite{KS-75} and \cite{KT-79} (see also  \cite{La-97} and
 \cite{MS-99}).
 \item
 The multimomentum phase space
 $J^1\pi^*\equiv\Lambda_1^m\Tan^*E/\Lambda_0^m\Tan^*E$
 (where $\Lambda_0^m\Tan^*E$ is the
 bundle of $\pi$-semibasic $m$-forms in $E$)
 has been studied in \cite{CCI-91} and used, later on, in
 \cite{Ka-97a}, \cite{LMM-95} and \cite{LMM-96}
 for the analysis of different aspects of Hamiltonian Field Theories.
 \item
  Finally, in \cite{EM-92}, \cite{GMS-95}, \cite{GMS-97},
 \cite{GMS-99},
  \cite{Sd-95}, \cite{SZ-92c} and \cite{SZ-93}
  the basic choice is the bundle
 $\Pi\equiv\pi^*\Tan M\otimes{\rm
 V}^*(\pi)\otimes\pi^*\Lambda^m\Tan^* M$ (here ${\rm V}^*(\pi)$
 denotes the dual bundle of the $\pi$-vertical subbundle ${\rm
 V}(\pi)$ of $\Tan E$) which, in turn, is canonically related to
 $J^1E^*\equiv\pi^*\Tan M\otimes\Tan^*E\otimes\pi^*\Lambda^m\Tan^* M$.
 \een

 Although in \cite{MS-99} (and later papers by these authors),
 a covariant Hamiltonian formalism is constructed in ${\cal M}\pi$,
 in most of the works, this multimomentum bundle
 is not really used in order to establish
 a Hamiltonian formalism on ${\cal M}\pi$, but just for defining canonical
 differential structures which, translated to $J^1E$ and
 $J^1\pi^*$, are used for setting the
 Lagrangian and Hamiltonian formalisms, respectively.
 The choice of $J^1\pi^*$ or $\Pi$ as multimomentum phase
 space allows us to state covariant Hamiltonian
 formalisms for Field Theories. Nevertheless, none of them have
 canonical structures, so the Hamiltonian
 forms of the Hamiltonian formalism must be obtained
 from the canonical forms of the {\sl multicotangent bundle}
 $\Lambda^m\Tan^*E$. This is done by using
 sections of the projection ${\cal M}\pi\to J^1\pi^*$,
 (or $J^1E^*\to \Pi$) which are called {\sl Hamiltonian sections}
 \cite{CCI-91}, or the so-called
 {\sl Hamiltonian densities} \cite{GMS-97}, \cite{Sd-95}.
 To our knowledge, a rigorous analysis comparing these
 formulations and their equivalence has not been
 done. The aim of this work is to carry out a comparative
 study of some of these Hamiltonian formulations, establishing
 the equivalence between them. In every
 case, the geometrical structures
 needed for setting the field equations in the
 Hamiltonian formalism are introduced,
 as well as the corresponding {\sl Legendre maps}
 when the multimomentum bundles are related to a Lagrangian system.

 The question of whether the use of connections in the bundle
 $\pi\colon E\to M$ is needed for the construction of
 the covariant formalisms in Field theories is studied. It was
 analized for the first time in \cite{CCI-91},
 where a connection was used to define {\sl Hamiltonian densities}
 in the Hamiltonian formalism,
 and in \cite{EMR-96} for the case of
 the {\sl density of Lagrangian energy} in the Lagrangian formalism.
 In this work we make a deeper analysis on the role
 played by connections in the construction of Hamiltonian systems.

 An obvious subject of interest is the statement of
 the {\sl Hamiltonian field equations}.
 In all the multisymplectic models field equations are obtained
 by characterizing the critical sections which are solutions of the problem
 by means of the multisymplectic form \cite{AA-80},
 \cite{EMR-96}, \cite{Gc-73}, \cite{GS-73}.
 This characterization can be derived from a
 variational principle: the so-called
 {\sl Hamilton principle} in the Lagrangian formalism and
 {\sl Hamilton-Jacobi principle} in the Hamiltonian one.
 Nevertheless, this aspect of the theory is overlooked in many
 papers. We give an accurate derivation of the Hamiltonian equations
 starting from the Hamilton-Jacobi principle,
 and the role played by connections in the statement of
 covariant Hamiltonian equations is discussed.

 An important kind of Hamiltonian systems are those which are the Hamiltonian
 counterpart of Lagrangian systems. The construction of such systems
 starting from the Lagrangian formalism is carried out by using a
 {\sl Legendre map} associated with the
 {\sl Lagrangian density} and the corresponding multimomentum bundle.
 This problem has been studied
 by different authors in the {\sl (hyper) regular} case
 (see, for instance, \cite{CCI-91}, \cite{Sa-89}), and in the
 {\sl singular} ({\sl almost-regular\/}) case \cite{GMS-97},
 \cite{LMM-96}, \cite{Sd-95}. In this work we review some of these
 constructions, developing new methods,
 and giving a unified perspective of all of them.

 The structure of the work is as follows:

 Section 2 is devoted to a review of
 the main features of the Lagrangian formalism of Field theories,
 and afterwards the definition of the different multimomentum bundles
 for the Hamiltonian formalism, as well as the construction and
 characterization of the canonical forms with which some of them are endowed.
 Furthermore, when these multimomentum bundles are related with
 a Lagrangian system, the corresponding Legendre maps are introduced
 for both the {\sl(hyper)-regular} and the {\sl almost-regular} cases.

 In section 3, we undertake the construction of {\sl Hamiltonian systems}
 in the multimomentum bundle $\Pi$.
 As a first step, we will define the
 {\sl Hamiltonian forms} which allow us to set the field equations
 in an intrinsic way. Since $\Pi$ has
 no canonical geometric form, we must
 use the canonical forms with which ${\cal M}\pi$ and $J^1E^*$ are
 endowed. Ways of constructing Hamiltonian systems
 are studied and compared, and in this multimomentum bundle
 we make a careful deduction of the Hamiltonian equations from
 the variational principle.
 In addition, the Hamiltonian formalism associated to a Lagrangian system
 is developed,  both for the hyper-regular and almost-regular
 cases. Finally, the equivalence between the Lagrangian and
 Hamiltonian formalisms is proved (for the hyper-regular case).

 The construction of {\sl Hamiltonian systems}
 in the multimomentum bundle $J^1\pi^*$ is stated
 and analyzed in Section 4,
 following the same pattern as in the above section,
 and proving the equivalence between the formalisms
 developed for both multimomentum bundles.

 As typical examples, time-dependent mechanics and the electromagnetic field
 are analyzed (in this context) in Section 5.

 An appendix describing the basic geometrical structures in
 first-order jet bundles is included.

 All manifolds are real, paracompact, connected and $C^\infty$. All
 maps are $C^\infty$. Sum over crossed repeated indices is
 understood. Throughout this paper
 $\pi\colon E\to M$ will be a fiber bundle
 ($\dim\, M=m$, $\dim\, E=N+m$),
 where $M$ is an oriented manifold with volume form
 $\omega\in\df^m(M)$,
 and $\pi^1\colon J^1E\to E$ will be the
 jet bundle of local sections of $\pi$.
 The map $\bar\pi^1 = \pi \circ \pi^1\colon J^1E
 \longrightarrow M$ defines another structure of differentiable
 bundle. We denote by ${\rm V}(\bar\pi^1 )$ the
 vertical bundle associated with $\bar\pi^1$, that is,
 ${\rm V}(\bar\pi^1)=\Ker\Tan\bar\pi^1$, and by
 $\vf^{{\rm V}(\bar\pi^1 )}(J^1E)$
 the corresponding sections or vertical
 vector fields. Finally, $(x^\nu,y^A,v^A_\nu)$
 (with $\nu = 1,\ldots,m$; $A= 1,\ldots,N$) will be
 natural local systems of coordinates in $J^1E$
 adapted to the bundle $\pi\colon E\to M$, and such that
 $\omega=\d x^1\wedge\ldots\wedge\d x^m\equiv\d^mx$.

 \section{Geometrical background of the Lagrangian and
  Hamiltonian formalisms}

\subsection{Lagrangian systems}

  From the Lagrangian point of view,
 a {\sl first-order classical Field Theory} is described by its
 {\sl configuration bundle} $\pi\colon E\to M$, and a
 {\sl Lagrangian density} which is a $\bar\pi^1$-semibasic $m$-form
 on $J^1E$ (see the appendix for notation and terminology).
 A Lagrangian density is usually written as
 $\Lag =\lag (\bar\pi^{1^*}\omega)$,
 where $\lag\in\Cinfty (J^1E)$ is the {\sl
 Lagrangian function} associated with $\Lag$ and $\omega$. The
 {\sl Poincar\'e-Cartan $m$ and $(m+1)$-forms} associated with the
 Lagrangian density $\Lag$ are defined using the {\sl vertical
 endomorphism} ${\cal V}$ of the bundle $J^1E$:
 $$
 \Theta_{\Lag}:=\inn({\cal V})\Lag+\Lag\equiv\theta_{\Lag}+\Lag\in\df^{m}(J^1E)
 \quad ;\quad
 \Omega_{\Lag}:= -\d\Theta_{\Lag}\in\df^{m+1}(J^1E) $$
 In a natural chart in $J^1E$ we have
 \beann
  \Theta_{\Lag}&=&\derpar{\lag}{v^A_\nu}\d
y^A\wedge\d^{m-1}x_\nu - \left(\derpar{\lag}{v^A_\nu}v^A_\nu
-\lag\right)\d^mx
\\
\Omega_{\Lag}&=& -\frac{\partial^2\lag}{\partial v^B_\eta\partial
v^A_\nu} \d v^B_\eta\wedge\d y^A\wedge\d^{m-1}x_\nu
-\frac{\partial^2\lag}{\partial y^B\partial v^A_\nu}\d y^B\wedge
\d y^A\wedge\d^{m-1}x_\nu + \nonumber \\ & &
\frac{\partial^2\lag}{\partial v^B_\eta\partial v^A_\nu}v^A_\nu \d
v^B_\eta\wedge\d^mx  + \left(\frac{\partial^2\lag}{\partial
y^B\partial v^A_\nu}v^A_\nu -\derpar{\lag}{y^B}+
\frac{\partial^2\lag}{\partial x^\nu\partial v^B_\nu} \right)\d
y^B\wedge\d^mx
 \eeann
 (See, for instance, \cite{BSF-88}, \cite{EMR-96},
\cite{Gc-73}, \cite{GS-73}, \cite{Sa-87} and \cite{Sa-89}, for
details).
 Then a {\sl Lagrangian system} is a couple $\ls$.

As we can see, the factor
 \dst{\rm E}_{\lag}\equiv\derpar{\lag}{v^A_\nu}v^A_\nu -\lag\)
 appears in the local expression of the Poincar\'e-Cartan $(m+1)$-form, and it
is recognized as the classical expression of the {\sl Lagrangian
energy} associated with the Lagrangian function $\lag$. In fact,
the existence of such a function as a global object,
and by extension a {\sl density of Lagrangian energy},
 is closely related to the existence of a
connection in the bundle $\pi\colon E\to M$, in the same way that
happens in non-autonomous mechanics \cite{EMR-sdtc}. As
shown in \cite{EMR-96}, we can define the density of
Lagrangian energy using the vertical endomorphisms in $J^1E$. In
fact, given a connection $\nabla$ in $\pi\colon E\to M$, we can
identify ${\rm V}^*(\pi )$ as a subbundle of $\Tan^*E$. Then the
operation ${\cal S}^\nabla-{\cal V}$ makes sense, where ${\cal S}$ and
${\cal V}$ are the {\sl vertical endomorphisms} of the bundle
$J^1E$, and ${\cal S}^\nabla$ denotes the action of ${\cal S}$ followed
by the injection of ${\rm V}^*(\pi)$ in $\Tan^*E$ induced
by $\nabla$ (see the appendix). Therefore:

\begin{definition}
Let $\ls$ be a Lagrangian system and $\nabla$ a connection in the
bundle $\pi\colon E\to M$. The density of Lagrangian energy
associated with the Lagrangian density $\Lag$ and the connection
$\nabla$ is given by
 $$
{\cal E}^{\nabla}_{\Lag}=\inn ({\cal S}^\nabla-{\cal V})\d\Lag -\Lag=
\inn ({\cal S}^\nabla)\d\Lag -\Theta_\Lag\equiv
\Theta_\Lag^\nabla-\Theta_\Lag
 $$
 It is a $\bar\pi^1$-vertical $m$-form in $J^1E$.
Hence, we can write ${\cal E}^{\nabla}_{\Lag}={\rm
E}^{\nabla}_{\Lag}(\bar\pi^{1*}\omega)$, where $\fel\in\Cinfty
(J^1E)$ is the {\rm Lagrangian energy function} associated with
$\Lag$, $\nabla$ and $\omega$. 
\label{energ}
\end{definition}

 {\bf Remark}:
\bit
\item
 Note that every connection $\nabla$ in $\pi\colon E\to M$ allows us
 to split the Poincar\'e-Cartan forms as 
 $$
 \Theta_\Lag=\Theta_\Lag^\nabla-\del
 \quad , \quad
 \Omega_\Lag=-\d\Theta_\Lag^\nabla+\d\del\equiv\Omega_\Lag^\nabla+\d\del
 $$
\eit

Using natural systems of coordinates, and
${\mit\Gamma}^A_\nu$ being the component functions of the connection, we
have the following local expressions $$ {\cal E}^{\nabla}_{\Lag}=
\left(\derpar{\lag}{v^A_\nu} (v^A_\nu-{\mit\Gamma}^A_\nu
)-\lag\right)\d^mx \quad ;\quad {\rm E}_{\Lag}^\nabla=
\derpar{\lag}{v^A_\nu} (v^A_\nu-{\mit\Gamma}^A_\nu )-\lag $$
Observe also that if we take a local connection with
${\mit\Gamma}^A_\nu =0$, then the Lagrangian energy associated
with this natural connection has the classical local expression
given above.

 A variational problem can be posed from the Lagrangian density
$\Lag$, which is called the {\sl Hamilton principle} of the
Lagrangian formalism: the states of the field are the sections of
$\pi$ which are critical for the functional ${\bf
L}\colon\Gamma_c(M,E)\to\Real$ defined by
 $$
{\bf L}(\phi):=\int_M(j^1\phi)^*\Lag
\quad , \quad
\mbox{\rm , for every $\phi\in\Gamma_c(M,E)$}
$$
 where $\Gamma_c(M,E)$ is the set of
compact supported sections of $\pi$. These (compact-supported)
critical sections can be characterized in several equivalent ways.
In fact (see \cite{EMR-96}, \cite{Gc-73}, \cite{LMM-95} and
\cite{Sa-89}):

\begin{teor}
The critical sections of the Hamilton's principle are sections
$\phi\colon M\to E$ whose canonical liftings $j^1\phi\colon M\to
J^1E$ satisfy the following equivalent conditions: \ben
\item
\dst \frac{\d}{\d t}\Big\vert_{t=0}\int_M(j^1\phi_t)^*\Lag =0\) ,
being $\phi_t=\sigma_t\circ\phi$, where $\{\sigma_t\}$ denotes a
local one-parameter group of any $\pi$-vertical vector field
$Z\in\vf (E)$.
\item
\dst\int_M(j^1\phi)^*\Lie (j^1Z)\Lag = 0\) ,
for every $Z\in\vf^{{\rm V}(\pi)}(E)$.
\item
\dst\int_M(j^1\phi)^*\Lie (j^1Z)\Theta_{\Lag} = 0\) ,
for every $Z\in\vf^{{\rm V}(\pi)}(E)$.
\item
$(j^1\phi)^*\inn (j^1Z)\Omega_{\Lag} = 0$,
for every $Z\in\vf^{{\rm V}(\pi)}(E)$.
\item
\dst(j^1\phi)^*\inn (X)\Omega_{\Lag} = 0\) , for every $X\in\vf
(J^1E)$.
\item
The coordinates of $\phi$ satisfy the {\sl Euler-Lagrange
equations}: \dst \derpar{\lag}{y^A}\Big\vert_{j^1\phi}-
\derpar{}{x^\nu}\derpar{\lag}{v_\nu^A}\Big\vert_{j^1\phi} = 0 \)
 (for $A=1,\ldots ,N$).
\een \label{importantlag}
\end{teor}

 \subsection{Multimomentum bundles and Legendre maps}
 \protect\label{mmblm}

 (See \cite{EMR-99} for a more detailed study of all
 these constructions).

 Let $\bar y\in J^1E$, with $\bar
 y\stackrel{\pi^1}{\mapsto}y\stackrel{\pi}{\mapsto}x$. We have that
 $\Tan_{\bar y}J^1_yE={\rm V}_{\bar y}\pi^1$ is canonically
 isomorphic to $\Tan^*_xM\otimes{\rm V}_y\pi$, by means of the
 directional derivatives; therefore ${\rm
 V}(\pi^1)=\bar\pi^{1*}\Tan^*M\otimes_{J^1E}\pi^{1*}{\rm V}(\pi)$.
 Moreover, if ${\cal D}\subset\Tan J^1E$ denotes the subbundle of
 total derivatives (which in a system of natural coordinates in
 $J^1E$, is generated by
 \dst\left\{\derpar{}{x^\nu}+v^A_\nu\derpar{}{y^A}\right\}\) ), we
 have that $\pi^{1*}\Tan E=\pi^{1*}{\rm V}(\pi)\oplus{\cal D}$ with
 $\Tan_yE\big\vert_{\bar y}= {\rm V}_y\pi\big\vert_{\bar
 y}\oplus{\cal D}_{\bar y}$ (see \cite{Sa-89} for details). Hence
 there is a natural projection $\sigma\colon\pi^{1*}\Tan
 E\to\pi^{1*}{\rm V}(\pi)$ and its dual injection
 $\sigma^*\colon\pi^{1*}{\rm V}^*(\pi)\to\pi^{1*}\Tan^*E$ and so we
 can consider the projection
 $$
 {\rm Id}\otimes\sigma\colon
\bar\pi^{1*}\Tan^*M\otimes\pi^{1*}\Tan E\longrightarrow
\bar\pi^{1*}\Tan^*M\otimes\pi^{1*}{\rm V}(\pi)={\rm V}(\pi^1)
 $$

In a natural chart $(x^\nu,y^A)$ adapted to the bundle
 $\pi\colon E\to M$, the
local expression of this mapping is $$
\sigma\left(\left(\derpar{}{x^\nu}\right)_y\Big\vert_{\bar
y}\right)= -v^A_\nu(\bar
y)\left(\derpar{}{y^A}\right)_y\Big\vert_{\bar y} \quad ;\quad
\sigma\left(\left(\derpar{}{y^A}\right)_y\Big\vert_{\bar
y}\right)= \left(\derpar{}{y^A}\right)_y\Big\vert_{\bar y} $$ and,
if $\{\zeta^A\}$ is the dual basis of
\dst\left\{\derpar{}{y^A}\right\}\) in ${\rm V}^*(\pi)$, we have
that $ \sigma^*(\zeta^A)=\d y^A-v^A_\nu\d x^\nu $.

Now let $\ls$ be a Lagrangian system,
and consider the restriction
$\Lag_y\colon J^1_yE\to\Lambda^m\Tan_x^*M$. Its differential
map at $\bar y\in J^1_yE$ is $$ D_{\bar y}\Lag_y\colon \Tan_{\bar
y}J^1_yE\longrightarrow\Tan_{\Lag_y(\bar y)}\Lambda^m\Tan_x^*M $$
(which, bearing in mind that $\Lambda^m\Tan_x^*M$ is a vector
space, it is just the vertical differential of $\Lag$). Thus,
using the defined projection $\sigma$, we have
 $$
\begin{array}{cccc}
\begin{picture}(80,60)(0,0)
\put(0,49){\mbox{$\Tan_{\bar y}J^1_yE={\rm V}_{\bar y}\pi^1\simeq$}}
\end{picture}&
\begin{picture}(70,60)(0,0)
\put(0,49){\mbox{$(\Tan_x^*M\otimes{\rm V}_y\pi)_{\bar y}$}}
\put(50,25){\mbox{${\rm Id}\otimes\sigma_{\bar y}$}}
\put(35,15){\vector(0,1){25}}
\put(0,0){\mbox{$(\Tan_x^*M\otimes\Tan_yE)_{\bar y}$}}
\end{picture}
&
\begin{picture}(70,60)(0,0)
\put(0,50){\vector(1,0){70}}
\put(25,55){\mbox{$D_{\bar y}\Lag_y$}}
\put(0,3){\vector(2,1){70}}
\end{picture}
&
\begin{picture}(20,60)(0,0)
\put(0,49){\mbox{$(\Lambda^m\Tan_x^*M)_{\bar y}$}}
\end{picture}
\end{array}
$$

\begin{definition}
\ben
\item
The bundle (over $E$) $$ J^1E^*:=\pi^*\Tan
M\otimes_E\Tan^*E\otimes_E\pi^*\Lambda^m\Tan^*M $$ is called the
{\rm generalized multimomentum bundle} associated with the bundle
$\pi\colon E\to M$. We denote the natural projections by
$\hat\rho^1\colon J^1E^*\to E$ and
$\bar{\hat\rho}^1:=\pi\circ\hat\rho^1\colon J^1E^*\to M$.
\item
The {\rm generalized Legendre map}
 associated with a Lagrangian density $\Lag$
 is the $\Cinfty$-map $$
\begin{array}{ccccc}
\widehat{{\rm F}\Lag} & \colon & J^1E &$\rightarrowfill$ & J^1E^*
\\
& &\bar y & \mapsto &
D_{\bar y}\Lag_y\circ ({\rm Id}\otimes\sigma)_{\bar y}
\end{array}
$$
\een
\end{definition}

(We have departed a little from the notation in this definition,
because $\sigma$ acts on $\pi^{1*}\Tan E$, and not on $\Tan E$.
Then, given $\bar y\in J^1E$, the right way consists in taking
$\Tan_yE$ and lifting it to $\bar y$).

 Natural coordinates in $J^1E^*$ will be denoted by
 $(x^\nu,y^A,{\rm p}_\nu^\eta,{\rm p}_A^\nu)$,
 and for every ${\bf y}\in J^1E^*$, with
 ${\bf y}\stackrel{\hat\rho^1}{\to}y\stackrel{\pi}{\to}x$,
 we have
 $$
 {\bf y}=\derpar{}{x^\nu}\Big\vert_y\otimes
 ({\rm p}^\nu_\eta({\bf y})\d x^\eta+
 {\rm p}_A^\nu({\bf y})\d y^A)_y\otimes\d^mx\vert_y
  $$
 and the local expression of the generalized Legendre map is
 \beq
\widehat{{\rm F}\Lag}^*x^\nu=x^\nu \quad ,\quad \widehat{{\rm
F}\Lag}^*y^A=y^A \quad ,\quad \widehat{{\rm F}\Lag}^*{\rm
p}^\nu_\eta=-v^A_\eta\derpar{\lag}{v^A_\nu} \quad ,\quad
\widehat{{\rm F}\Lag}^*{\rm p}^\nu_A=\derpar{\lag}{v^A_\nu}
\label{coorglt}
 \eeq

Now, let $\bar y\in J^1E$ with \dst\bar y\stackrel{\pi^1}{\mapsto}
y\stackrel{\pi}{\mapsto} x\) . We define the map $\Lag_y\colon
J_y^1E\longrightarrow \Lambda^m\Tan_x^*M$ as
\dst\Lag_y:=\Lag\vert_{J^1_yE}\) . It is a $\Cinfty$-map of the
affine space $J^1_yE$, modeled on $\Tan^*_xM\otimes{\rm V}_y(\pi
)$, with values on $\Lambda^m\Tan_x^*M$. Then, the tangent map
$\Tan_{\bar y}\Lag_y$ allows us to construct the following diagram
(where the vertical arrows are canonical isomorphisms given by the
directional derivatives) $$
\begin{array}{ccc}
\Tan_{\bar y}J^1_yE & $\rightarrowfill$ &
\Tan_{\Lag_y(\bar y )}\Lambda^m\Tan_x^*M
\\
& \Tan_{\bar y}\Lag_y &
\\
\simeq\ \Big\updownarrow & & \Big\updownarrow\ \simeq
\\
& \tilde\Tan_{\bar y}\Lag_y &
\\
\Tan_x^*M\otimes{\rm V}_y\pi  & $\rightarrowfill$ &
\Lambda^m\Tan_x^*M
\end{array}
$$
Hence, taking into account these identifications,
we have that $\tilde\Tan_{\bar y}\Lag_y$ is an element of
$\Tan_xM\otimes{\rm V}_y^*(\pi )\otimes\Lambda^m\Tan_x^*M$, and so,
bearing in mind the analogy with classical mechanics,
we define:

 \begin{definition}
 \ben
 \item
 The bundle (over $E$)
 $$
 \Pi:=\pi^*\Tan M\otimes_E{\rm V}^*(\pi )\otimes_E \pi^*\Lambda^m\Tan^*M
 $$
 is called the {\rm reduced multimomentum bundle}
 associated with the bundle $\pi\colon E\to M$.
 We denote the natural projections by $\rho^1\colon \Pi\to E$
 and $\bar\rho^1:=\pi\circ\rho^1\colon \Pi\to M$.
 \item
 The {\rm reduced Legendre map}
 associated with a Lagrangian density $\Lag$ is the $\Cinfty$-map
 $$
\begin{array}{ccccc}
{\rm F}\Lag & \colon & J^1E &$\rightarrowfill$ & \Pi
\\
& &\bar y & \mapsto & \tilde\Tan_{\bar y}\Lag_y
\end{array}
 $$
 \een
\end{definition}

 Natural coordinates in $\Pi$ are denoted by
 $(x^\nu ,y^A,{\rm p}_A^\nu )$, and for every $\tilde y\in \Pi$ with
 $\tilde y\stackrel{\rho^1}{\to}y\stackrel{\pi}{\to}x$,
 $$
 \tilde y = {\rm p}^\nu_A(\tilde y)\derpar{}{x^\nu}\otimes\zeta^A\otimes\d^mx
 \Big\vert_y
 $$
 (We have departed from the
notation by denoting the momentum coordinates in $\Pi$ and $J^1E^*$
with the same symbol, ${\rm p}^\nu_A$,. This departure will be
repeated frequently throughout the work).

The local expression of the reduced Legendre map is
 \beq
  {\rm F}\Lag^*x^\nu = x^\nu \quad ,
\quad {\rm F}\Lag^*y^A = y^A \quad , \quad
 {\rm F}\Lag^*{\rm p}_A^\nu = \derpar{\lag}{v^A_\nu}
  \label{redlt}
 \eeq

 If we recall that
 $J^1E^*:=\pi^*\Tan M\otimes\Tan^*E\otimes\pi^*\Lambda^m\Tan^*M$,
 then the natural projection
 $\Tan^*E\to{\rm V}^*(\pi)$ (which is the transpose of
 the natural injection ${\rm V}(\pi)\hookrightarrow\Tan E$) induces
 another one
 $$
 \delta\colon J^1E^*\longrightarrow \Pi
 $$

\begin{prop}
The natural map $\delta$ is onto, and ${\rm
F}\Lag=\delta\circ\widehat{{\rm F}\Lag}$.
\end{prop}

 Furthermore, we can introduce the following map:

 \begin{definition}
 The {\rm canonical contraction} in $J^1E^*$ is the map $$
 \iota\colon J^1E^*\equiv
 \pi^*\Tan M\otimes\Tan^*E\otimes\pi^*\Lambda^m\Tan^*M
 \longrightarrow
 \Lambda^m\Tan^*E $$ defined as follows:
  $\iota ({\bf y}):=\alpha^k\wedge\pi^*\inn (u_k)\beta $,
  for every ${\bf y}=u_k\otimes\alpha^k \otimes\beta\in J^1E^*$.
 \end{definition}

 In a chart of natural coordinates in $J^1E^*$, we have that
 \beq
 \iota ({\bf y}) = ({\rm p}^\nu_\eta({\bf y})\d x^\eta+{\rm p}_A^\nu({\bf y})
 \d y^A)_y\wedge \inn\left(\derpar{}{x^\nu}\right)\d^mx\Big\vert_y=
 ({\rm p}^\nu_\nu({\bf y})\d^mx+{\rm p}_A^\nu({\bf y})\d y^A
 \wedge\d^{m-1}x_\nu)_y
 \label{iota}
 \eeq
 (let us recall that ${\rm p}^\nu_\nu$ denotes
 \dst\sum_{\nu=1}^m{\rm p}^\nu_\nu\) ).

 For every $y\in E$, we have that
 $$
 \iota
 (J^1E^*)_y =\{\gamma\in\Lambda^m\Tan_y^*E \ ;\ \inn (u_1)\inn
 (u_2)\gamma =0\ ,\ u_1,u_2\in{\rm V}_y(\pi)\} \equiv
 \Lambda_1^m\Tan_y^*E
 $$
 We will denote $\iota_0\colon J^1E^*\to\iota (J^1E^*)=
 \Lambda^m_1\Tan^*E=\bigcup_{y\in E} \{ (y,\alpha ) \ ;\
\alpha\in\Lambda_1^m\Tan_y^*E\}$
 the restriction of $\iota$ onto its image.

\begin{definition}
\ben
\item
The bundle (over $E$) $$ {\cal M}\pi:=\Lambda_1^m\Tan^*E $$ will
be called the {\rm extended multimomentum bundle} associated with
the bundle $\pi\colon E\to M$. We denote the natural projections
by $\hat\tau^1\colon{\cal M}\pi\to E$ and
$\bar{\hat\tau}^1\colon{\cal M}\pi\to M$.
\item
The {\rm (first) extended Legendre map}
 associated with a Lagrangian density $\Lag$
 is the $\Cinfty$-map $$
\widehat{{\cal F}\Lag}:=\iota_0\circ\widehat{{\rm F}\Lag} $$ The
{\rm (second) extended Legendre map} is the $\Cinfty$-map ${\cal
F}\Lag\colon J^1E \to {\cal M}\pi$ given by
 $$ \widetilde{{\cal
F}\Lag}=\widehat{{\cal F}\Lag}+\pi^*\Lag $$ \een
\end{definition}

Natural coordinates in ${\cal M}\pi$ are denoted by
$(x^\nu,y^A,p,p_A^\nu)$, and for every ${\bf y}\in J^1E^*$
we have
$$
\iota\colon {\bf y}\equiv (x^\nu,y^A,{\rm p}_A^\nu,{\rm p}_\nu^\eta)
\mapsto \hat y\equiv (x^\nu,y^A,{\rm p}_A^\nu,{\rm p}=p_\nu^\nu)
$$
 The local expressions of the extended Legendre maps are
 \beq
\begin{array}{ccccccc}
\widehat{{\cal F}\Lag}^*x^\nu = x^\nu &\ ,\ &
\widehat{{\cal F}\Lag}^*y^A = y^A &\ , \ &
\widehat{{\cal F}\Lag}^*p_A^\nu = \derpar{\lag}{v^A_\nu} &\ , \ &
\widehat{{\cal F}\Lag}^*p=-v^A_\nu\derpar{\lag}{v^A_\nu}
\\
\widetilde{{\cal F}\Lag}^*x^\nu = x^\nu &\  ,\ &
\widetilde{{\cal F}\Lag}^*y^A = y^A &\  , \ &
\widetilde{{\cal F}\Lag}^*p_A^\nu = \derpar{\lag}{v^A_\nu}  &\  , \ &
\widetilde{{\cal F}\Lag}^*p = \lag -v^A_\nu\derpar{\lag}{v^A_\nu}
\end{array}
\label{extlt}
 \eeq
{\bf Remarks}:
 \bit
 \item
 It can be proved \cite{CCI-91}, \cite{EMR-99} that
 ${\cal M}\pi\equiv\Lambda_1^m\Tan^*E$
 is canonically isomorphic to ${\rm Aff}(J^1E,\Lambda^m\Tan^*M)$.
 \item
It is interesting to point out that, as $\Theta_{\Lag}$ and
$\theta_{\Lag}$ can be thought of as $m$-forms on $J^1E$ along the
projection $\pi^1\colon J^1E\to E$, the extended Legendre maps can
be defined as
 \beann
 (\widehat{{\cal F}\Lag}(\bar y))
 (\moment{Z}{1}{m})&=& (\theta_{\Lag})_{\bar y}(\moment{\bar Z}{1}{m})
\\
(\widetilde{{\cal F}\Lag}(\bar y))(\moment{Z}{1}{m})&=&
(\Theta_{\Lag})_{\bar y}(\moment{\bar Z}{1}{m}) \eeann
 where $\bar y\in J^1E$,
 $\moment{Z}{1}{m}\in\Tan_{\pi^1(\bar y)}E$, and $\moment{\bar
 Z}{1}{m}\in\Tan_{\bar y}J^1E$ are such that
 $\Tan_{\bar y}\pi^1\bar Z_\nu=Z_\nu$.

 In addition, the (second) extended Legendre map can also be defined
 as the ``first order  vertical Taylor approximation to
 $\lag$'' \cite{CCI-91}, \cite{GIMMSY-mm}.
 \eit

 For the construction of the last multimomentum bundle,
 observe that the sections of the bundle
 $\pi^*\Lambda^m\Tan^*M\to E$ are the $\pi$-semibasic $m$-forms on
 $E$; therefore we introduce the notation
 $\Lambda_0^m\Tan^*E\equiv\pi^*\Lambda^m\Tan^*M$, and then:

\begin{definition}
\ben
\item
 The bundle (over $E$)
 $$
 J^1\pi^*:=
 \Lambda_1^m\Tan^*E/\Lambda_0^m\Tan^*E\equiv{\cal M}\pi/\Lambda_0^m\Tan^*E
 $$
 will be called the {\rm restricted multimomentum bundle}
 associated with the bundle $\pi\colon E\to M$.
 We denote the natural projections by
 $\tau^1\colon J^1\pi^*\to E$ and
 $\bar\tau^1:=\pi\circ\tau^1\colon J^1\pi^*\to M$.
\item
 The {\rm restricted Legendre map}
 associated with a Lagrangian density $\Lag$
 is the $\Cinfty$-map
 $$
 {\cal F}\Lag:=\mu\circ\widehat{{\cal F}\Lag}=
 \mu\circ\widetilde{{\cal F}\Lag}
 $$
 where $\mu\colon{\cal M}\pi\longrightarrow J^1\pi^*$ is
 the natural projection. \een
\end{definition}

Natural coordinates in $J^1\pi^*$ will also be denoted as
$(x^\nu,y^A,p_A^\nu)$. As is evident in this system, the local
expression of the restricted Legendre map is
 \beq
 {\cal F}\Lag^*x^\nu = x^\nu \quad , \quad {\cal F}\Lag^*y^A = y^A
 \quad , \quad {\cal F}\Lag^*p_A^\nu = \derpar{\lag}{v^A_\nu}
 \label{restlt}
 \eeq

\begin{teor}
The multimomentum bundles $J^1\pi^*$ and $\Pi$ are canonically
diffeomorphic as vector bundles over $E$, and denoting this
diffeomorphism by ${\mit\Psi}\colon J^1\pi^*\to \Pi$, therefore
${\rm F}\Lag ={\cal F}\Lag\circ{\mit\Psi}$.
\end{teor}
 \proof
 Consider the diagram
$$
\begin{array}{ccc}
 J^1E^* &
 \begin{picture}(60,10)(0,0)
 \put(0,3){\vector(1,0){60}}
 \put(20,7){\mbox{$\mu\circ\iota_0$}}
 \end{picture} &
 J^1\pi^*
 \\
  \begin{picture}(10,60)(0,0)
 \put(3,60){\vector(0,-1){60}}
 \put(6,30){\mbox{$\delta$}}
 \end{picture} &
 \begin{picture}(60,60)(0,0)
 \put(0,60){\vector(1,-1){60}}
 \end{picture} &
 \begin{picture}(10,60)(0,0)
 \put(3,60){\vector(0,-1){60}}
 \put(6,30){\mbox{$\tau^1$}}
 \end{picture}
 \\
 \Pi &
 \begin{picture}(60,10)(0,0)
 \put(0,3){\vector(1,0){60}}
 \put(25,7){\mbox{$\rho^1$}}
 \end{picture} &
 E
\end{array}
$$
 We have that the maps $\mu\circ\iota_0$ and $\delta$ are
sobrejective, linear on the fibers, and restrict to the identity on
the base. On the other hand, for every $y\in E$, we have that
$\ker\,\delta_y=\ker\, (\mu\circ\iota_0)_y$ (as can be shown from
the corresponding expressions in coordinates). Hence we conclude
that $J^1\pi^*$ and $\Pi$ are canonically isomorphic as vector
bundles over $E$.

(See \cite{EMR-99} for another version of this proof,
and an explicit construction of ${\mit\Psi}$).
 \qed

 $\Pi$ and $J^1\pi^*$ are fiber bundles over
 $E$, then ${\mit\Psi}$ is a fiber-diffeomorphism
 (it is the identity on the base), whose
 local expression in natural coordinates in $J^1\pi^*$ and $\Pi$ is
 $$
 {\mit\Psi}^*x^\nu=x^\nu \ ,\
 {\mit\Psi}^*y_A=y_A \ ,\
 {\mit\Psi}^*{\rm p}^\nu_A=p^\nu_A \quad ;\quad (\forall \nu,A)
 $$

\subsection{Canonical forms}

 As is known \cite{CIL-98}, the {\sl multicotangent
 bundle} $\Lambda^m\Tan^*E$ is endowed with canonical forms: ${\bf
 \Theta}\in\df^m(\Lambda^m\Tan^*E)$ and the multisymplectic form
 ${\bf  \Omega}:=-\d{\bf \Theta}\in\df^{m+1}(\Lambda^m\Tan^*E)$.
 Then:

 \begin{definition}
 The {\rm canonical $m$ and $(m+1)$ forms} of $J^1E^*$ are
 $$
 \hat\Theta=\iota^*{\bf \Theta}\in\df^m(J^1E^*)
 \quad ,\quad
 \hat\Omega :=-\d\hat\Theta=\iota^*{\bf \Omega}\in\df^{m+1}(J^1E^*)
 $$
 \label{def4}
 \end{definition}

 On the other hand,
 observe that ${\cal M}\pi\equiv\Lambda^m_1\Tan^*E$ is a subbundle
 of the multicotangent bundle $\Lambda^m\Tan^*E$. Let
 $$
 \lambda\colon\Lambda^m_1\Tan^*E\hookrightarrow\Lambda^m\Tan^*E
 $$
 be the natural imbedding (hence $\lambda\circ\iota_0=\iota$). Then:

 \begin{definition}
 The {\rm canonical $m$ and $(m+1)$ forms} of ${\cal M}\pi$
 ({\rm multimomentum Liouville $m$ and $(m+1)$ forms} of ${\cal M}\pi$)
 are
 $$
 \Theta :=\lambda^* {\bf \Theta}\in\df^m({\cal M}\pi)
 \quad ,\quad
 \Omega =-\d\Theta=\lambda^*{\bf \Omega}\in\df^{m+1}({\cal M}\pi)
 $$
 \label{def7}
 \end{definition}

\bit
\item
 Of course, $\hat\Theta=\iota_0^*\Theta$ and
 $\hat\Omega=\iota_0^*\Omega$
\item
 $\Omega$ is $1$-nondegenerate, and hence $({\cal M}\pi,\Omega )$
 is a multisymplectic manifold.
\eit

 The canonical forms $\hat\Theta$ and $\Theta$
 can also be characterized as follows
 (see \cite{EMR-99}):
\bit
\item
 $\hat\Theta$ is the only $m$-form in $J^1E^*$, such that
 if ${\bf y}\in J^1E^*$, and $\moment{X}{1}{m}\in\Tan_{\bf y}J^1E^*$,
 then
 $$
 \hat\Theta({\bf y};X_1,\ldots ,X_m)=
 \iota ({\bf y})[\Tan_{\bf y}\hat\rho^1(X_1),\ldots ,
 \Tan_{\bf y}\hat\rho^1(X_m)]
 $$
\item
 In turn, considering the natural projection
$\hat\kappa^1\colon\Lambda^m_1\Tan^*E\to E$, then
 $$
 \Theta
((y,\alpha );\moment{X}{1}{m}):= \alpha
(y;\Tan_{(y,\alpha)}\hat\kappa^1(X_1),\ldots ,
\Tan_{(y,\alpha)}\hat\kappa^1 (X_m))
 $$
 for every $(y,\alpha)\in\Lambda^m_1\Tan^*E$ (where $y\in E$ and
$\alpha\in\Lambda^m_1\Tan_y^*E$), and $X_i\in\vf(\Lambda^m_1\Tan^*E)$.
\eit

 Bearing in mind the following diagram
$$
\begin{array}{ccc}
 & & {\cal M}\pi
 \\ &
 \begin{picture}(60,60)(0,0)
 \put(0,0){\vector(1,1){60}}
 \put(25,35){\mbox{$\iota_0$}}
 \end{picture} &
 \begin{picture}(10,60)(0,0)
 \put(3,60){\vector(0,-1){60}}
 \put(6,30){\mbox{$\hat\tau^1$}}
 \end{picture}
 \\
 J^1E^* &
 \begin{picture}(60,10)(0,0)
 \put(0,3){\vector(1,0){60}}
 \put(25,7){\mbox{$\hat\rho^1$}}
 \end{picture} &
 E
\end{array}
$$
 we observe that the map $\iota_0\colon J^1E^*\to{\cal M}\pi$
 is a {\sl form along the projection}
 $\hat\rho^1\colon J^1E^*\to E$. Then:

 \begin{lem}
 $\hat\Theta=\iota_0^*\Theta=\hat\rho^{1*}\iota_0$.
\label{lemauno}
 \end{lem}
\proof
 Let ${\bf y}\in J^1E^*$, and
 $\moment{X}{1}{m}\in\Tan_{\bf y}J^1E^*$.
 We have
\beann
 \iota_0^*\Theta({\bf y};\moment{X}{1}{m}) &=&
 \Theta(\iota_0({\bf y});
 \Tan_{\bf y}\iota_0(X_1),\ldots ,\Tan_{\bf y}\iota_0(X_m))
\\ &=&
 (\iota_0({\bf y})[\Tan_{\bf y}(\hat\tau^1\circ\iota_0)(X_1),\ldots ,
 \Tan_{\bf y}(\hat\tau^1\circ\iota_0)(X_m)]
\\ &=&
 (\iota_0({\bf y})(\Tan_{\bf y}\hat\rho^1(X_1),\ldots ,
 \Tan_{\bf y}\rho^1(X_m))=
(\hat\rho^{1*}\iota)({\bf y};\moment{X}{1}{m})
\eeann
\qed

 In natural coordinates in $J^1E^*$ and ${\cal M}\pi$,
 the local expressions of these forms are
  \beann
 \hat\Theta =
 {\rm p}^\eta_\eta\d^mx+{\rm p}^\nu_A\d y^A\wedge\d^{m-1}x_\nu
 &\ , \ &
 \hat\Omega = -\d {\rm p}^\eta_\eta\wedge\d^mx-\d {\rm
 p}^\nu_A\wedge\d y^A\wedge\d^{m-1}x_\nu
 \\
 \Theta=p\d^mx+p^\nu_A\d y^A\wedge\d^{m-1}x_\nu
 &\ , \ &
 \Omega=-\d p\wedge\d^mx-\d p^\nu_A\wedge\d y^A\wedge\d^{m-1}x_\nu
 \eeann

 \begin{prop}
 Let $\ls$ be a Lagrangian system.
 Let $\widehat{{\rm F}\Lag}$ be the generalized Legendre map,
 and $\widehat{{\cal F}\Lag}$ and $\widetilde{{\cal F}\Lag}$
 the extended Legendre maps. Then
  \beann
 \widehat{{\rm F}\Lag}^*\hat\Theta=\Theta_{\Lag}-\Lag =\theta_{\Lag}
 &\quad ;\quad&
 \widehat{{\rm }{\rm F}\Lag}^*\hat\Omega=\Omega_{\Lag}-\d\Lag=-\d\theta_\Lag
 \\
 \widehat{{\cal F}\Lag}^*\Theta=\Theta_{\Lag}-\Lag =
 \theta_{\Lag}
  &\quad ;\quad&
 \widehat{{\cal F}\Lag}^*\Omega=\Omega_{\Lag}-\d\Lag=-\d\theta_\Lag
 \\
 \widetilde{{\cal F}\Lag}^*\Theta=\Theta_{\Lag}
 &\quad ;\quad&
 \widetilde{{\cal F}\Lag}^*\Omega=\Omega_{\Lag}
 \eeann
 \label{propositio5}
 \end{prop}

\subsection{Regular and singular systems}

\begin{definition}
Let $\ls$ be a Lagrangian system.
 \ben
\item
$\ls$ is said to be a {\rm regular} or {\rm non-degenerate}
Lagrangian system if ${\cal F}\Lag$, and hence, ${\rm F}\Lag$ are
local diffeomorphisms.

As a particular case, $\ls$ is said to be a {\rm hyper-regular}
Lagrangian system if ${\cal F}\Lag$, and hence ${\rm F}\Lag$, are
global diffeomorphisms.
\item
Elsewhere $\ls$ is said to be a {\rm singular} or {\rm degenerate}
Lagrangian system.
 \een
\end{definition}

The matrix of the tangent maps ${\cal F}\Lag_*$ and
 ${\rm F}\Lag_*$ in a natural coordinate system is
 \beq
 \left(\matrix{{\rm Id} & 0 & 0 \cr 0 & {\rm Id} & 0 \cr
 \frac{\partial^2\lag}{\partial x^\nu\partial v^A_\mu} &
 \frac{\partial^2\lag}{\partial y^B\partial v^A_\mu} &
 \frac{\partial^2\lag}{\partial v^B_\nu\partial v^A_\mu}\cr}\right)
 \label{matrixfl}
 \eeq
 where the sub-matrix
 \dst\left(\frac{\partial^2\lag}{\partial v^B_\nu\partial v^A_\mu}\right)\)
 is the {\sl partial Hessian matrix} of $\lag$.
 Then, the regularity condition is equivalent
 to demanding that this matrix is regular everywhere in $J^1E$.
 This fact establishes the relation to the concept of regularity
 given in an equivalent way by saying that
 a Lagrangian system $\ls$ is {\sl regular} if
 $\Omega_{\Lag}$ is $1$-nondegenerate.
 (See also \cite{CCI-91} for a different definition of this concept).

 \begin{prop}
{\rm (See \cite{EMR-99} and \cite{LMM-96})}. Let $\ls$ be a
hyper-regular Lagrangian system. Then
 \ben
\item
$\widehat{{\rm F}\Lag}(J^1E)$ is a $m^2$-codimensional imbedded
submanifold of $J^1E^*$, which is transverse to the projection
$\delta$.
\item
 $\widehat{{\cal F}\Lag}(J^1E)$ and
 $\widetilde{{\cal F}\Lag}(J^1E)$ are 1-codimensional imbedded submanifolds
 of ${\cal M}\pi$, which are transverse to the projection $\mu$.
\item
 The manifolds $J^1\pi^*$, $\widehat{{\cal F}\Lag}(J^1E)$,
 $\widetilde{{\cal F}\Lag}(J^1E)$, $\widehat{{\rm F}\Lag}(J^1E)$
 and $\Pi$ are diffeomorphic.

Hence, $\widehat{{\rm F}\Lag}$, $\widehat{{\cal F}\Lag}$ and
$\widetilde{{\cal F}\Lag}$ are diffeomorphisms on their images;
and the maps $\mu$, restricted to $\widehat{{\cal F}\Lag}(J^1E)$
or to $\widetilde{{\cal F}\Lag}(J^1E)$, and $\iota_0$ and
$\delta$, restricted to $\widehat{{\rm F}\Lag}(J^1E)$, are also
diffeomorphisms. \een \label{hrprop}
\end{prop}

In this way we have the following  diagram
 \beq
\begin{array}{ccccccccc}
\begin{picture}(15,180)(0,0)
\put(0,85){\mbox{$J^1E$}}
\end{picture}
&
\begin{picture}(65,180)(0,0)
\put(20,137){\mbox{${\cal F}\Lag$}}
\put(35,118){\mbox{$\widetilde{{\cal F}\Lag}$}}
\put(36,88){\mbox{$\widehat{{\cal F}\Lag}$}}
\put(36,51){\mbox{$\widehat{{\rm F}\Lag}$}}
\put(20,30){\mbox{${\rm F}\Lag$}}
 \put(0,93){\vector(1,1){70}}
\put(0,89){\vector(2,1){70}}
 \put(0,85){\vector(1,0){70}}
\put(0,81){\vector(3,-1){70}}
 \put(0,77){\vector(1,-1){70}}
\end{picture}
&
\begin{picture}(35,180)(0,0)
\put(16,0){\mbox{$\Pi$}}
 \put(9,42){\mbox{$J^1E^*$}}
\put(12,85){\mbox{${\cal M}\pi$}}
 \put(12,129){\mbox{${\cal M}\pi$}}
 \put(11,170){\mbox{$J^1\pi^*$}}
\put(20,142){\vector(0,1){25}}
 \put(6,98){\vector(0,1){68}}
\put(20,98){\vector(0,1){25}}
 \put(20,56){\vector(0,1){25}}
\put(20,36){\vector(0,-1){25}}
 \put(11,152){\mbox{$\mu$}}
\put(-2,140){\mbox{$\mu$}}
 \put(9,109){\mbox{$\mu'$}}
\put(11,64){\mbox{$\iota_0$}}
 \put(11,20){\mbox{$\delta$}}
\end{picture}
&
\begin{picture}(10,180)(0,0)
\put(0,165){\vector(0,-1){155}}
 \put(5,109){\mbox{${\mit\Psi}$}}
\end{picture}
&
 \begin{picture}(65,180)(0,0)
 \put(22,135){\mbox{$\lambda$}}
 \put(0,133){\vector(1,0){65}}
 \put(0,88){\vector(1,0){65}}
 \put(22,90){\mbox{$\lambda$}}
 \put(0,48){\vector(2,1){65}}
 \put(40,55){\mbox{$\iota$}}
\end{picture}
&
\begin{picture}(40,180)(0,0)
\put(0,129){\mbox{$\Lambda^m\Tan^*E$}}
\put(0,85){\mbox{$\Lambda^m\Tan^*E$}}
\end{picture}
\end{array}
\label{diag1}
 \eeq
 where the map
 $\mu'\colon{\cal M}\pi\to{\cal M}\pi$ is defined by the relation
 $$
 \mu':=\widetilde{{\cal F}\Lag}\circ{\cal F}\Lag^{-1}\circ\mu
 $$
 and it satisfies that $\mu\circ\mu'=\mu$.
 Observe also that the restriction
 $\mu'\colon \widehat{{\cal F}\Lag}(J^1E)\subset{\cal M}\pi
 \to \widetilde{{\cal F}\Lag}(J^1E)\subset{\cal M}\pi$,
 is a diffeomorphism, which is also defined by the relation
 $\widetilde{{\cal F}\Lag}=\mu'\circ\widehat{{\cal F}\Lag}$.

For dealing with singular Lagrangians, we must assume minimal
``regularity'' conditions. Hence we introduce the following
terminology:

 \begin{definition}
 A singular Lagrangian system $\ls$ is said to be
 {\rm almost-regular} if:
 \ben
  \item
 ${\cal P}:={\cal F}\Lag (J^1E)$ and $P:={\rm F}\Lag (J^1E)$ are
 closed submanifolds of $J^1\pi^*$ and $\Pi$, respectively.

 (We will denote the corresponding imbeddings by
 $\jmath_0\colon {\cal P}\hookrightarrow J^1\pi^*$
 and  $\j_0\colon P\hookrightarrow \Pi$).
 \item
 ${\cal F}\Lag$, and hence ${\rm F}\Lag$, are submersions onto
 their images (with connected fibers).
 \item
 For every $\bar y\in J^1E$, the fibers ${\cal F}\Lag^{-1}({\cal
 F}\Lag (\bar y))$ and hence ${\rm F}\Lag^{-1}({\rm F}\Lag (\bar
 y))$ are connected submanifolds of $J^1E$.
  \een
 \end{definition}

(This definition is equivalent to that in reference \cite{LMM-96},
but slightly different from that in references \cite{GMS-97} and
\cite{Sd-95}).

Let $\ls$ be an almost-regular Lagrangian system. Denote
 $$
 \hat{\cal P}:=\widehat{{\cal F}\Lag}(J^1E) \quad ,\quad
 \tilde{\cal P}:=\widetilde{{\cal F}\Lag}(J^1E) \quad ,\quad
 \hat P:=\widehat{{\rm F}\Lag}(J^1E)
 $$
 Let
 $\hat\jmath_0\colon\hat{\cal P}\hookrightarrow{\cal M}\pi$,
 $\tilde\jmath_0\colon\tilde{\cal P}\hookrightarrow{\cal M}\pi$,
 $\hat{\j}_0\colon\hat P\hookrightarrow J^1E^*$
 be the canonical inclusions, and
 $$
\hat\mu\colon\hat{\cal P}\to{\cal P} \quad ,\quad
\tilde\mu\colon\tilde{\cal P}\to{\cal P} \quad ,\quad
\hat\iota_0\colon\hat P\to\hat{\cal P} \quad ,\quad
\hat\delta\colon \hat P\to P \quad ,\quad
{\mit\Psi}_0\colon\hat{\cal P}\to P
 $$
 the restrictions of the
maps $\mu$, $\iota_0$, $\delta$ and the diffeomorphism
${\mit\Psi}$, respectively. Finally, define the restriction mappings
 $$
 {\cal F}\Lag_0\colon J^1E\to{\cal P} \ ,\
\widetilde{{\cal F}\Lag}_0\colon J^1E\to\tilde{\cal P} \ ,\
\widehat{{\cal F}\Lag}_0\colon J^1E\to\hat{\cal P} \ ,\
\widehat{{\rm F}\Lag}_0\colon J^1E\to\hat P \ ,\
 {\rm F}\Lag_0\colon J^1E\to P
 $$

\begin{prop}
{\rm (See \cite{EMR-99},\cite{LMM-96},\cite{LMM-96b})}.
 Let $\ls$ be an almost-regular Lagrangian system. Then:
 \ben
 \item
The maps ${\mit\Psi}_0$ and $\tilde\mu$ are diffeomorphisms.
\item
For every $\bar y\in J^1E$, \beq \widetilde{{\cal
F}\Lag_0}^{-1}(\widetilde{{\cal F}\Lag_0}(\bar y))= {\cal
F}\Lag_0^{-1}({\cal F}\Lag_0(\bar y))= {\rm F}\Lag_0^{-1}({\rm
F}\Lag_0(\bar y)) \label{uno} \eeq
 \item
$\tilde{\cal P}$ and  $\hat{\cal P}$ are submanifolds of ${\cal
M}\pi$, $\hat P$ is a submanifold of $J^1E^*$, and
$\tilde\jmath_0\colon\tilde{\cal P}\hookrightarrow{\cal M}\pi$,
$\hat\jmath_0\colon\hat{\cal P}\hookrightarrow{\cal M}\pi$,
$\hat{\j}_0\colon\hat P\hookrightarrow J^1E^*$ are imbeddings.
\item
The restriction mappings $\widetilde{{\cal F}\Lag}_0$,
$\widehat{{\cal F}\Lag}_0$ and $\widehat{{\rm F}\Lag}_0$ are
submersions with connected fibers. \een \label{arprop}
\end{prop}

Thus we have the  diagram
 \beq
\begin{array}{ccccccc}
\begin{picture}(15,180)(0,0)
\put(0,85){\mbox{$J^1E$}}
\end{picture}
&
 \begin{picture}(65,180)(0,0)
 \put(20,137){\mbox{${\cal F}\Lag_0$}}
 \put(35,118){\mbox{$\widetilde{{\cal F}\Lag_0}$}}
 \put(36,91){\mbox{$\widehat{{\cal F}\Lag_0}$}}
 \put(36,51){\mbox{$\widehat{{\rm F}\Lag_0}$}}
 \put(20,30){\mbox{${\rm F}\Lag_0$}}
 \put(0,93){\vector(1,1){70}}
 \put(0,89){\vector(2,1){70}}
 \put(0,87){\vector(1,0){70}}
 \put(0,81){\vector(3,-1){70}}
 \put(0,77){\vector(1,-1){70}}
\end{picture}
&
\begin{picture}(25,180)(0,0)
\put(17,0){\mbox{$P$}}
 \put(17,42){\mbox{$\hat P$}}
\put(17,85){\mbox{$\hat{\cal P}$}}
 \put(17,129){\mbox{$\tilde{\cal P}$}}
 \put(17,170){\mbox{${\cal P}$}}
\put(20,142){\vector(0,1){25}}
 \put(20,98){\vector(0,1){25}}
\put(5,98){\vector(0,1){68}}
 \put(20,56){\vector(0,1){25}}
\put(20,36){\vector(0,-1){25}}
 \put(11,152){\mbox{$\tilde\mu$}}
\put(-5,140){\mbox{$\hat\mu$}}
 \put(7,109){\mbox{$\hat\mu'$}}
\put(11,64){\mbox{$\hat\iota_0$}}
 \put(11,20){\mbox{$\hat\delta$}}
\end{picture}
&
\begin{picture}(6,180)(0,0)
\put(0,165){\vector(0,-1){155}} \put(3,65){\mbox{${\mit\Psi}_0$}}
\end{picture}
&
 \begin{picture}(50,180)(0,0)
 \put(0,171){\vector(1,0){50}}
 \put(22,175){\mbox{$\jmath_0$}}
 \put(0,131){\vector(1,0){50}}
 \put(22,135){\mbox{$\tilde\jmath_0$}}
 \put(0,90){\vector(1,0){50}}
 \put(22,94){\mbox{$\hat\jmath_0$}}
 \put(0,46){\vector(1,0){50}}
 \put(22,50){\mbox{$\hat{\j}_0$}}
 \put(0,2){\vector(1,0){50}}
 \put(22,6){\mbox{$\j_0$}}
\end{picture}
&
\begin{picture}(35,180)(0,0)
\put(16,0){\mbox{$\Pi$}}
 \put(9,42){\mbox{$J^1E^*$}}
\put(12,85){\mbox{${\cal M}\pi$}}
 \put(12,129){\mbox{${\cal M}\pi$}}
 \put(11,170){\mbox{$J^1\pi^*$}}
\put(20,142){\vector(0,1){25}}
 \put(6,98){\vector(0,1){68}}
 \put(20,56){\vector(0,1){25}}
\put(20,36){\vector(0,-1){25}}
 \put(11,152){\mbox{$\mu$}}
\put(-2,140){\mbox{$\mu$}}
\put(11,64){\mbox{$\iota_0$}}
 \put(11,20){\mbox{$\delta$}}
\end{picture}
&
\begin{picture}(10,180)(0,0)
\put(0,165){\vector(0,-1){155}}
 \put(5,85){\mbox{${\mit\Psi}$}}
\end{picture}
\end{array}
\label{diag2}
 \eeq
 where $\hat\mu'\colon \hat{\cal P}\to\tilde{\cal P}$
 is defined by the relation
 $\hat\mu':=\tilde\mu^{-1}\circ\hat\mu$.

The maps $\hat\mu$ and $\hat\mu'$ are not diffeomorphisms in
general, since ${\rm rank}\, \widehat{{\cal F}\Lag}_0\geq {\rm
rank}\, \widetilde{{\cal F}\Lag}_0= {\rm rank}\, {\cal F}\Lag_0$,
as is evident from the analysis of the corresponding Jacobian
matrices.

\begin{prop}
 Let $\ls$ be an almost-regular Lagrangian system. Then:
 $$
\ker\, {\rm F}\Lag_*=\ker\, {\cal F}\Lag_*=\widetilde{\ker\, {\cal F}\Lag}_*=
\ker\,\Omega_\Lag\cap\vf^{{\rm V}(\pi^1)}(J^1E)
 $$
\label{kerfl}
\end{prop}
\proof
 The first two equalities are immediate,
 since $P:={\rm F}\Lag(J^1E)$, ${\cal P}:={\cal F}\Lag(J^1E)$
 and $\tilde P:=\widetilde{{\cal F}\Lag}(J^1E)$ are diffeomorphic.

 For the last equality,
 as ${\rm F}\Lag$, ${\cal F}\Lag$ and $\widetilde{{\cal F}\Lag}$
 are the identity on the basis $E$ of
 the bundle $\pi^1\colon J^1E\to E$, first we have that
 $$
 \ker\, {\rm F}\Lag_*=\ker\, {\cal F}\Lag_*=
 \ker\,\widetilde{{\cal F}\Lag}\subset\vf^{{\rm V}(\pi^1)}(J^1E)
 $$
 (and this relation holds also for the other Legendre maps).
 Then, for every
 $X\in\ker\,\widetilde{{\cal F}\Lag}_*$ we have
$$
\inn(X)\Omega_\Lag=\inn(X)(\widetilde{{\cal F}\Lag}^*\Omega)=
\widetilde{{\cal F}\Lag}^*[\inn(\widetilde{{\cal F}\Lag}_*X)\Omega]=0
$$
and hence
 $$
 \ker\,\widetilde{{\cal F}\Lag}_*=\ker\, {\cal F}\Lag_*=\ker\, {\rm F}\Lag_*
 \subset\ker\,\Omega_\Lag\cap\vf^{{\rm V}(\pi^1)}(J^1E)
 $$
Conversely, if $X\in\vf^{{\rm V}(\pi^1)}(J^1E)$,
in a natural system of coordinates in $J^1E$ we have
\dst X=f^B_\nu\derpar{}{v^B_\nu}\) , and if, in addition,
 $X\in{\rm ker}\,\Omega_\Lag$, we obtain
$$
 0=\inn(X)\Omega_\Lag=
 -\frac{\partial^2\lag}{\partial v^B_\eta\partial v^A_\nu}
 f^B_\eta\d y^A\wedge\d^{m-1}x_\nu+
 \frac{\partial^2\lag}{\partial v^B_\eta\partial v^A_\nu}v^A_\nu f^B_\eta\d^mx
$$
 this is equivalent to demanding that
 \dst f^B_\eta\frac{\partial^2\lag}{\partial v^B_\eta\partial v^A_\nu}=0\) ,
 and this is the condition which characterizes locally the vector
 fields belonging to
 $\ker\,{\rm F}\Lag_*=\ker\, {\cal F}\Lag_*=\ker\,\widetilde{{\cal F}\Lag}_*$
 (see (\ref{matrixfl})).
\qed

\section{Hamiltonian formalism in the reduced multimomentum bundle}

 \subsection{Hamiltonian systems}
 \protect\label{dsc}

(Compare this presentation with references \cite{GMS-97},
\cite{GMS-99} and \cite{Sd-95}).

 The more standard way for constructing Hamiltonian systems in $\Pi$
 consists in using sections of the projection $\delta$, and it
 is similar to that developed in \cite{CCI-91}
 for the Hamiltonian formalism in $J^1\pi^*$
 (which we will review later). Thus:

 \begin{definition}
 Consider the bundle $\bar\rho^1\colon\Pi\to M$.
\ben
\item
 A section $h_\delta\colon\Pi\to J^1E^*$ of the projection
 $\delta$ is called
 a {\rm Hamiltonian section of $\delta$}.
\item
 The differentiable forms
 $$
 \Theta_{h_\delta}:=h_\delta^*\hat\Theta=
 (\iota_0\circ h_\delta)^*\Theta
 \quad ,\quad
 \Omega_{h_\delta}:=-\d\Theta_{h_\delta}=h_\delta^*\hat\Omega=
 (\iota_0\circ h_\delta)^*\Omega
 $$
 are called the {\rm Hamilton-Cartan $m$ and $(m+1)$ forms} of $\Pi$
 associated with the Hamiltonian section $h_\delta$.
\item
 The couple $\hspi$ is said to be a {\rm Hamiltonian system}.
\een
 \end{definition}

 Using charts of natural coordinates in $\Pi$ and $J^1E^*$,
 a Hamiltonian section is specified by a set of local functions
 $H^\eta_\nu\in\Cinfty (U)$, $U\subset\Pi$, such that
 \beq
 h_\delta(x^\nu,y^A,{\rm p}^\nu_A)\equiv
 (x^\nu,y^A,{\rm p}^\eta_\nu=-H^\eta_\nu(x^\gamma,y^B,{\rm p}^\gamma_B),
 {\rm p}^\nu_A)
 \label{hdelta}
 \eeq
 Then, the local expressions of these Hamilton-Cartan forms are

 \bea
 \Theta_{h_\delta} &=&
 {\rm p}_A^\nu\d y^A\wedge\d^{m-1}x_\nu-H_{h_\delta}\d^mx
 \nonumber
 \\
 \Omega_{h_\delta} &=& -\d{\rm p}_A^\nu\wedge\d y^A\wedge\d^{m-1}x_\nu
 +\d H_{h_\delta}\wedge\d^mx
 \label{omegaH0}
 \eea
 where $H_{h_\delta}\equiv H^\nu_\nu$ is a
 {\sl local Hamiltonian function} associated with the Hamiltonian
 section $h_\delta$.

 As $\iota_0$ is a submersion, we can state the following:

 \begin{definition}
 There is a natural equivalence relation in the set of Hamiltonian
 sections of $\delta$, which is defined as follows:
 two Hamiltonian sections $h^1_\delta,h^2_\delta$ are
 {\rm equivalent} if
 $\iota_0\circ h_\delta^1=\iota_0\circ h_\delta^2$.
 We denote by $\{ h_\delta\}$ the equivalence classes of this
 relation.
 \label{def16}
 \end{definition}

 {\bf Remarks}:
\bit
\item
 Of course, Hamiltonian sections belonging to the same
 equivalence class give the same Hamilton-Cartan forms,
 and hence the same Hamiltonian system.
\item
 Observe that all the Hamiltonian sections
 of the same equivalence class have the same local
 Hamiltonian function $H_{h_\delta}\equiv H^\nu_\nu$
 (in the same open set $U\subset \Pi$).
\eit

 There is a relation between sections of $\mu$ and of $\delta$. In fact:

 \begin{prop}
 There is a bijective correspondence between the set of sections
 of the projection $\mu\colon{\cal M}\pi\to J^1\pi^*$
 and the set of equivalence classes of sections of the projection
 $\delta\colon J^1E^*\to \Pi$.
 \label{firstprop}
 \end{prop}
 \proof
 In fact, this correspondence is established by the commutativity
 of the following diagram
$$
\begin{picture}(160,60)(0,0)
\put(-5,0){\mbox{$J^1\pi^*$}}
\put(10,37){\vector(0,-1){25}}
\put(-23,20){\mbox{${\rm Id}_{J^1\pi^*}$}}
\put(-5,39){\mbox{$J^1\pi^*$}}
\put(20,5){\vector(2,1){25}}
\put(40,5){\mbox{$h_\mu$}}
\put(47,27){\vector(-2,1){25}}
\put(40,35){\mbox{$\mu$}}
\put(50,20){\mbox{${\cal M}\pi$}}
\put(97,22){\vector(-1,0){25}}
\put(83,26){\mbox{$\iota_0$}}
\put(101,20){\mbox{$J^1E^*$}}
\put(125,5){\mbox{$h_\delta$}}
\put(155,5){\vector(-2,1){25}}
\put(127,32){\mbox{$\delta$}}
\put(130,27){\vector(2,1){25}}
\put(160,0){\mbox{$\Pi$}}
\put(160,39){\mbox{$\Pi$}}
\put(170,20){\mbox{${\rm Id}_\Pi$}}
\put(165,37){\vector(0,-1){25}}
\put(25,45){\vector(1,0){125}}
\put(85,47){\mbox{${\mit\Psi}$}}
\put(82,2){\mbox{${\mit\Psi}$}}
\put(25,0){\vector(1,0){125}}
\end{picture}
$$
 that is, a section
 $h_\mu\colon J^1\pi^*\to{\cal M}\pi$ and a class
 $\{ h_\delta\}\colon\Pi\to  J^1E^*$ are in correspondence if,
 and only if,
 $$
 h_\mu=\iota_0\circ h_\delta\circ{\mit\Psi}
\quad , \quad \mbox{\rm for every $h_\delta\in\{ h_\delta\}$}
 $$
 and this correspondence is one-to-one.
 \qed

 Now we can study the structure of the set of Hamilton-Cartan forms,
 and hence of Hamiltonian systems.
 So, for every Hamiltonian section
 $h_\delta$ of $\delta$, consider the  diagram
$$
 \begin{array}{ccc}
 & & {\cal M}\pi
 \\ &
 \begin{picture}(140,60)(0,0)
 \put(0,0){\vector(2,1){140}}
 \put(30,35){\mbox{$\iota_0\circ h_\delta$}}
 \put(90,0){\vector(1,1){55}}
 \put(100,27){\mbox{$\iota_0$}}
 \end{picture}
 &
 \begin{picture}(20,60)(0,0)
 \put(8,55){\vector(0,-1){55}}
 \put(12,25){\mbox{$\hat\tau^1$}}
 \end{picture}
 \\
 \Pi &
 \begin{picture}(50,12)(0,0)
 \put(0,7){\vector(1,0){50}}
 \put(20,12){\mbox{$h_\delta$}}
 \put(50,3){\vector(-1,0){50}}
 \put(25,-10){\mbox{$\delta$}}
 \end{picture}
 \quad  J^1E^* \quad
 \begin{picture}(50,10)(0,0)
 \put(0,3){\vector(1,0){50}}
 \put(25,9){\mbox{$\hat\rho^1$}}
 \end{picture} &
 E
 \\ &
 \begin{picture}(150,1)(0,0)
 \put(0,1){\vector(1,0){150}}
 \put(70,-12){\mbox{$\rho^1$}}
  \end{picture} &
\\ &
 \begin{picture}(140,60)(0,0)
 \put(0,60){\vector(3,-1){140}}
 \put(70,17){\mbox{$\bar\rho^1$}}
 \end{picture}
 &
 \begin{picture}(20,50)(0,0)
 \put(8,50){\vector(0,-1){35}}
 \put(12,30){\mbox{$\pi$}}
 \put(4,0){\mbox{$M$}}
 \end{picture}
\end{array}
$$

 \begin{lem}
 Let $h^1_\delta,h^2_\delta\colon\Pi\to J^1E^*$ be two
 Hamiltonian sections of $\delta$, then:
 \ben
 \item
 $ h_\delta^{1*}\hat\Theta-h_\delta^{2*}\hat\Theta=
 (\iota_0\circ h_\delta^1)^*\Theta-(\iota_0\circ h_\delta^2)^*\Theta=
 \rho^{1*}(\iota_0\circ h_\delta^1-\iota_0\circ h_\delta^2)$.
 \item
 $\rho^{1*}(\iota_0\circ h_\delta^1-\iota_0\circ h_\delta^2)$
 is a $\bar\rho^1$-semibasic form in $\Pi$.
 \een
 \label{hcfsect}
 \end{lem}
 \proof
 \ben
 \item
 For every Hamiltonian section $h_\delta$,
 the map $\iota_0\circ h_\delta$
 is a form along the map $\hat\rho^1\circ h_\delta =\rho^1$.
 Therefore, following the same pattern as in Lemma \ref{lemauno},
 we obtain that
$$
 (\iota_0\circ h_\delta)^*\Theta=\rho^{1*}(\iota_0\circ h_\delta)
$$
 and hence the result is immediate.
 \item
 As ${\mit\Psi}\circ\mu\circ\iota_0=\delta$, for every section $h_\delta$
 we have that ${\mit\Psi}\circ\mu\circ\iota_0\circ h_\delta={\rm Id}_\Pi$,
 then $\mu\circ(\iota_0\circ h_\delta^1-\iota_0\circ h_\delta^2)=0_\Pi$,
 and therefore
 ${\rm Im}(\iota_0\circ h_\delta^1-\iota_0\circ h_\delta^2)\in
 \ker\,\mu=\Lambda_m^0\Tan^*E$ (that is, the $\bar\rho^1$-semibasic
 forms in $\Pi$.)
 \een
\qed

 From the local expressions (\ref{iota}) and (\ref{hdelta}),
 for every $\tilde y\in U\subset\Pi$, we have
 $$
 (\iota_0\circ h_\delta^1-\iota_0\circ h_\delta^2)(\tilde y)=
 (H_{h_\delta^1}-H_{h_\delta^2})(\tilde y) \d^mx\vert_{\tilde y}
 $$
 which is the local expression (at $\tilde y$) of a
 $\bar\rho^1$-semibasic form in $\Pi$.
 In addition, this is also the local expression of the form
 \beq
(\iota_0\circ h_\delta^1)^*\Theta-(\iota_0\circ h_\delta^2)^*\Theta=
 \Theta_{h_\delta^1}- \Theta_{h_\delta^2}\equiv{\mbox{\es H}}
 \label{basicsplit}
 \eeq

\begin{definition}
 A $\bar\rho^1$-semibasic form ${\mbox{\es H}}\in\df^m(\Pi)$
 is said to be a {\rm Hamiltonian density} in $\Pi$.

 It can be written as
 ${\mbox{\es H}}={\rm H}(\bar\rho^{1^*}\omega)$,
 where ${\rm H}\in\Cinfty (\Pi)$ is the
 {\sl global Hamiltonian function} associated with
 ${\mbox{\es H}}$ and $\omega$.
\end{definition}

 In this way, we have proved that two Hamiltonian systems generated by
 two Hamiltonian sections of $\delta$ belonging to different
 equivalence classes, are related by means of a Hamiltonian density. 
 We can state this result as follows:

\begin{teor}
 The set of Hamilton-Cartan $m$-forms associated with Hamiltonian sections
 of $\delta$ is an affine space modelled on the set of
 Hamiltonian densities in $\Pi$.
 \label{afinHC}
\end{teor}

 {\bf Remark}:
\bit
\item
 If $\hspi$ is a Hamiltonian system, taking into account (\ref{basicsplit})
 we have that every Hamiltonian section $h'_\delta$ (such that
 $h'_\delta\not\in\{h_\delta\}$) allows us
 to split globally the Hamilton-Cartan forms as
 \beq
 \Theta_{h_\delta}=\Theta_{h'_\delta}-{\mbox{\es H}}
 \quad ; \quad
  \Omega_{h_\delta}= \Omega_{h'_\delta}+\d{\mbox{\es H}}
 \label{splitomega0}
 \eeq
 If $(x^\nu ,y^A,{\rm p}_A^\nu )$ is a natural system of
 coordinates in $\Pi$, such that $\bar\rho^{1*}\omega=\d^mx$,
 and $H_{h'_\delta}(x^\nu,y^A,{\rm p}_A^\nu )$ is the
 local Hamiltonian function associated with the Hamiltonian section
 $h'_\delta$, and
 ${\mbox{\es H}}={\rm H}(x^\nu,y^A,{\rm p}_A^\nu )\d^mx$, then
 \bea
 \Theta_{h_\delta} &=&
 {\rm  p}^\nu_A\d y^A\wedge \d^{m-1}x_\nu-({\rm H}+H_{h'_\delta})\d^mx
 \nonumber
\\
 \Omega_{h_\delta} &=&
 -\d{\rm  p}^\nu_A\wedge\d y^A\wedge
 \d^{m-1}x_\nu+\d ({\rm H}+H_{h'_\delta})\wedge\d^mx
 \label{omegahpi0}
 \eea
 If $H_{h_\delta}$ is the local Hamiltonian function
 associated with the Hamiltonian section $h_\delta$, we have the relation
 ${\rm H}=H_{h_\delta}-H_{h'_\delta}$
 (in an open set $U$). Hence, taking this into account,
 the local expressions (\ref{omegaH0})
 and (\ref{omegahpi0}) are really the same thing.
\eit

\subsection{Hamiltonian sections, Hamiltonian densities and connections}
 \protect\label{hshdc}

 In order to obtain a Hamiltonian density using
 two Hamiltonian sections, it is usual for one of them
 to be a linear section induced by a connection.
 This is a natural procedure for different reasons. For instance,
 when we construct the Hamiltonian
 formalism associated with a Lagrangian system, the
 Hamiltonian density must be related with the density of Lagrangian
 energy and, as this last is defined by using a connection,
 this same connection must be used for constructing the
 related Hamiltonian density (see sections \ref{hsahrs} and \ref{vp}).

 Next, we are going to show how to define the linear Hamiltonian section
 induced by a connection. Hence, suppose that a connection $\nabla$
 has been chosen in $\pi\colon E\to M$. It
 allows us to identify ${\rm V}^*(\pi )$ as a subbundle of
 $\Tan^*E$. So, if $v_{\nabla}^*\colon {\rm V}^*(\pi )\to\Tan^*E$
 is the dual injection of the vertical projection $v_{\nabla}$
 induced by $\nabla$. Then:

 \begin{definition}
 The linear Hamiltonian section of $\delta$ induced by the
 connection $\nabla$ is the map
 $$
 \begin{array}{ccccc}
  h_\delta^\nabla&\colon& \Pi &\rightarrow&
 \pi^*\Tan M\otimes\Tan^*E\otimes\pi^*\Lambda^m\Tan^*M:=J^1E^*
 \\
 & &u_k\otimes\alpha^k\otimes\beta&\mapsto& u_k\otimes
 v^*_{\nabla}(\alpha^k)\otimes\beta
 \end{array}
 $$
 that is, $h_\delta^\nabla:=
 {\rm Id}_{\pi^*\Tan M}\otimes v^*_\nabla\otimes
 {\rm Id}_{\pi^*\Lambda^m\Tan^*M}$.
 \end{definition}

{\bf Remark}:
 \bit
 \item
 Two linear sections
 $ h_\delta^{\nabla_1}$ and $ h_\delta^{\nabla_2}$ induced by two
 different connections $\nabla_1$ and $\nabla_2$ cannot belong
 to the same equivalence class of Hamiltonian sections,
 as can be proved comparing their coordinate expressions.
 \eit

 If $\hat\Theta$ is the canonical $m$-form in $J^1E^*$, the forms
 \beq
 \Theta_{h_\delta^\nabla}:=(\iota_0\circ h_\delta^\nabla)^*{\bf\Theta}=
 h_\delta^{\nabla*}\Theta\in\df^m(\Pi)
 \quad ,\quad
 \Omega_{h_\delta^\nabla}:=-\d\Theta_{h_\delta^\nabla}=
 h_\delta^{\nabla*}\Omega\in\df^{m+1}(\Pi)
 \label{def12}
 \eeq
 are the {\sl Hamilton-Cartan $m$ and $m+1$ forms} of $\Pi$
 associated with $\nabla$.

{\bf Remark}:
\bit
\item
 It can be proved \cite{EMR-99} that
 the  Hamilton-Cartan $m$-form associated with
 a connection $\nabla$ is the unique form
 $\Theta_{h_\delta^\nabla}\in\df^m(\Pi)$ such that, if $\tilde y\in \Pi$ and
 $\moment{w}{1}{m}\in\Tan_{\tilde y}\Pi$, then
 \bea
 \Theta_{h_\delta^\nabla} (\tilde y;\moment{w}{1}{m}) &:=&
 (\iota_0\circ h_\delta^\nabla)(\tilde y)(\rho^1 (\tilde y);
 \Tan_{\tilde y}\rho^1(w_1),\ldots , \Tan_{\tilde y}\rho^1(w_m))
 \nonumber \\ &=&
 [\rho^{1*}(\iota_0\circ h_\delta^\nabla)](\tilde y;w_1,\ldots ,w_m)
 \label{caracteriza}
 \eea
 that is,  $\Theta_{h_\delta^\nabla}=\rho^{1*}(\iota_0\circ h_\delta^\nabla)$
\eit

 If $(x^\nu ,y^A,{\rm p}^\nu_A)$ is a system of natural
 coordinates in $\Pi$, and
 \dst\tilde y={\rm p}^\nu_A(\tilde y)\derpar{}{x^\nu}\otimes\zeta^A\otimes
 \d^mx\in\Pi\), taking into account the local expression of
 $v^*_{\nabla}$ which,  for a connection
 \dst\nabla=\d x^\nu\otimes\left(
 \derpar{}{x^\nu}+{\mit\Gamma}_\nu^A\derpar{}{y^A}\right)\),
 is $ v^*_{\nabla}(\zeta^A)=\d y^A-{\mit\Gamma}^A_\nu\d x^\nu$,
 we have that
  \beann
  h_\delta^\nabla(\tilde y) &=&
  h_\delta^\nabla\left({\rm p}^\nu_A(\tilde y)\derpar{}{x^\nu}
 \otimes\zeta^A\otimes\d^mx\Big\vert_{\rho^1(\tilde y)}\right)=
 {\rm p}^\nu_A(\tilde y)\derpar{}{x^\nu}\otimes
 (\d y^A-{\mit\Gamma}^A_\eta\d x^\eta )
 \otimes\d^mx\Big\vert_{\rho^1(\tilde y)}
 \\
 (\iota_0\circ h_\delta^\nabla)(\tilde y) &=&
 {\rm p}^\nu_A(\tilde y)
 (\d y^A\wedge\d^{m-1}x_\nu-{\mit\Gamma}^A_\nu\d^mx)
 \Big\vert_{\rho^1(\tilde y)}
 \eeann
 Observe that $\iota_0$ restricted to the image
 of $h_\delta^\nabla$ is injective. Therefore
 \bea
 \Theta_{h_\delta^\nabla} &=& {\rm p}^\nu_A(\d y^A-{\mit\Gamma}^A_\eta\d x^\eta )
 \wedge\d^{m-1}x_\nu =
 {\rm  p}^\nu_A\d y^A\wedge
 \d^{m-1}x_\nu-{\rm p}^\nu_A{\mit\Gamma}^A_\nu\d^mx
  \nonumber\\
 \Omega_{h_\delta^\nabla} &=&
 -\d {\rm p}^\nu_A\wedge\d y^A\wedge\d^{m-1}x_\nu+
 {\mit\Gamma}^A_\nu\d {\rm p}^\nu_A\wedge
 \d^mx+ {\rm p}^\nu_A\d{\mit\Gamma}^A_\nu\wedge\d^mx
 \label{Lforms}
 \eea

 Now, given a connection $\nabla$ and a
 Hamiltonian section $h_\delta$, from Lemma \ref{hcfsect} we have that
 $$
 \rho^{1*}(\iota_0\circ h_\delta^\nabla-\iota_0\circ h_\delta)=
 (\iota_0\circ h_\delta^\nabla)^*\Theta-(\iota_0\circ h_\delta)^*\Theta=
 \Theta_{h_\delta^\nabla}-\Theta_{h_\delta}=
 h_\delta^{1*}\hat\Theta-h_\delta^{2*}\hat\Theta:=
 {\mbox{\es H}}^\nabla_{h_\delta}
 $$
 is a $\bar\rho^1$-semibasic $m$-form in $\Pi$.
 It is usually written as
 ${\mbox{\es H}}^\nabla_{h_\delta}=
{\rm H}^\nabla_{h_\delta}(\bar\rho^{1^*}\omega)$,
 where ${\rm H}^\nabla_{h_\delta}\in\Cinfty (\Pi)$ is the
 {\sl global Hamiltonian function} associated with
 ${\mbox{\es H}}^\nabla_{h_\delta}$ and $\omega$.

 Therefore, given a Hamiltonian system $\hspi$,
 taking into account (\ref{splitomega0}),
 we have that every connection $\nabla$ in $\pi\colon E\to M$ allows us
 to split globally the Hamilton-Cartan forms as
 \beq
 \Theta_{h_\delta}=\Theta_{h_\delta^\nabla}-{\mbox{\es H}}^\nabla_{h_\delta}
 \quad , \quad
 \Omega_{h_\delta}=-\d\Theta_{h_\delta}=
 \Omega_{h_\delta^\nabla}+\d{\mbox{\es H}}^\nabla_{h_\delta}
 \label{splitomega}
 \eeq
 In a natural system of coordinates in $\Pi$,
 such that $\bar\rho^{1*}\omega=\d^mx$,
 we write ${\mbox{\es H}}^\nabla_{h_\delta}=
 {\rm H}^\nabla_{h_\delta}(x^\nu,y^A,{\rm p}_A^\nu )\d^mx$, and
 \bea
 \Theta_{h_\delta} &=&
 {\rm  p}^\nu_A\d y^A\wedge
 \d^{m-1}x_\nu-({\rm H}^\nabla_{h_\delta} +
 {\rm p}^\nu_A{\mit\Gamma}^A_\nu)\d^mx
 \nonumber
\\
 \Omega_{h_\delta} &=&
 -\d{\rm  p}^\nu_A\wedge\d y^A\wedge
 \d^{m-1}x_\nu+\d ({\rm H}^\nabla_{h_\delta}+
{\rm p}^\nu_A{\mit\Gamma}^A_\nu)\wedge\d^mx
 \label{omegahpi}
 \eea

\begin{prop}
 A couple $(\{ h_\delta\},\nabla)$ in $\Pi$
 is equivalent to a couple
 $({\mbox{\es H}},\nabla)$ (that is, given a connection $\nabla$,
 classes of Hamiltonian sections of $\delta$
 and Hamiltonian densities in $\Pi$ are in one-to-one correspondence).
 \label{otraprop}
\end{prop}
\proof
 Given a connection in $\pi\colon E\to M$,
 we have just seen that
 all the Hamiltonian sections belonging to the same
 equivalence class $\{ h_\delta\}$ define a unique
 Hamiltonian density ${\mbox{\es H}}^\nabla_{h_\delta}$ and, hence, the same
 Hamilton-Cartan forms.

 Conversely, given a Hamiltonian density ${\mbox{\es H}}$
 and a connection $\nabla$,
 we can construct an equivalence class of Hamiltonian sections
 $\{ h_\delta\}$ (which leads to the same Hamilton-Cartan forms),
 since, as ${\mbox{\es H}}\colon\Pi\to{\cal M}\pi$
 takes values in $\bar{\hat\rho}^{1*}\Lambda^m\Tan^*M$,
 we have a map
 $\iota_0\circ h_\delta^\nabla-{\mbox{\es H}}\colon\Pi\to{\cal M}\pi$.
 From the local expression of this map, it is easy to prove that
 there exists a local section $h_\delta$ of $\delta$, such that
 $\iota_0\circ h_\delta=\iota_0\circ h_\delta^\nabla-{\mbox{\es H}}$.
 Then, using a partition of unity we can construct a global section fulfilling
 this condition, and hence a family of sections $\{ h_\delta\}$
 defined by the relation
 $\iota_0\circ h_\delta=\iota_0\circ h_\delta^\nabla-{\mbox{\es H}}$.
 \qed

 As a direct consequence of this proposition,
 we have another way of obtaining a Hamiltonian system,
 which consists in giving a couple $({\mbox{\es H}},\nabla)$. In fact:

\begin{prop}
 Let $\nabla$ be a connection in $\pi\colon E\to M$,
 and  ${\mbox{\es H}}$ a Hamiltonian density.
 There exist a unique class $\{ h_\delta\}$ of Hamiltonian sections
 of $\delta$ such that
  $$
 \Theta_{h_\delta}=\Theta_{h_\delta^\nabla}-{\mbox{\es H}} \quad ,\quad
 \Omega_{h_\delta}=-\d\Theta_{h_\delta}=
 \Omega_{h_\delta^\nabla}+\d{\mbox{\es H}}
 $$
\end{prop}

 {\bf Remark}:
\bit
\item
If $\pi\colon Eo M$ is a trivial bundle, then there is a natural
connection (the trivial one). So, in this case,
there is a bijective correspondence between Hamiltonian systems
and Hamiltonian densities.

This is the situation in classical non-autonomous mechanics
\cite{CLM-94}, \cite{EMR-91}, \cite{EMR-sdtc}.
\eit

 \subsection{Variational principle and field equations}
 \protect\label{vpfe}

 Now we can establish the field equations for Hamiltonian systems.
 First we need to introduce the notion of {\sl prolongation} of
 diffeomorphisms and vector fields from $E$ to $\Pi$.

 \begin{definition}
 Let $\Phi\colon E\to E$ be a diffeomorphism of $\pi$-fiber bundles
 and $\Phi_M\colon M\to M$ the induced diffeomorphism in $M$. The
 {\rm prolongation} of $\Phi$ to $\Pi$ is the diffeomorphism
 $j^{1*}\Phi\colon \Pi\to\Pi$ defined by
 $j^{1*}\Phi:=(\Phi_M)_*\otimes\Phi^{*-1}\otimes
 \stackrel{m}{\wedge}\Phi_M^{*-1}$.
 \label{proldif}
 \end{definition}

 \begin{prop}
 Let $\Phi\colon E\to E$ be a diffeomorphism of fiber bundles,
 $\Phi_M\colon M\to M$ its restriction to $M$ and $j^{1*}\Phi$ its
 prolongation to $\Pi$. Then
 \ben
 \item
 $\rho^1\circ j^{1*}\Phi =\Phi\circ\rho^1$,
 $\bar\rho^1\circ j^{1*}\Phi =\Phi_M\circ\bar\rho^1$.
 \item
 If $\Psi\colon E\to E$ is another fiber bundle diffeomorphism, then
 $$
 j^{1*}(\Psi\circ\Phi )=j^{1*}\Psi\circ j^{1*}\Phi
 $$
 \item
 $j^{1*}\Phi$ is a diffeomorphism of $\rho^1$-bundles and
 $\bar\rho^1$-bundles, and $(j^{1*}\Phi )^{-1}=j^{1*}\Phi^{-1}$.
 \een
 \label{ppc}
 \end{prop}

 \begin{definition}
 Let $Z\in\vf (E)$ be a $\pi$-projectable vector field.
 The {\rm prolongation} of $Z$ to $\Pi$ is the vector field $j^{1*}Z$
 whose local one-parameter group of diffeomorphisms are the
 extensions $\{ j^{1*}\sigma_t\}$ of the local one-parameter group
 of diffeomorphisms $\{\sigma_t\}$ of $Z$.
 \label{prolvf}
 \end{definition}

 Now we can state:

 \begin{definition}
 Let $\hspi$ be a Hamiltonian system.
 Let $\Gamma_c(M,\Pi)$ be
 the set of compact-supported sections of $\bar\rho^1$,
 and $\psi\in\Gamma_c(M,\Pi)$. Consider the map
 $$
 \begin{array}{ccccc}
 {\bf H}&\colon&\Gamma_c(M,\Pi)&\longrightarrow&\Real
 \\
 & &\psi&\mapsto&\int_M\psi^*\Theta_{h_\delta}
 \end{array}
 $$
 The {\rm variational problem} for this Hamiltonian system
 is the search of the critical (or
 stationary) sections of the functional ${\bf H}$,
 with respect to the variations of $\psi$ given
 by $\psi_t =j^{1*}\sigma_t\circ\psi$, where $\{\sigma_t\}$ is a
 local one-parameter group of every $Z\in\vf^{{\rm V}(\pi)}(E)$
 (the module of $\pi$-vertical vector fields in $E$).
 $$
 \frac{\d}{\d t}\Big\vert_{t=0}\int_M\psi_t^*\Theta_{h_\delta} = 0
 $$

 This is the so-called {\rm Hamilton-Jacobi principle}
 of the Hamiltonian formalism.
 \label{hjvp}
 \end{definition}

 \begin{teor}
 Let $\hspi$ be a Hamiltonian system. The following assertions on a
 section $\psi\in\Gamma_c(M,\Pi)$ are equivalent:
 \ben
 \item
 $\psi$ is a critical section for the variational problem posed by
 $\Theta_{h_\delta}$.
 \item
 \dst\int_M\psi^*\Lie (j^{1*}Z)\Theta_{h_\delta}= 0\) ,
 for every $Z\in\vf^{{\rm V}(\pi)}(E)$.
 \item
 $\psi^*\inn (j^{1*}Z)\Omega_{h_\delta}= 0$, for every
 $Z\in\vf^{{\rm V}(\pi)}(E)$.
 \item
 $\psi^*\inn (X)\Omega_{h_\delta}= 0$, for every
 $X\in\vf (\Pi)$.
 \item
 If $(U;x^\nu ,y^A,{\rm p}_A^\nu )$ is a natural system of
 coordinates in $\Pi$, then
 $\psi =(x^\nu ,y^A(x^\eta),{\rm p}^\nu_A(x^\eta))$ in $U$
 satisfies the system of equations
 \beq
 \derpar{y^A}{x^\nu}\Big\vert_{\psi}=
 \derpar{H_{h_\delta}}{{\rm p}^\nu_A}
 \Big\vert_{\psi} \quad ;\quad
 \derpar{{\rm p}_A^\nu}{x^\nu}\Big\vert_{\psi}=
 - \derpar{H_{h_\delta}}{y^A}\Big\vert_{\psi}
 \label{HDWnocov}
 \eeq
 which are known as the
 {\rm Hamilton-De Donder-Weyl equations} of the Hamiltonian formalism.
 \een
 \label{equics}
 \end{teor}
 \proof
 \quad ($1\ \Leftrightarrow\ 2$)\quad
 If $Z\in\vf^{{\rm V}(\pi)}(E)$ and $\sigma_t$ is a one-parameter
 local group of $Z$, we have
 \beann
 \frac{\d}{\d t}\Big\vert_{t=0}
 \int_M(j^{1*}\sigma_t\circ\psi)^*\Theta_{h_\delta}
 &=& \lim_{t\to 0}\frac{1}{t}
 \left(\int_M((j^{1*}\sigma_t\circ\psi )^*\Theta_{h_\delta}-
 \int_M\psi^*\Theta_{h_\delta}\right)
 \\ &=& \lim_{t\to 0}\frac{1}{t}
 \left(\int_M\psi^*(j^{1*}\sigma_t)^*\Theta_{h_\delta}-
 \int_M\psi^*\Theta_{h_\delta}\right)
 \\ &=&\lim_{t\to 0}\frac{1}{t}
 \left(\int_M\psi^*[(j^1\sigma_t)^*\Theta_{h_\delta} -
 \Theta_{h_\delta}] \right)=
 \int_M\psi^*\Lie (j^1Z)\Theta_{h_\delta}
 \eeann
 and the results follows immediately.

 \quad\quad ($2\ \Leftrightarrow\ 3$)\quad
 Taking into account that
 $$
 \Lie (j^{1*}Z)\Theta_{h_\delta} =
 \d\inn (j^{1*}Z)\Theta_{h_\delta}+
 \inn (j^{1*}Z)\d\Theta_{h_\delta} =
 \d\inn (j^{1*}Z)\Theta_{h_\delta}-
 \inn (j^{1*}Z)\Omega_{h_\delta}
 $$
 we obtain that
 $$
 \int_M \psi^*\Lie (j^{1*}Z)\Theta_{h_\delta} =
 \int_M\psi^*\d\inn (j^{1*}Z)\Theta_{h_\delta} -
 \int_M\psi^*\inn (j^{1*}Z)\Omega_{h_\delta}
 $$
 and, as $\psi$ has compact support, using Stoke's theorem we have
 $$
 \int_M\psi^*\d\inn (j^{1*}Z)\Theta_{h_\delta}=
 \int_M\d\psi^*\inn (j^{1*}Z)\Theta_{h_\delta} =0
 $$
 hence
 \dst\int_M\psi^*\Lie (j^{1*}Z)\Theta_{h_\delta} =0\)
 (for every $Z\in\vf^{{\rm V}(\pi)}(E)$)
 if, and only if,
 \dst\int_M\psi^*\inn (j^{1*}Z)\Omega_{h_\delta} =0\) ,
 and, according to the fundamental theorem
 of variational calculus, this is equivalent to
 $$
 \psi^*\inn (j^{1*}Z)\Omega_{h_\delta}=0
 $$

 \quad\quad ($3\ \Leftrightarrow\ 5$)\quad
 Suppose that $\psi$ is a section verifying that
 $\psi^*\inn (j^{1*}Z)\Omega_{h_\delta} =0$, for every
 $Z\in\vf^{{\rm V}(\pi)}(E)$. In a natural chart in $\Pi$, if
 \dst Z= \beta^A\derpar{}{y^A}\) , then
 \dst j^{1*}Z= \beta^A\derpar{}{y^A}
 -{\rm p}^\nu_B\derpar{\beta^B}{y^A}\derpar{}{{\rm p}^\nu_A}\) .
 Taking into account the local expression of $\Omega_{h_\delta}$
 given in (\ref{omegaH0}), we have
 $$
 \inn (j^{1*}Z)\Omega_{h_\delta} =
  \beta^A\left(\d {\rm p}^\nu_A\wedge\d^{m-1}x_\nu +
 \derpar{H_{h_\delta}}{y^A} \d^mx\right) +
 {\rm p}^\nu_B\derpar{\beta^B}{y^A}
 \left(\d y^A\wedge\d^{m-1}x_\nu -
 \derpar{H_{h_\delta}}
 {{\rm p}^\nu_A}
 \d {\rm p}^\nu_A\wedge\d^mx\right)
 $$
 As $\psi =(x^\nu ,f^A(x^\eta ),g^\nu_A(x^\eta ))$
 is a section of $\bar\rho^1$, then
 on the points of the image of $\psi$ we have
 $y^A=f^A(x^\eta )$, ${\rm p}^\nu_A=g^\nu_A(x^\eta )$, and we obtain
 \beann
 0 = \psi^*\inn (j^{1*}Z)\Omega_{h_\delta} &=&
 \beta^A\left(\derpar{g^\nu_A}{x_\nu}+
 \derpar{H_{h_\delta}}{y^A}\right)
 \d^mx
 +g^\nu_C\left(\derpar{\beta^C}{y^A}\derpar{f^A}{x_\nu} -\derpar{\beta^C}{y^A}
 \derpar{H_{h_\delta}}
 {{\rm p}^\nu_A} \right)\d^mx
 \\ &=&
 \left[\beta^A\left(\derpar{g^\nu_A}{x^\nu}+
 \derpar{H_{h_\delta}}
 {y^A}\right)+
  g^\nu_C\derpar{\beta^C}{y^A}\left(\derpar{f^A}{x^\nu}-
 \derpar{H_{h_\delta}}
 {{\rm p}^\nu_A} \right)\right]\d^mx
 \eeann
 and, as this holds for every $Z\in\vf^{{\rm V}(\pi)}(E)$,
 this is equivalent to demanding that
 $$
\beta^A\left(\derpar{g^\nu_A}{x^\nu}+
 \derpar{H_{h_\delta}}{y^A}\right) +
 g^\nu_C\derpar{\beta^C}{y^A}\left(\derpar{f^A}{x^\nu}-
 \derpar{H_{h_\delta}}
 {{\rm p}^\nu_A} \right)=0
 $$
 for every $\beta^A(x^\nu,y^A)$. Therefore
 $$
 \beta^A\left(\derpar{g^\nu_A}{x^\nu}+
 \derpar{H_{h_\delta}}{y^A}\right)
  = 0
 \quad ; \quad
 g^\nu_C\derpar{\beta^C}{y^A}\left(\derpar{f^A}{x^\nu}-
 \derpar{H_{h_\delta}}
 {{\rm p}^\nu_A} \right) = 0
 $$
 From the first equalities we obtain the first group
 of the Hamiltonian equations. For the second ones,
 let $(W; x^\nu,y^A,{\rm p}_A^\nu)$ a natural chart, $U=\bar\rho^1(W)$,
 and $\psi$ a critical section. Then, for every $x\in U$ we have
 $$
 g^\nu_C(x)\derpar{\beta^C}{y^A}\Big\vert_{(x,f^A(x))}
 \left(\derpar{f^A}{x^\nu}-
 \derpar{H_{h_\delta}}
 {{\rm p}^\nu_A} \right)\Big\vert_{\psi(x)} = 0
 $$
 but, as there are critical sections passing through every point in $W$,
 we obtain that
 $$
 \derpar{\beta^C}{y^A}\Big\vert_{(x,f^A(x))}
 \left(\derpar{f^A}{x^\nu}-
 \derpar{H_{h_\delta}}
 {{\rm p}^\nu_A} \right)\Big\vert_{\psi(x)} = 0
 \qquad \mbox{\rm (for every $B,\nu$)}
 $$
 Now we can choose $\beta^B$ such that \dst\derpar{\beta^B}{y^A}\)
 take arbitrary values, and then
 $$
 \derpar{f^A}{x^\nu}-
 \derpar{H_{h_\delta}}
 {{\rm p}^\nu_A} = 0
 $$
 which is the second group of the Hamiltonian equations.
 The converse is trivial.

 \quad\quad ($4\ \Leftrightarrow\ 5$)\quad
 Suppose that $\psi$ is a section verifying that
 $\psi^*\inn (  X)\Omega_{h_\delta} =0$, for every $X\in\vf (\Pi)$.
 If \dst X=\alpha^\nu\derpar{}{x^\nu}+
 \beta^A\derpar{}{y^A}+\gamma^\nu_A\derpar{}{{\rm p}^\nu_A}\) ,
 taking into account (\ref{omegaH0}), we have
 \beann
 \inn (X)\Omega_{h_\delta} &=&
 (-1)^\eta\alpha^\eta\left(\d {\rm p}^\nu_A\wedge
 \d y^A\wedge\d^{m-2}x_{\eta\nu}-
 \derpar{H_{h_\delta}}
 {{\rm p}^\nu_A} \d {\rm p}^\nu_A \wedge\d^{m-1}x_\eta\right)
 \\ &+&
 \beta^A\left(\d {\rm p}^\nu_A\wedge\d^{m-1}x_\nu +
 \derpar{H_{h_\delta}}{y^A}\d^mx \right) +
 \gamma^\nu_A \left(-\d y^A\wedge\d^{m-1}x_\nu +
 \derpar{H_{h_\delta}}
 {{\rm p}^\nu_A} \d {\rm p}^\nu_A\wedge\d^mx\right)
 \eeann
 but as $\psi =(x^\nu ,f^A(x^\eta ),g^\nu_A(x^\eta ))$
 is a section of $\bar\rho^1$, then
 on the points of the image of $\psi$ we have
 $y^A=f^A(x^\eta )$, ${\rm p}^\nu_A=g^\nu_A(x^\eta )$, and
 \beann
 0 = \psi^*\inn (X)\Omega_{h_\delta}  &=&
 (-1)^{\eta+\nu}\alpha^\eta\left(\derpar{f^A}{x^\nu}-
 \derpar{H_{h_\delta}}
 {{\rm p}^\nu_A} \right) \derpar{g^\nu_A}{x^\eta} \d^mx +
 \\ & &
 \beta^A\left(\derpar{g^\nu_A}{x_\nu}+
 \derpar{H_{h_\delta}}{y^A}\right)
 \d^mx +
 \gamma^\nu_A\left(-\derpar{f^A}{x_\nu}+
 \derpar{H_{h_\delta}}
 {{\rm p}^\nu_A} \right)\d^mx
 \eeann
 and, as this holds for every $X\in\vf (\Pi)$,
 we obtain the Hamilton-De Donder-Weyl equations.
 The converse is trivial.
 \qed

 {\bf Remark}:
 \bit
 \item
 In relation to the equations (\ref{HDWnocov}),
 it is important to point out that they are not covariant, since
 the Hamiltonian function $H_{h_\delta}$ is defined only locally,
 and hence it is not intrinsically defined.

 In order to write a set of covariant Hamiltonian
 equations we must use a global Hamiltonian function,
 which can be obtained by introducing another Hamiltonian section
 $h'_\delta$, with $h'_\delta\not\in\{h_\delta\}$
 (as we have seen in section \ref{dsc}).
 It is usual to take the section induced by a connection
 $\nabla$ in $\pi\colon E\to M$, and hence
 we have the splitting given in
 (\ref{splitomega}) for the form $\Omega_{h_\delta}$. Then,
 if ${\mit\Gamma}^B_\eta$ are the local component functions of $\nabla$
 in $U\subset\Pi$,
 starting from the local expression (\ref{omegahpi}),
 and following the same pattern as in the proof of the last item,
 we obtain for a critical section
 $\psi =(x^\nu ,y^A(x^\eta),{\rm p}^\nu_A(x^\eta))$ in $U$ the
 covariant {\sl Hamiltonian equations}:
 $$
 \derpar{y^A}{x^\nu}\Big\vert_{\psi}=
 \left(\derpar{{\rm H}^\nabla_{h_\delta}}{{\rm p}^\nu_A}+
 {\mit\Gamma}^A_\nu\right)
 \Big\vert_{\psi} \quad ;\quad
 \derpar{{\rm p}_A^\nu}{x^\nu}\Big\vert_{\psi}=
 - \left(\derpar{{\rm H}^\nabla_{h_\delta}}{y^A}+
 {\rm p}^B_\eta\derpar{{\mit\Gamma}^B_\eta}{y^A}\right)\Big\vert_{\psi}
 $$
 Observe that, as
 ${\rm H}^\nabla_{h_\delta}=H_{h_\delta}-{\rm p}_A^\nu{\mit\Gamma}^A_\nu$
 (on each open set $U\subset\Pi$, where
 $H_{h_\delta}$ is the corresponding local Hamiltonian function),
 then from these last equations we recover the
 Hamilton-De Donder-Weyl equations.
 (See \cite{CCI-91} for comments on this subject).

 \eit

 \subsection{Hamiltonian system
  associated with a hyper-regular Lagrangian system}
 \protect\label{hsahrs}

 It is evident that different choices of equivalence
 classes of Hamiltonian sections of $\delta$ lead to
 different Hamiltonian systems in $\Pi$.
 The question now is how to associate (if possible) a
 Hamiltonian system with a Lagrangian system. The answer to this
 question is closely related to the regularity of the
 Lagrangian system.

 First, let $\ls$ be a hyper-regular Lagrangian system. Then:

 \begin{lem}
 For every section ${\rm h}_\delta\colon \Pi\to J^1E^*$ of $\delta$,
 the relation
 $$
{\rm h}_\mu:=\mu'\circ\iota_0\circ{\rm h}_\delta\circ{\mit\Psi}
 $$
 defines a unique section of $\mu$, which is just
 ${\rm h}_\mu=\widetilde{{\cal F}\Lag}\circ{\cal F}\Lag^{-1}$.
 \label{lemcero}
 \end{lem}
 \proof
 We have the diagram
 \beq
\begin{array}{ccccc}
\begin{picture}(15,180)(0,0)
\put(0,85){\mbox{$J^1E$}}
\end{picture}
&
\begin{picture}(65,180)(0,0)
\put(20,137){\mbox{${\cal F}\Lag$}}
\put(35,118){\mbox{$\widetilde{{\cal F}\Lag}$}}
\put(36,88){\mbox{$\widehat{{\cal F}\Lag}$}}
\put(36,51){\mbox{$\widehat{{\rm F}\Lag}$}}
\put(20,30){\mbox{${\rm F}\Lag$}}
 \put(0,93){\vector(1,1){70}}
\put(0,89){\vector(2,1){70}}
 \put(0,85){\vector(1,0){70}}
\put(0,81){\vector(3,-1){70}}
 \put(0,77){\vector(1,-1){70}}
\end{picture}
&
\begin{picture}(35,180)(0,0)
\put(16,0){\mbox{$\Pi$}}
 \put(9,42){\mbox{$J^1E^*$}}
\put(12,85){\mbox{${\cal M}\pi$}}
 \put(12,129){\mbox{${\cal M}\pi$}}
 \put(11,170){\mbox{$J^1\pi^*$}}
\put(20,142){\vector(0,1){25}}
 \put(6,98){\vector(0,1){68}}
\put(20,98){\vector(0,1){25}}
 \put(20,56){\vector(0,1){25}}
\put(20,36){\vector(0,-1){25}}
 \put(11,152){\mbox{$\mu$}}
\put(-2,140){\mbox{$\mu$}}
 \put(9,109){\mbox{$\mu'$}}
\put(11,64){\mbox{$\iota_0$}}
 \put(11,20){\mbox{$\delta$}}
\end{picture}
&
 \begin{picture}(25,180)(0,0)
 \put(0,172){\vector(1,0){25}}
 \put(25,168){\vector(-1,-1){25}}
 \put(13,143){\mbox{${\rm h}_\mu$}}
 \put(1,177){\mbox{${\rm Id}_{J^1\pi^*}$}}
 \put(0,4){\vector(1,0){25}}
 \put(25,12){\vector(-1,1){25}}
 \put(15,27){\mbox{${\rm h}_\delta$}}
 \put(4,8){\mbox{${\rm Id}_\Pi$}}
 \end{picture}
  &
 \begin{picture}(10,180)(0,0)
 \put(0,170){\mbox{$J^1\pi^*$}}
 \put(7,168){\vector(0,-1){155}}
 \put(10,85){\mbox{${\mit\Psi}$}}
 \put(3,0){\mbox{$\Pi$}}
 \end{picture}
  \end{array}
 \label{diagn}
 \eeq
 Then, taking into account the commutativity of this diagram, we have
 \beann
 {\rm h}_\mu&=&\mu'\circ\iota_0\circ{\rm h}_\delta\circ{\mit\Psi}=
 \widetilde{{\cal F}\Lag}\circ{\cal F}\Lag^{-1}\circ\mu
 \circ\iota_0\circ{\rm h}_\delta\circ{\mit\Psi}
 \\ &=&
 \widetilde{{\cal F}\Lag}\circ{\cal F}\Lag^{-1}
 \circ{\mit\Psi}^{-1}\circ{\mit\Psi}=
 \widetilde{{\cal F}\Lag}\circ{\cal F}\Lag^{-1}
 \eeann
 So ${\rm h}_\mu$ is independent of ${\rm h}_\delta$.
 \qed

 {\bf Remarks}:
 \bit
 \item
 This result is to be expected, since
 $\widetilde{{\cal F}\Lag}(J^1E)$ is a 1-codimensional submanifold
 of ${\cal M}\pi$, transverse to the projection $\mu$, and
 hence it defines a section ${\rm h}_\mu$ of $\mu$. This is just the
 natural section used in \cite{CCI-91} and \cite{LMM-96}
 for associating a Hamiltonian system
 to a hyper-regular Lagrangian one (see section \ref{hslsjpi1}).
 \item
 Observe that a natural section ${\rm h}_\delta$ of $\delta$ can be
 selected by making
 $$
 {\rm h}_\delta:=\widehat{{\rm F}\Lag}\circ{\rm F}\Lag^{-1}
 $$
 or, what is equivalent, its associated class can be defined by
 $$
 \iota_0\circ h_\delta:= \widehat{{\cal F}\Lag}\circ{\rm F}\Lag^{-1}
\quad , \quad
\mbox{\rm for every $h_\delta\in\{{\rm h}_\delta\}$}
 $$
 Observe that this section ${\rm h}_\delta$ is just the inverse of
 $\delta$ restricted to $\widehat{{\rm F}\Lag}(J^1E)$,
 and that $\iota_0\circ h_\delta$ is a diffeomorphism.
 \eit

 \begin{definition}
 Given a section ${\rm h}_\delta\colon \Pi\to J^1E^*$ of $\delta$,
 we define the Hamilton-Cartan forms
 $$
 \Theta_{{\rm h}_\delta}:=(\mu'\circ\iota_0\circ{\rm h}_\delta)^*\Theta
 \quad ; \quad
 \Omega_{{\rm h}_\delta}:=(\mu'\circ\iota_0\circ{\rm h}_\delta)^*\Omega
 $$
 \label{delpro0}
 \end{definition}
 
 \begin{prop}
 The Hamilton-Cartan forms are independent of the section
 ${\rm h}_\delta$, and
 \beq
 {\rm F}\Lag^*\Theta_{{\rm h}_\delta}= \Theta_\Lag
 \quad , \quad
 {\rm F}\Lag^*\Omega_{{\rm h}_\delta}=\Omega_\Lag
 \label{aux00}
 \eeq
 Then, $\hslpi$ is the (unique) Hamiltonian system
 which is associated with the hyper-regular Lagrangian system $\ls$.
 \end{prop}
 \proof
 The independence of the section ${\rm h}_\delta$ is a consequence of
 lemma \ref{lemcero}. Then, 
 taking into account the commutativity of diagram (\ref{diagn}),
 and proposition \ref{propositio5}, for every section ${\rm h}_\delta$,
 we have
 $$
{\rm F}\Lag^*\Theta_{{\rm h}_\delta}=
{\rm F}\Lag^*(\mu'\circ\iota_0\circ{\rm h}_\delta)^*\Theta=
(\mu'\circ\iota_0\circ{\rm h}_\delta\circ{\rm F}\Lag)^*\Theta=
\widetilde{{\cal F}\Lag}^*\Theta=\Theta_\Lag
 $$
and the same result follows for $\Omega_{{\rm h}_\delta}$.
 \qed

 Using charts of natural coordinates in
 $\Pi$ and $J^1E^*$, and the expressions (\ref{coorglt})
 and (\ref{redlt}) of the Legendre maps,
 we have that the natural Hamiltonian section
 ${\rm h}_\delta=\widehat{{\rm F}\Lag}\circ{\rm F}\Lag^{-1}$
 has associated the local Hamiltonian function
 \beq
  H_{{\rm h}_\delta}(x^\nu,y^A,{\rm p}^\nu_A)= {\rm F}\Lag^{-1*}
 \left(v^A_\nu\derpar{\lag}{v^A_\nu}-\lag\right)=
 {\rm p}^\nu_A{\rm F}\Lag^{-1^*}v_\nu^A-{\rm F}\Lag^{-1^*}\lag
 \label{lochamf0}
 \eeq
 and for the Hamilton-Cartan forms:
 \beann
 \Theta_{{\rm h}_\delta} &=& {\rm p}_A^\nu\d y^A\wedge\d^{m-1}x_\nu -
 ({\rm p}^\nu_A{\rm F}\Lag^{-1^*}v_\nu^A-{\rm F}\Lag^{-1^*}\lag )\d^mx
\\
 \Omega_{{\rm h}_\delta} &=& -\d {\rm p}_A^\nu\wedge\d y^A\wedge\d^{m-1}x_\nu
 +\d ({\rm p}^\nu_A{\rm F}\Lag^{-1^*}v_\nu^A-{\rm F}\Lag^{-1^*}\lag )
 \wedge\d^mx
 \eeann

There is another way of obtaining this Hamiltonian system.
In fact, suppose that a connection $\nabla$ is given in $\pi\colon E\to M$,
and let $h_\delta^\nabla\colon\Pi\to J^1E^*$ be
the induced linear section of $\delta$.
If we have used $\nabla$ for constructing the associated
density of Lagrangian energy $\del\in\df^m(J^1E)$
(see definition \ref{energ}), the key is to define a Hamiltonian density
${\mbox{\es H}}^\nabla\in\df^m(\Pi)$ which is ${\rm F}\Lag$-related with
$\del$. We can make this construction in two ways:

 \begin{prop}
 \ben
 \item
 The $m$-form  ${\rm F}\Lag^*\Theta_{h_\delta^\nabla}-\Theta_{\Lag}$
 is $\bar\pi^1$-semibasic and
 \beq
 {\rm F}\Lag^*\Theta_{h_\delta^\nabla}-\Theta_{\Lag} ={\cal E}_\Lag^{\nabla}
 \label{form1}
 \eeq
 \item
 There exists a unique Hamiltonian density
 ${\mbox{\es H}}^\nabla\in\df^m(\Pi)$ such that
 \beq
 {\rm F}\Lag^*{\mbox{\es H}}^\nabla=
 {\rm F}\Lag^*\Theta_{h_\delta^\nabla}-\Theta_{\Lag}=\del
 \label{form2}
 \eeq
 Let ${\mbox{\es H}}^\nabla={\rm H}^\nabla(\bar\rho^{1*}\omega)$,
 with ${\rm H}^\nabla\in\Cinfty (\Pi)$.
 Then,  ${\mbox{\es H}}^\nabla$ and ${\rm H}^\nabla$ are called the
 {\rm Hamiltonian density} and the {\sl Hamiltonian function}
 associated with the Lagrangian system,
 the connection $\nabla$ and $\omega$.
 \item
 The Hamilton-Cartan forms of definition \ref{delpro0} split as
  \beq
 \Theta_{{\rm h}_\delta}= \Theta_{h_\delta^\nabla}-{\mbox{\es H}}^\nabla
 \quad , \quad
 \Omega_{{\rm h}_\delta}=
 -\d\Theta_{{\rm h}_\delta}=\Omega_{h_\delta^\nabla}+\d{\mbox{\es H}}^\nabla
 \label{form3}
  \eeq
 \een
 \label{delpro}
 \end{prop}
 \proof
 \ben
 \item
 Once again, it suffices to see it in a natural local system
 $(x^\nu,y^A,v^A_\nu )$. Then, if $\Lag =\lag\d^mx$,
 taking into account the corresponding local expressions we have that
 $$
 {\rm F}\Lag^*\Theta_{h_\delta^\nabla}-\Theta_{\Lag}=
 \left(\derpar{\lag}{v_\nu^A}(v^A_\nu -{\mit\Gamma}^A_\nu)-
 \lag\right)\d^mx
 $$
 and the result holds. The last part follows,
 recalling the local expression of the density of Lagrangian
 energy. Thus this form is $\bar\pi^1$-semibasic.
 \item
 It is immediate, as ${\rm F}\Lag$ is a diffeomorphism.
 \item
 From (\ref{form1}) and (\ref{form2}) we obtain that
 $$
  {\rm F}\Lag^*(\Theta_{h_\delta^\nabla}-{\mbox{\es H}}^\nabla)=
 {\rm F}\Lag^*\Theta_{h_\delta^\nabla}-{\rm F}\Lag^*\Theta_{h_\delta^\nabla}+
 \Theta_\Lag= \Theta_\Lag={\rm F}\Lag^*\Theta_{{\rm h}_\delta}
 $$
 and therefore ${\rm F}\Lag^*\Omega_{{\rm h}_\delta}=\Omega_\Lag$ too.
 Then, the result follows because ${\rm F}\Lag$ is a diffeomorphism.
 \een
 \qed

 {\bf Remark}:
\bit
\item
Notice that the item 1 holds even if ${\rm F}\Lag$ is not a diffeomorphism.
\eit

 In a system of natural coordinates we have
 $$
 {\rm H}^\nabla(x^\nu ,y^A,{\rm p}_A^\nu )=
 {\rm p}^\nu_A({\rm F}\Lag^{-1^*}v_\nu^A-{\mit\Gamma}^A_\nu)
 -{\rm F}\Lag^{-1^*}\lag
 $$
 and thus ${\rm F}\Lag^*{\rm H}^\nabla={\rm E}^\nabla_\Lag$.

 An alternative way is to obtain this Hamiltonian density
 using only Hamiltonian sections.

 \begin{prop}
 Consider the Hamiltonian section
 ${\rm h}_\delta= \widehat{{\rm F}\Lag}\circ{\rm F}\Lag^{-1}$,
 and a connection $\nabla$. Then we have that
 $$
 (\iota_0\circ h_\delta^\nabla)^*\Theta-
 (\mu'\circ\iota_0\circ{\rm h}_\delta)^*\Theta =
 {\mbox{\es H}}^\nabla
 $$
 and hence the splitting (\ref{form3}) holds
 \label{noseque}
 \end{prop}
 \proof
 We have the following  diagram
 $$
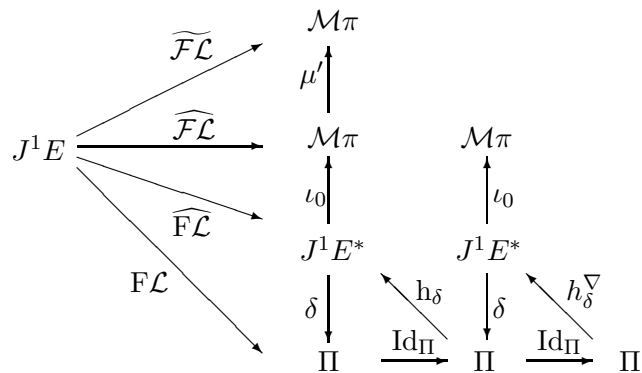
\begin{array}{ccccccc}
\begin{picture}(15,140)(0,0)
\put(0,80){\mbox{$J^1E$}}
\end{picture}
&
\begin{picture}(65,140)(0,0)
\put(35,118){\mbox{$\widetilde{{\cal F}\Lag}$}}
\put(36,88){\mbox{$\widehat{{\cal F}\Lag}$}}
\put(36,51){\mbox{$\widehat{{\rm F}\Lag}$}}
\put(20,30){\mbox{${\rm F}\Lag$}}
\put(0,89){\vector(2,1){70}}
 \put(0,85){\vector(1,0){70}}
\put(0,81){\vector(3,-1){70}}
 \put(0,77){\vector(1,-1){70}}
\end{picture}
&
\begin{picture}(30,140)(0,0)
\put(16,0){\mbox{$\Pi$}}
 \put(9,42){\mbox{$J^1E^*$}}
\put(12,85){\mbox{${\cal M}\pi$}}
 \put(12,129){\mbox{${\cal M}\pi$}}
  \put(20,56){\vector(0,1){25}}
\put(20,36){\vector(0,-1){25}}
\put(20,98){\vector(0,1){25}}
\put(11,64){\mbox{$\iota_0$}}
 \put(11,20){\mbox{$\delta$}}
 \put(9,109){\mbox{$\mu'$}}
\end{picture}
 &
 \begin{picture}(25,140)(0,0)
 \put(0,4){\vector(1,0){25}}
 \put(25,12){\vector(-1,1){25}}
 \put(13,27){\mbox{${\rm h}_\delta$}}
 \put(4,8){\mbox{${\rm Id}_\Pi$}}
 \end{picture}
 &
 \begin{picture}(10,140)(0,0)
 \put(5,37){\vector(0,-1){25}}
 \put(5,56){\vector(0,1){25}}
 \put(7,20){\mbox{$\delta$}}
 \put(7,64){\mbox{$\iota_0$}}
 \put(0,0){\mbox{$\Pi$}}
 \put(-7,42){\mbox{$J^1E^*$}}
 \put(-5,85){\mbox{${\cal M}\pi$}}
 \end{picture}
 &
 \begin{picture}(25,140)(0,0)
 \put(0,4){\vector(1,0){25}}
 \put(25,12){\vector(-1,1){25}}
 \put(15,27){\mbox{$h_\delta^\nabla$}}
 \put(4,8){\mbox{${\rm Id}_\Pi$}}
 \end{picture}
 &
 \begin{picture}(10,140)(0,0)
 \put(0,0){\mbox{$\Pi$}}
 \end{picture}
 \end{array}
 $$
 Observe that
 $ \mu'\circ \iota_0\circ{\rm h}_\delta=
 \widetilde{{\cal F}\Lag}\circ{\rm F}\Lag^{-1}$.
 Therefore, taking into account definition
 \ref{def4}, Proposition \ref{propositio5}, (\ref{def12})
 and (\ref{form2}), we have
 $$
 (\iota_0\circ h_\delta^\nabla)^*\Theta-
 (\mu'\circ\iota_0\circ{\rm h}_\delta)^*\Theta =
  \Theta_{h_\delta^\nabla}-({\rm F}\Lag^{-1})^*\widetilde{{\cal F}\Lag}^*
 \Theta=
 \Theta_{h_\delta^\nabla}-({\rm F}\Lag^{-1})^*\Theta_\Lag=
 {\mbox{\es H}}^\nabla
 $$
 Then the result for the splittings of
 the Hamilton-Cartan forms follows straightforwardly.
 \qed

{\bf Remark}:
\bit
\item
 Note that the use of both extended Legendre maps
 is necessary for obtaining the Hamiltonian density in this way.
\eit

 As a final remark,
 all the results stated in section \ref{vpfe}
 in relation to the variational principle and the characterization of
 critical sections are true. In particular, field equations are
 the Hamilton-De Donder-Weyl equations
 (\ref{HDWnocov}), where the local Hamiltonian function
 $H_{{\rm h}_\delta}$ is given by (\ref{lochamf0}).

 \subsection{Hamiltonian system
  associated with an almost-regular Lagrangian system}
 \protect\label{hspiar}

 Now, let $\ls$ be an almost-regular Lagrangian system.
 Bearing in mind diagram (\ref{diag2}),
 first observe that the submanifold
 $\j_0\colon P\hookrightarrow\Pi$,
 is a fiber bundle over $E$ (and $M$), and
 the corresponding projections will be denoted
 $\kappa^1_0\colon P\to E$ and $\bar\kappa^1_0\colon P\to M$,
 satisfying that $\kappa^1\circ\j_0=\kappa^1_0$ and
 $\bar\kappa^1\circ\j_0=\bar\kappa^1_0$.

 \begin{prop}
 The Lagrangian forms $\Theta_{\Lag}$ and $\Omega_{\Lag}$,
 are ${\rm F}\Lag$-projectable.
 \label{flproj1}
 \end{prop}
 \proof
 By Proposition \ref{kerfl}, we have that
 $\ker\,{\rm F}\Lag_*=\ker\,\Omega_\Lag\cap\vf^{{\rm V}(\pi^1)}(J^1E)$.
 Then, for every $X\in{\rm ker}\, {\rm F}\Lag_*$ we have that
 $\inn (X)\Theta_{\Lag}=0$, since $\Theta_\Lag$ is a $\pi^1$-semibasic
 $m$-form, and in the same way
 $ \Lie (X)\Theta_{\Lag}=0$. Therefore
 $\Theta_{\Lag}$ is ${\rm F}\Lag$-projectable.

 As a trivial consequence of this fact,
 $\inn (X)\Omega_{\Lag}=0$, and
 $\Lie(X)\Omega_{\Lag}=0$, and therefore
 $\Omega_{\Lag}$ is also ${\rm F}\Lag$-projectable.
 \qed

 \begin{definition}
 Given a section
 $\hat{\rm h}_\delta\colon P\to\hat P$ of $\hat\delta$,
 we define the Hamilton-Cartan forms
 $$
\Theta^0_{\hat{\rm h}_\delta}:=
 (\tilde\jmath_0\circ\hat\mu'\circ\hat\iota_0\circ\hat{\rm h}_\delta)^*\Theta
 \quad ; \quad
 \Omega^0_{\hat{\rm h}_\delta}:=
 (\tilde\jmath_0\circ\hat\mu'\circ\hat\iota_0\circ\hat{\rm h}_\delta)^*\Omega
  $$
 \label{faltaba}
 \end{definition}

 \begin{prop}
 The Hamilton-Cartan forms $\Theta^0_{\hat{\rm h}_\delta}$
 and $\Omega^0_{\hat{\rm h}_\delta}$ are independent of the section
 $\hat{\rm h}_\delta$ of $\hat\delta$, and
 \beq
 {\rm F}\Lag_0^*\Theta_{\hat{\rm h}_\delta}^0=\Theta_\Lag
 \quad , \quad
 {\rm F}\Lag_0^*\Omega_{\hat{\rm h}_\delta}^0=\Omega_\Lag
 \label{proprel2}
 \eeq
 Then $\hspio$ is the unique Hamiltonian system
 associated with the almost-regular Lagrangian system $\ls$.
 \end{prop}
\proof
 We have the following diagram
 $$
\begin{array}{ccccc}
\begin{picture}(15,140)(0,0)
\put(0,80){\mbox{$J^1E$}}
\end{picture}
&
 \begin{picture}(70,140)(0,0)
 \put(28,120){\mbox{$\widetilde{{\cal F}\Lag_0}$}}
 \put(36,91){\mbox{$\widehat{{\cal F}\Lag_0}$}}
 \put(36,51){\mbox{$\widehat{{\rm F}\Lag_0}$}}
 \put(20,30){\mbox{${\rm F}\Lag_0$}}
 \put(0,91){\vector(2,1){70}}
 \put(0,87){\vector(1,0){70}}
 \put(0,81){\vector(3,-1){70}}
 \put(0,77){\vector(1,-1){70}}
\end{picture}
&
 \begin{picture}(16,140)(0,0)
 \put(7,0){\mbox{$P$}}
 \put(7,46){\mbox{$\hat P$}}
 \put(7,89){\mbox{$\hat{\cal P}$}}
 \put(7,130){\mbox{$\tilde{\cal P}$}}
 \put(12,103){\vector(0,1){23}}
 \put(12,59){\vector(0,1){25}}
 \put(12,43){\vector(0,-1){29}}
 \put(-1,109){\mbox{$\hat\mu'$}}
 \put(0,64){\mbox{$\hat\iota_0$}}
 \put(0,20){\mbox{$\hat\delta$}}
 \end{picture}
 &
 \begin{picture}(25,140)(0,0)
 \put(0,135){\vector(1,0){25}}
 \put(10,120){\mbox{$\tilde\jmath_0$}}
 \put(0,4){\vector(1,0){25}}
 \put(25,12){\vector(-1,1){25}}
 \put(15,27){\mbox{$\hat{\rm h}_\delta$}}
 \put(4,8){\mbox{${\rm Id}_\Pi$}}
 \end{picture}
 &
 \begin{picture}(10,140)(0,0)
 \put(0,130){\mbox{${\cal M}\pi$}}
 \put(0,0){\mbox{$\Pi$}}
 \end{picture}
 \end{array}
 $$
 Then, taking into account the commutativity of this diagram,
 and proposition \ref{propositio5}, for every section
 $\hat{\rm h}_\delta$ of $\hat\delta$ we have that
\beann
{\rm F}\Lag_0^*\Theta_{\hat{\rm h}_\delta}^0&=&
{\rm F}\Lag_0^*
(\tilde\jmath_0\circ\hat\mu'\circ\hat\iota_0\circ\hat{\rm h}_\delta)^*\Theta=
(\tilde\jmath_0\circ\hat\mu'\circ\hat\iota_0\circ\hat{\rm h}_\delta\circ
{\rm F}\Lag_0)^*\Theta
 \\ &=&
(\tilde\jmath_0\circ\widetilde{{\cal F}\Lag}_0)^*\Theta=
\widetilde{{\cal F}\Lag}^*\Theta=\Theta_\Lag
\eeann
and the same result follows for $\Omega_{{\rm h}_\delta}^0$.
 \qed

 {\bf Remarks}:
 \bit
 \item
 Following the terminology of sections above, we have that all the sections
 $\hat{\rm h}_\delta$ belong to the same equivalence class.
  \item
 In the particular situation that
 ${\rm rank}\,\widehat{{\rm F}\Lag}_0={\rm rank}\,\widehat{{\cal F}\Lag}_0$,
 we have that $\hat\iota_0$ is a diffeomorphism and, as the fibers of
 $\hat\delta$ are also the fibers of $\hat\iota_0$, then so is $\hat\delta$.
 In this case there is only one map $\hat{\rm h}_\delta$,
 which is just $\hat\delta^{-1}$.
 \eit

 As in the hyper-regular case, we can construct this Hamiltonian system
 using a connection. Thus, let $\nabla$ be
 connection in $\pi\colon E\to M$, and
 $h_\delta^\nabla\colon\Pi\to J^1E^*$ the induced linear section of $\delta$.
 Let $\del\in\df^m(J^1E)$ be the density of Lagrangian energy
 associated with $\nabla$ (see definition \ref{energ}).
 Then:

\begin{prop}
 \ben
 \item
 The density of Lagrangian energy $\del$ is ${\rm F}\Lag$-projectable.
 \item
 The $\bar\pi^1$-semibasic $m$-form
 ${\rm F}\Lag^*\Theta_{h_\delta^\nabla}-\Theta_{\Lag}$
 is ${\rm F}\Lag$-projectable and denoting
 $\Theta_{h_\delta^\nabla}^0=\j^*_0\Theta_{h_\delta^\nabla}$, we have
 $$
 \del= {\rm F}\Lag^*\Theta_{h_\delta^\nabla}-\Theta_{\Lag}=
 {\rm F}\Lag_0^*\Theta_{h_\delta^\nabla}^0-\Theta_{\Lag}
 $$
 \item
 There exists a unique $\bar\rho^1_0$-semibasic form
 ${\mbox{\es H}}_0^\nabla\in\df^m(P)$,
 such that
 $$
{\rm F}\Lag^*_0{\mbox{\es H}}^\nabla_0=\del
 $$
 Let ${\mbox{\es H}}^\nabla_0={\rm H}^\nabla_0(\bar\rho_0^{1*}\omega)$,
 with ${\rm H}^\nabla_0\in\Cinfty (P)$. Then
  ${\mbox{\es H}}^\nabla_0$ and ${\rm H}^\nabla_0$ are called
 the {\rm Hamiltonian density} and  the {\rm Hamiltonian function}
 associated with the Lagrangian system,
 the connection $\nabla $ and $\omega$.
 Obviously we have that
 ${\rm F}\Lag^*_0{\rm H}^\nabla_0={\rm E}^\nabla_\Lag$.
 \item
 The Hamilton-Cartan forms of definition \ref{faltaba} split as
 \bea
 \Theta_{\hat{\rm h}_\delta}^0&=&
 \j_0^*\Theta_{h_\delta^\nabla}-{\mbox{\es H}}^\nabla_0 =
 \Theta_{h_\delta^\nabla}^0-{\mbox{\es H}}^\nabla_0
 \nonumber \\
 \Omega_{\hat{\rm h}_\delta}^0&=&
 -\d\Theta_{\hat{\rm h}_\delta}^0=\j_0^*\Omega_{h_\delta^\nabla}+
 \d{\mbox{\es H}}^\nabla_0= \Omega_{h_\delta^\nabla}^0+
\d{\mbox{\es H}}^\nabla_0
 \label{splitar}
 \eea
  \een
 \label{proalmost}
 \end{prop}
 \proof
 \ben
 \item
 As $\del=\fel(\bar\pi^{1*}\omega)$, it suffices to prove that the
 Lagrangian energy $\fel$ is ${\rm F}\Lag$-projectable. Then,
 for every $X\in\ker\,{\rm F}\Lag_*$,
 using natural coordinates we have
 \beann
 \Lie (X)\fel &=&
 \Lie\left( f^B_\eta\derpar{}{v^B_\eta}\right)
 \left(\derpar{\lag}{v^A_\nu}(v^A_\nu-{\mit\Gamma}^A_\nu)-\lag\right)
\\ &=&
 f^B_\eta\frac{\partial^2\lag}{\partial v^B_\eta\partial v^A_\nu}
 (v^A_\nu-{\mit\Gamma}^A_\nu) +
 f^B_\eta\derpar{\lag}{v^A_\nu}\delta^A_B\delta^\eta_\nu-
 f^B_\eta\derpar{\lag}{v^B_\eta}=0
 \eeann
 therefore $\fel$ is ${\rm F}\Lag$-projectable, and so is $\del$.
 \item
 It is immediate, taking into account that $\Theta_\Lag$ is
 ${\rm F}\Lag$-projectable, and the first item of Proposition \ref{delpro}.
 \item
 The existence is assured, since $\del$ is ${\rm F}\Lag$-projectable
 and the uniqueness because ${\rm F}\Lag_0$ is a submersion.

 Next we prove that ${\mbox{\es H}}^\nabla_0$ is $\bar\rho^1_0$-semibasic.
 As ${\rm F}\Lag_0$ is a submersion,
 for every $y\in J^1E$ and
 $\tilde u\in{\rm V}_{{\rm F}\Lag_0(y)}(\bar\rho^1_0)$,
 there exist $u\in\Tan_y(J^1E)$ such that
 $\tilde u=\Tan_y{\rm F}\Lag_0(u)$ and, in addition,
 $u\in{\rm V}_y(\bar\pi^1)$ because
 $$
 \Tan_y\bar\pi^1(u)=
 (\Tan_{{\rm F}\Lag_0(\bar y)}\bar\rho^1_0\circ\Tan_y{\rm F}\Lag_0)(u) =
 \Tan_{{\rm F}\Lag_0(\bar y)}\bar\rho^1_0(\tilde u)=0
 $$
 Furthermore, $\del$ is $\bar\pi^1$-semibasic, and hence
 $$
 0=\inn (u)[\del (\bar y)]=
 \inn (u)[({\rm F}\Lag_0^*{\mbox{\es H}}^\nabla_0)(y)]=
 ({\rm F}\Lag_0)_{{\rm F}\Lag_0(\bar y)}^*
 [\inn (\tilde u)({\mbox{\es H}}^\nabla_0({\rm F}\Lag_0(\bar y))]
 $$
 then, for every $y\in J^1E$ and
 $\tilde u\in{\rm V}_{{\rm F}\Lag_0(\bar y)}(\bar\rho^1_0)$
 we have
 $\inn (\tilde u)({\mbox{\es H}}^\nabla_0({\rm F}\Lag_0(\bar y))
 \in {\rm ker}\,({\rm F}\Lag_0)_{{\rm F}\Lag_0(\bar y)}^*=\{ 0\}$,
 since ${\rm F}\Lag_0$ is a submersion.
 So ${\mbox{\es H}}^\nabla_0$ is $\bar\rho^1_0$-semibasic.
 \item
 Taking into account items 3 and 2, we obtain
 $$
 {\rm F}\Lag_0^*(\Theta_{h_\delta^\nabla}^0-{\mbox{\es H}}^\nabla_0)=
 {\rm F}\Lag_0^*\Theta_{h_\delta^\nabla}^0-
 {\rm F}\Lag_0^*{\mbox{\es H}}^\nabla_0=
 {\rm F}\Lag_0^*\Theta_{h_\delta^\nabla}^0-\del=
 \Theta_\Lag={\rm F}\Lag_0^*\Theta_{\hat{\rm h}_\delta}^0
 $$
 and therefore $\Omega_\Lag={\rm F}\Lag_0^*\Omega_{\hat{\rm h}_\delta}^0$
 too. Then the result follows because ${\rm F}\Lag$ is a submersion.
 \een
 \qed

 We can construct the above Hamiltonian density in an alternative way,
 as follows:
 
 \begin{prop}
 Let $\hat{\rm h}_\delta\colon P\to\hat P$ be a section of $\hat\delta$,
 and $\nabla$ a connection.
 Then, $h_\delta^\nabla$ induces a map
 $\hat h_\delta^\nabla\colon P\to\hat P$
 defined by the relation
 $\hat\j_0\circ\hat h_\delta^\nabla= h_\delta^\nabla\circ\j_0$.
 Therefore
 $$
 (\hat\jmath_0\circ\hat\iota_0\circ\hat h^\nabla_\delta)^*\Theta-
 (\tilde\jmath_0\circ\hat\mu'\circ\hat\iota_0\circ\hat{\rm h}_\delta)^*
 \Theta= {\mbox{\es H}}^\nabla_0
 $$
  and hence the splitting (\ref{splitar}) holds.
 \label{once}
 \end{prop}
 \proof
 We have the following diagram
 $$
\begin{array}{cccccc}
\begin{picture}(15,140)(0,0)
\put(0,80){\mbox{$J^1E$}}
\end{picture}
&
 \begin{picture}(70,140)(0,0)
 \put(28,120){\mbox{$\widetilde{{\cal F}\Lag_0}$}}
 \put(36,91){\mbox{$\widehat{{\cal F}\Lag_0}$}}
 \put(36,51){\mbox{$\widehat{{\rm F}\Lag_0}$}}
 \put(20,30){\mbox{${\rm F}\Lag_0$}}
 \put(0,91){\vector(2,1){70}}
 \put(0,87){\vector(1,0){70}}
 \put(0,81){\vector(3,-1){70}}
 \put(0,77){\vector(1,-1){70}}
\end{picture}
&
 \begin{picture}(160,140)(0,0)
 \put(7,0){\mbox{$P$}}
 \put(7,46){\mbox{$\hat P$}}
 \put(7,89){\mbox{$\hat{\cal P}$}}
 \put(7,130){\mbox{$\tilde{\cal P}$}}
 \put(12,103){\vector(0,1){23}}
 \put(12,59){\vector(0,1){25}}
 \put(12,43){\vector(0,-1){29}}
 \put(-1,109){\mbox{$\hat\mu'$}}
 \put(0,64){\mbox{$\hat\iota_0$}}
 \put(0,20){\mbox{$\hat\delta$}}
 \put(25,4){\vector(1,0){25}}
 \put(50,14){\vector(-1,1){25}}
 \put(38,27){\mbox{$\hat{\rm h}_\delta$}}
 \put(28,8){\mbox{${\rm Id}_P$}}
 \put(65,36){\vector(0,-1){23}}
 \put(65,52){\vector(0,1){25}}
 \put(65,97){\vector(0,1){25}}
 \put(67,20){\mbox{$\hat\delta$}}
 \put(67,64){\mbox{$\hat\iota_0$}}
 \put(67,109){\mbox{$\hat\mu'$}}
 \put(60,0){\mbox{$P$}}
 \put(60,38){\mbox{$\hat P$}}
 \put(60,81){\mbox{$\hat{\cal P}$}}
 \put(60,125){\mbox{$\tilde{\cal P}$}}
 \put(75,4){\vector(1,0){25}}
 \put(100,11){\vector(-1,1){24}}
 \put(95,20){\mbox{$\hat h_\delta^\nabla$}}
 \put(76,8){\mbox{${\rm Id}_P$}}
 \put(105,0){\mbox{$P$}}
 \put(25,139){\vector(1,0){130}}
 \put(75,126){\vector(1,0){80}}
 \put(100,129){\mbox{$\tilde\jmath_0$}}
 \put(25,94){\vector(1,0){130}}
 \put(75,81){\vector(1,0){80}}
 \put(100,85){\mbox{$\hat\jmath_0$}}
 \put(25,50){\vector(1,0){130}}
 \put(75,38){\vector(1,0){80}}
 \put(110,42){\mbox{$\hat{\j}_0$}}
 \put(120,2){\vector(1,0){35}}
 \put(133,6){\mbox{$\j_0$}}
\end{picture}
&
\begin{picture}(20,140)(0,0)
 \put(2,0){\mbox{$\Pi$}}
 \put(-3,42){\mbox{$J^1E^*$}}
 \put(0,85){\mbox{${\cal M}\pi$}}
 \put(0,129){\mbox{${\cal M}\pi$}}
 \put(5,56){\vector(0,1){25}}
 \put(5,37){\vector(0,-1){25}}
 \put(8,64){\mbox{$\iota_0$}}
 \put(8,20){\mbox{$\delta$}}
\end{picture}
 &
 \begin{picture}(25,140)(0,0)
 \put(0,4){\vector(1,0){25}}
 \put(25,12){\vector(-1,1){25}}
 \put(15,27){\mbox{$h_\delta^\nabla$}}
 \put(4,8){\mbox{${\rm Id}_\Pi$}}
 \end{picture}
 &
 \begin{picture}(10,140)(0,0)
 \put(0,0){\mbox{$\Pi$}}
 \end{picture}
 \end{array}
 $$
 Taking into account the commutativity of this diagram,
 we have that every section $\hat{\rm h}_\delta$ of $\hat\delta$
 satisfies that
$$
 \hat\mu'\circ\hat\iota_0\circ\hat{\rm h}_\delta\circ{\rm F}\Lag_0=
 \widetilde{{\cal F}\Lag}_0
$$
 Then, bearing in mind the second item of Proposition \ref{proalmost}
 and (\ref{proprel2}), we have that
 \beann
  {\rm F}\Lag_0^*
 [(\hat\jmath_0\circ\hat\iota_0\circ\hat h^\nabla_\delta)^*\Theta-
 (\tilde\jmath_0\circ\hat\mu'\circ\hat\iota_0\circ\hat{\rm h}_\delta)^*
 \Theta]  &=&
 (\hat\jmath_0\circ\hat\iota_0
 \circ\hat h^\nabla_\delta\circ{\rm F}\Lag_0)^*\Theta-
 (\tilde\jmath_0\circ\hat\mu'\circ\hat\iota_0
 \circ\hat{\rm h}_\delta\circ{\rm F}\Lag_0)^* \Theta
 \\ &=&
 (\iota_0 \circ\hat\j_0\circ\hat h^\nabla_\delta\circ{\rm F}\Lag_0)^*\Theta-
 (\tilde\jmath_0\circ\hat\mu'\circ\hat\iota_0
 \circ\hat{\rm h}_\delta\circ{\rm F}\Lag_0)^* \Theta
 \\ &=&
 (\iota_0\circ h^\nabla_\delta\circ\j_0\circ{\rm F}\Lag_0)^* \Theta
 -(\tilde\jmath_0\circ\widetilde{{\cal F}\Lag}_0)^*\Theta
 \\ &=&
 (\j_0\circ{\rm F}\Lag_0)^* \Theta_{h_\delta^\nabla}
 -(\tilde\jmath_0\circ\widetilde{{\cal F}\Lag}_0)^*\Theta
 \\ &=&
 {\rm F}\Lag^* \Theta_{h_\delta^\nabla}-{\cal F}\Lag^*\Theta =
 \del+\Theta_\Lag-\Theta_\Lag = \del
 \eeann
 and the result follows as a consequence of
 the third item of Proposition \ref{proalmost}.

 The statement for the splittings of $\Theta^0_{\hat{\rm h}_\delta}$ and
 $\Omega^0_{\hat{\rm h}_\delta}$ is immediate.
 \qed

 Note that, once again, the use of both extended Legendre maps
 is necessary to obtain the Hamiltonian density in this way.

 Finally, in the almost-regular case, the Hamilton-Jacobi variational
 principle of definition \ref{hjvp} is stated in the same way, now
 using sections of $\bar\rho^1_0\colon P\to M$, and the form
 $\Theta_{\hat{\rm h}_\delta}^0$.
 So we look for sections $\psi_0\in\Gamma_c(M,P)$ which are
 stationary with respect to the variations given by
 $\psi_{0t}=\sigma_t\circ\psi_0$, where $\{\sigma_t\}$ is a local
 one-parameter group of any $\bar\rho_0^1$-vertical vector field
 $Z\in\vf (P)$, such that
 $$
 \frac{\d}{\d t}\Big\vert_{t=0}
\int_M\psi_{0t}^*\Theta_{\hat{\rm h}_\delta}^0= 0
 $$
 Then these critical sections will be characterized by the condition
 (analogous to Theorem \ref{equics})
 \beq
 \psi_0^*\inn (\tilde X^0)\Omega_{\hat{\rm h}_\delta}^0 = 0
 \quad ; \quad
 \mbox{\rm for every $\tilde X^0\in\vf (P)$}
 \label{equics2}
 \eeq

 \subsection{Equivalence between the Lagrangian and Hamiltonian formalisms}
 \protect\label{vp}

One expects that both the Lagrangian and Hamiltonian formalism
must be equivalent. As in mechanics, this equivalence
 can be proved by using the (reduced) Legendre map.

First, using the Legendre map, we can lift sections of $\pi$ from
$E$ to $\Pi$ as follows:

\begin{definition}
Let $\ls$ be a hyper-regular Lagrangian system, ${\rm F}\Lag$ the
induced Legendre transformation, $\phi\colon M\to E$ a section of
$\pi$ and $j^1\phi\colon M\to J^1E$ its canonical prolongation to
$J^1E$. The {\rm Lagrangian prolongation} of $\phi$ to $\Pi$ is
the section $$ j^{1*}\phi:={\rm F}\Lag\circ j^1\phi\colon M\to \Pi
$$ If $\ls$ is an almost-regular Lagrangian system, the {\sl
Lagrangian prolongation} of a section
 $\phi\colon M\to E$ to ${\rm P}$ is
 $$
 j_0^{1*}\phi:={\rm F}\Lag_0\circ j^1\phi\colon M\to {\rm P}
 $$
\end{definition}

\begin{teor}
{\rm (Equivalence theorem for sections)} Let $\ls$ and $\hslpi$ be
the Lagrangian and Hamiltonian descriptions of a hyper-regular
system.

If a section $\phi\in\Gamma_c(M,E)$ is a solution of the
Lagrangian variational problem (Hamilton principle) then the
section $\psi\equiv j^{1*}\phi:={\rm F}\Lag\circ
j^1\phi\in\Gamma_c(M,\Pi)$ is a solution of the Hamiltonian
variational problem (Hamilton-Jacobi principle).

Conversely, if a section $\psi\in\Gamma_c(M,\Pi)$ is a solution of
the Hamiltonian variational problem, then the section
$\phi\equiv\rho^1\circ\psi\in\Gamma_c(M,E)$ is a solution of the
Lagrangian variational problem. \label{equiv1}
\end{teor}
\proof Bearing in mind the diagram
 \beq
\begin{array}{ccc}
J^1E &
\begin{picture}(135,10)(0,0)
\put(63,6){\mbox{${\rm F}\Lag$}} \put(0,3){\vector(1,0){135}}
\end{picture}
& \Pi
\\ &
\begin{picture}(135,100)(0,0)
\put(34,82){\mbox{$\pi^1$}} \put(90,82){\mbox{$\rho^1$}}
\put(30,55){\mbox{$j^1\phi$}} \put(96,55){\mbox{$\psi$}}
\put(75,30){\mbox{$\pi$}} \put(55,30){\mbox{$\phi$}}
\put(63,55){\mbox{$E$}} \put(65,0){\mbox{$M$}}
\put(0,100){\vector(3,-2){55}} \put(135,100){\vector(-3,-2){55}}
\put(55,13){\vector(-2,3){55}} \put(80,13){\vector(2,3){55}}
\put(65,13){\vector(0,1){30}} \put(71,43){\vector(0,-1){30}}
\end{picture} &
\end{array}
\label{dia} \eeq
 If $\phi$ is a solution of the Lagrangian
variational problem then $(j^1\phi)^*\inn (X)\Omega_{\Lag}=0$, for
every $X\in\vf (J^1E)$ (Theorem \ref{importantlag}); therefore, as
${\rm F}\Lag$ is a local diffeomorphism,
 \beann
0&=&(j^1\phi)^*\inn (X)\Omega_{\Lag}= (j^1\phi)^*\inn (X)({\rm
F}\Lag^*\Omega_{{\rm h}_\delta})
\\ &=&
(j^1\phi)^*{\rm F}\Lag^*(\inn ({\rm F}\Lag_*^{-1}X)\Omega_{{\rm h}_\delta})
= ({\rm F}\Lag\circ j^1\phi)^*(\inn (X')\Omega_{{\rm h}_\delta})
\eeann
 which holds for every $X'\in\vf (\Pi)$ and thus, by Theorem
\ref{equics}, $\psi\equiv {\rm F}\Lag\circ j^1\phi$ is a
solution of the Hamiltonian variational problem. (This proof holds
also for the almost-regular case).

Conversely, let $\psi\in\Gamma_c(M,\Pi)$ be a solution of the
Hamiltonian variational problem. Reversing the above
reasoning we obtain that $({\rm F}\Lag^{-1}\circ\psi)^*\inn
(X)\Omega_{\Lag}=0$, for every $X\in\vf (J^1E)$, and hence
$\sigma\equiv {\rm F}\Lag^{-1}\circ\psi\in\Gamma_c(M,J^1E)$ is a
critical section for the Lagrangian variational problem. Then, as
we are in the hyper-regular case, $\sigma$ must be a
holonomic section \cite{EMR-98}, \cite{LMM-95}, \cite{Sa-89},
 $\sigma =j^1\phi$, and since
(\ref{dia}) is commutative, $\phi
=\rho^1\circ\psi\in\Gamma_c(M,E)$, necessarily. \qed

 {\bf Remarks}:
 \bit
\item
 Observe that every section $\psi\colon M\to \Pi$ which is solution
of the Hamilton-Jacobi variational principle is necessarily a
Lagrangian prolongation of a section $\phi\colon M\to E$.
 \item
 In the almost-regular case, if $\phi$ is a critical section of
 the Lagrangian problem, then $\psi={\rm F}\Lag\circ j^1\phi$
 is a critical section of the Hamiltonian problem.
 Furthermore, ${\rm F}\Lag\colon j^1\phi(M)\to{\rm F}\Lag(j^1\phi(M))$
 is a diffeomorfism because sections of $\bar\pi^1$ are transversal
 to the fibres of ${\rm F}\Lag$.

 As critical sections are integral manifolds of
 multivector fields \cite{EMR-98}, \cite{EMR-99b}
 or, what is equivalent, Ehresmann connections
 \cite{LMM-95}, \cite{LMM-96b}, \cite{Sa-89},
 then critical sections through different points in the same fiber of
 ${\rm F}\Lag$ have the same image by ${\rm F}\Lag$.
 \eit

On the other hand, we can prove the equivalence between
the Lagrangian and Hamiltonian formalisms from the variational point of view.
 First, we need the following lemma:

\begin{lem}
Let $\beta\in\df^m(J^1E)$ and $f\in\Cinfty (J^1E)$. For every
differentiable section $\phi\colon U\subset M\to E$, the following
conditions are equivalent: \ben
\item
$(j^1\phi )^*[f(\bar\pi^{1*}\omega) ]=(j^1\phi )^*\beta$.
\item
\dst\int_{j^1\phi}f(\bar\pi^{1*}\omega) =\int_{j^1\phi}\beta\) . \een
\label{lema}
\end{lem}
\proof Trivially $1\ \Rightarrow\ 2$.

Conversely, if we suppose $1$ is not true, then there exists one
section $\phi\colon U\subset M\to E$ with $(j^1\phi )^*[f(\bar\pi^{1*}\omega)
-\beta ]\not= 0$ and hence there is $x\in U$ and a closed
neighbourhood $V$ of $x$ in $U$ such that, taking $\gamma\colon
V\to E$ with $\gamma =\phi\vert_V$, then
\dst\int_{j^1\gamma}[f(\bar\pi^{1*}\omega) -\beta ]\not=0\) , so $2$ is false.
\qed

Now, let $\nabla$ be a connection in $\pi\colon E\to M$,
and let $\del=\fel(\bar\pi^{1*}\omega)$ be the density of Lagrangian energy
 associated with $\nabla$ and $\Lag$. Then:

\begin{teor}
The Lagrangian energy function is the unique function in $J^1E$ verifying the
following condition:  for every section
 $\phi\colon U\subset M\to E$,
 $$ (j^1\phi )^*[{\rm E}^{\nabla}_{\Lag}(\bar\pi^{1*}\omega)]=
 (j^1\phi )^*({\rm F}\Lag^*\Theta_{h_\delta^\nabla}-\Lag ) $$
\label{condi}
\end{teor}
\proof (Uniqueness):\quad Let $f$ and $g$ be two functions
verifying this condition. Obviously
 $(j^1\phi )^*[(f-g)(\bar\pi^{1*}\omega)]=0$,
 but $0=(j^1\phi )^*[(f-g)(\bar\pi^{1*}\omega)]=
(f-g)(j^1\phi (x))(\bar\pi^{1*}\omega)$,
for every $x\in U$. Hence, $(f-g)(j^1\phi (x))=0$, and this
implies $f-g=0$, because every point in $J^1E$ is in the image of
some section $j^1\phi$.

\quad\quad (Existence):\quad
 From (\ref{form1}) we obtain
 \beann
 (j^1\phi )^*({\rm F}\Lag^*\Theta_{h_\delta^\nabla}-\Lag )&=&
(j^1\phi )^*[\Theta_{\Lag}+{\rm E}_{\Lag}^{\nabla}(\bar\pi^{1*}\omega) -\Lag]
\\ &=&
(j^1\phi )^*[\inn ({\cal V})\d\Lag +\Lag +{\rm
E}_{\Lag}^{\nabla}(\bar\pi^{1*}\omega) -\Lag] =
 (j^1\phi )^*[{\rm E}_{\Lag}^{\nabla}(\bar\pi^{1*}\omega)]
 \eeann
 since $(j^1\phi )^*(\inn ({\cal V})\d\Lag )=0$
 (as can be proved by using expressions in
 coordinates). So, the energy function introduced in definition
 \ref{energ} satisfies this condition.
 \qed

And this result leads to the following consequence:

\begin{teor}
Let $\ls$ and $\hslpi$ be the Lagrangian and Hamiltonian
descriptions of a hyper-regular system. Then,
the Hamilton variational principle of
the Lagrangian formalism and the Hamilton-Jacobi variational
principle of the Hamiltonian formalisms are equivalent.
That is, for every section $\phi\in\Gamma_c(M,E)$ we have that
$$
 \int_{j^1\phi}\Lag = \int_{j^{1*}\phi}\Theta_{{\rm h}_\delta}
$$
 \label{energteo}
\end{teor}
\proof
 The standpoint is the relation stated in Theorem \ref{condi}
 which, by Lemma \ref{lema}, is equivalent to
 $$
 \int_{j^1\phi}{\rm E}_{\Lag}^{\nabla}(\bar\pi^{1*}\omega) =
 \int_{j^1\phi}({\rm F}\Lag^*\Theta_{h_\delta^\nabla}-\Lag )
 $$
 therefore, from this equality, and using (\ref{form1}) and (\ref{aux00}),
 we obtain
 $$
 \int_{j^1\phi}\Lag =
 \int_{j^1\phi}[{\rm F}\Lag^*\Theta_{h_\delta^\nabla}-
 {\rm E}_{\Lag}^{\nabla}(\bar\pi^{1*}\omega)]=
\int_{j^1\phi}\Theta_\Lag= \int_{j^1\phi}{\rm F}\Lag^*\Theta_{{\rm h}_\delta}=
 \int_{{\rm F}\Lag\circ j^1\phi}\Theta_{{\rm h}_\delta}=
 \int_{j^{1*}\phi}\Theta_{{\rm h}_\delta}
 $$
 \qed

It is important to remark the essential role played by the
 Lagrangian energy function in the proof of this equivalence.

(The above results are generalizations of others in
non-autonomous mechanics \cite{EMR-sdtc}).

\section{Hamiltonian formalism in the restricted multimomentum bundle
         $J^1\pi^*$. Relation with the formalism in $\Pi$}

\subsection{Hamiltonian systems}
 \protect\label{hsjpsect}

 The construction of the Hamiltonian formalism in $J^1\pi^*$ is
 posed, for the first time,
 in \cite{CCI-91}, and the particular case of Hamiltonian
 systems associated with hyper-regular and almost-regular systems
 is stated in \cite{LMM-96}.
 The procedure is essentially similar to that developed in section
 \ref{dsc} for $\Pi$. Next we sketch this construction,
 relating it with the above one in $\Pi$.
 As we have proved the existence of the canonical diffeomorphism
 ${\mit\Psi}\colon\Pi\to J^1\pi^*$, we can use it to prove the equivalence
 between the Hamiltonian formalisms in $\Pi$ and $J^1\pi^*$.

 \begin{definition}
 Consider the bundle $\bar\tau^1\colon J^1\pi^*\to M$.
\ben
\item
 A section $h_\mu\colon J^1\pi^*\to{\cal M}\pi$ of the projection
 $\mu$ is called a {\rm Hamiltonian section} of $\mu$.
\item
 The differentiable forms
 $$
 \Theta_{h_\mu}:=h_\mu^*\Theta \quad ,\quad
 \Omega_{h_\mu}:=-\d\Theta_{h_\mu}=h_\mu^*\Omega
 $$
 are called the {\rm Hamilton-Cartan $m$ and $(m+1)$ forms} of $J^1\pi^*$
 associated with the Hamiltonian section $h_\mu$.
\item
 The couple $\hsjpi$ is said to be a {\rm Hamiltonian system}.
\een
\end{definition}

 In a local chart of natural coordinates,
 a Hamiltonian section is specified by a
 {\sl local Hamiltonian function}
 $H_{h_\mu}\in\Cinfty (U)$, $U\subset J^1\pi^*$, such that
 $H_{h_\mu}(x^\nu,y^A,p^\nu_A)\equiv
 (x^\nu,y^A,p=-H_{h_\mu}(x^\gamma,y^B,p_B^\eta),p^\nu_A)$.
 The local expressions of the Hamilton-Cartan forms
 associated with $h_\mu$ are similar to
 (\ref{omegaH0}), but changing $H_{h_\delta}$ by $H_{h_\mu}$,
 and ${\rm p}_A^\nu$ by $p_A^\nu$.

 For Hamiltonian sections of $\mu$,
 we have a similar result to that in Lemma \ref{hcfsect},
 and so, if $h^1_\mu,h^2_\mu$ are two sections of $\mu$, then
 \beq
 \Theta_{h_\mu^1}-\Theta_{h_\mu^2}=
 h_\mu^{1*}\Theta-h_\mu^{2*}\Theta=\tau^{1*}(h_\mu^1-h_\mu^2)\equiv{\cal H}
 \label{2sec}
 \eeq
 is a $\bar\tau^1$-semibasic $m$-form in $J^1\pi^*$.

 \begin{definition}
 A $\bar\tau^1$-semibasic form ${\cal H}\in\df^m(J^1\pi^*)$
 is said to be a {\rm Hamiltonian density} in $J^1\pi^*$.

 It can be written as
 ${\cal H}={\rm H}(\bar\tau^{1^*}\omega)$,
 where ${\rm H}\in\Cinfty (J^1\pi^*)$ is the
 {\rm global Hamiltonian function} associated with
 ${\cal H}$ and $\omega$.
 \end{definition}

 If (\ref{2sec}) holds, then the relation between the global
 Hamiltonian function ${\rm H}$ associated with ${\cal H}$,
 and the local Hamiltonian functions $H_{h^1_\mu}$, $H_{h^2_\mu}$
 associated with $h^1_\mu$ and $h^2_\mu$
 is ${\rm H}=H_{h^1_\mu}-H_{h^2_\mu}$ (in an open set $U$).

 In this way we have the analogous result as in Theorem \ref{afinHC}:

\begin{teor}
 The set of Hamilton-Cartan $m$-forms associated with Hamiltonian sections
 of $\mu$ is an affine space modelled on the set of
 Hamiltonian densities in $J^1\pi^*$.
\end{teor}

 Hence, if $\hsjpi$ is a Hamiltonian system,
 we have that every Hamiltonian section $h'_\mu\not=h_\mu$  allows us
 to split globally the Hamilton-Cartan forms as
$$
 \Theta_{h_\mu}=\Theta_{h'_\mu}-{\cal H}
 \quad , \quad
  \Omega_{h_\mu}= \Omega_{h'_\mu}+\d{\cal H}
$$
 The local expressions of these splittings are similar to
 (\ref{omegahpi0}), but changing $H_{h'_\delta}$
 by $H_{h'_\mu}$, and ${\rm p}_A^\nu$ by $p_A^\nu$.

 Now, if we have a connection $\nabla$ in $\pi\colon E\to M$,
 it induces a linear section $h_\delta^\nabla\colon\Pi\to J^1E^*$
 of $\delta$, and hence there exists another linear section
 $h_\mu^\nabla\colon J^1\pi^*\to{\cal M}\pi$ of $\mu$ given by
 $h^\nabla_\mu\circ{\mit\Psi}^{-1}=\iota_0\circ h^\nabla_\delta$
 (see \cite{CCI-91} for an alternative definition).
 Then, if $\Theta$ is the canonical $m$-form in $\df^m({\cal M}\pi)$, the forms
 $$
 \Theta_{h_\mu^\nabla}:= h_\mu^{\nabla*}\Theta\in\df^m(J^1\pi^*)
 \quad ,\quad
 \Omega_{h_\mu^\nabla}:=-\d\Theta_{h_\mu^\nabla}\in\df^{m+1}(J^1\pi^*)
 $$
 are the {\sl Hamilton-Cartan $m$ and $(m+1)$ forms} of $J^1\pi^*$
 associated with the connection $\nabla$.
 Of course, a characterization of $\Theta_{h_\mu^\nabla}$
 can be stated in the same way as in (\ref{caracteriza}).
 The local expression of these Hamilton-Cartan forms
 associated with $\nabla$ is similar to (\ref{Lforms}).

 Therefore, given a connection $\nabla$ and a
 Hamiltonian section $h_\mu$, from the above results we have that
 $$
 \tau^{1*}( h_\mu^\nabla-h_\mu)=
  h_\mu^{\nabla*}\Theta-h_\mu^*\Theta=
 \Theta_{h_\mu^\nabla}-\Theta_{h_\mu}:=
 {\cal H}^\nabla_{h_\mu}
 $$
 is a Hamiltonian density in $J^1\pi^*$, which is written as
 ${\cal H}^\nabla_{h_\mu}=
 {\rm H}^\nabla_{h_\mu}(\bar\tau^{1^*}\omega)$,
 where ${\rm H}^\nabla_{h_\mu}\in\Cinfty (J^1\pi^*)$ is the
 {\sl global Hamiltonian function} associated with
 ${\cal H}^\nabla_{h_\mu}$ and $\omega$. Then,
 the Hamilton-Cartan forms associated with $h_\mu$ split as
 $$
 \Theta_{h_\mu}=\Theta_{h_\mu^\nabla}-
 {\cal H}^\nabla_{h_\mu}
 \quad ,\quad
 \Omega_{h_\mu}=
 \Omega_{h_\mu^\nabla}+\d{\cal H}^\nabla_{h_\mu}
 $$
 The local expressions of these splittings are similar to
 (\ref{omegahpi}), but changing ${\rm H}^\nabla_{h_\delta}$
 by ${\rm H}^\nabla_{h_\mu}$, and ${\rm p}_A^\nu$ by $p_A^\nu$.

 If, conversely, we take a connection $\nabla$ and a
 Hamiltonian density ${\cal H}$, then making
 $ h_\mu^\nabla-{\cal H}$ we obtain
 a Hamiltonian section $h_\mu$, since
 ${\cal H}\colon J^1\pi^*\to{\cal M}\pi$
 takes values in $\pi^*\Lambda^m\Tan^*M$. Hence:

 \begin{prop}
 A couple $(h_\mu,\nabla)$ in $J^1\pi^*$ is equivalent to a couple
 $({\cal H},\nabla)$ (that is, given a connection $\nabla$,
 Hamiltonian sections of $\mu$
 and Hamiltonian densities in $J^1\pi^*$ are in one-to-one correspondence).
 \end{prop}

 Bearing in mind this last result, we have another way of
 obtaining a Hamiltonian system, which consists in giving a couple
 $({\cal H},\nabla)$. In fact:

 \begin{prop}
 Let $\nabla$ be a connection in $\pi\colon E\to M$,
 and  ${\cal H}$ a Hamiltonian density.
 There exists a unique Hamiltonian section $h_\mu$ of $\mu$ such that
 \beq
 \Theta_{h_\mu}=\Theta_{h_\mu^\nabla}-{\cal H} \quad ,\quad
 \Omega_{h_\mu}=-\d\Theta_{h_\mu}=
 \Omega_{h_\mu^\nabla}+\d{\cal H}
 \label{def19}
 \eeq
 \end{prop}

 Concerning field equations, observe that
 diffeomorphisms in $E$ (and hence vector fields in $E$)
 can be lifted to $J^1\pi^*$, for instance, lifting them to
 $\Pi$ (see definitions \ref{proldif} and \ref{prolvf}),
 and translating them to $J^1\pi^*$
 using the diffeomorphism ${\mit\Psi}$.
 Hence, for a Hamiltonian system $\hsjpi$, we can set the
 Hamilton-Jacobi variational principle as in definition \ref{hjvp}
 (but with the form $\Omega_{h_\mu}$ instead of
 $\Omega_{h_\delta}$), and state the same results and comments as
 in Theorem \ref{equics}.

 Hamiltonian systems in $\Pi$ and $J^1\pi^*$ are
 equivalent. In fact; as a first result we have:

 \begin{prop}
 Let $\nabla$ be a connection in $\pi\colon E\to M$,
 and ${\mbox{\es H}}$ and ${\cal H}$ Hamiltonian densities in $\Pi$
 and $J^1\pi^*$, respectively, such that
 ${\mit\Psi}^*\mbox{\es H}={\cal H}$. Then
 $$
 {\mit\Psi}^*\Omega_{h_\mu^\nabla}=\Omega_{h_\delta^\nabla} \quad ,\quad
 {\mit\Psi}^*\Omega_{h_\mu}=\Omega_{h_\delta}
 $$
 \label{relomega}
 \end{prop}
 \proof
 The proof is based in the following fact:
 $$
 \Theta_{h_\mu^\nabla}=h_\mu^{\nabla *}\Theta=
 (\iota_0\circ h_\delta^\nabla\circ{\mit\Psi})^*\Theta=
 {\mit\Psi}^*(\iota_0\circ h_\delta^\nabla)^*\Theta=
 {\mit\Psi}^*\Theta_{h_\delta^\nabla}
 $$
 and the result is immediate.
 \qed

 And therefore, as a direct consequence of Propositions
 \ref{firstprop} and \ref{relomega}, we can set the
 relation between the Hamiltonian systems in $\Pi$ and $J^1\pi^*$:

 \begin{teor}
 Every Hamiltonian system $\hspi$ is equivalent to a Hamiltonian system
 $\hsjpi$, and conversely.
 \end{teor}

 At this point, we can study the relation between the
 set of connections $\nabla$ in the bundle $\pi\colon E\to M$,
 and the sets of linear (Hamiltonian) sections of the projections
 $\mu\colon{\cal M}\pi\to J^1\pi^*$ and
 $\delta\colon J^1E^*\to\Pi$:

 \begin{teor}
 The map $\nabla\mapsto  h_\mu^\nabla$ is a bijective affine map
 between the set of connections in the bundle $\pi\colon E\to M$
 and the set of linear (Hamiltonian) sections of the projection
 $\mu\colon{\cal M}\pi\to J^1\pi^*$ or,
 what is equivalent,
 the set of linear (Hamiltonian) sections of the projection
 $\delta\colon J^1E^*\to\Pi$.
 \end{teor}
 \proof
 Let the projection
 $\mu\colon
 \Lambda^m\Tan^*E\to\Lambda_1^m\Tan^*E/\Lambda_0^m\Tan^*E$,
 and the set of linear sections of $\mu$
 $$
 \Gamma(\mu):=\{\ell\in\L(J^1\pi^*,\Lambda_1^m\Tan^*E)\ ,\
 \mu\circ\ell={\rm Id}_{J^1\pi^*}\}
 $$
 which is an affine bundle modeled on the vector bundle
 \beann
 (J^1\pi^*)^*\otimes\Lambda_0^m\Tan^*E&\simeq&
 (\pi^*\Tan M\otimes{\rm V}^*(\pi)\otimes\pi^*\Lambda^m\Tan^*M)^*
 \otimes\Lambda_0^m\Tan^*E
 \\ &=&
 \pi^*\Tan^*M\otimes{\rm V}(\pi)\otimes\pi^*\Lambda^m\Tan M
 \otimes\Lambda_0^m\Tan^*E\simeq
 \pi^*\Tan^*M\otimes{\rm V}(\pi)
 \eeann
 But this last bundle is just the vector bundle on which the affine
 bundle of the connection forms in $\pi\colon E\to M$
 is modeled. Then the result follows.

 Finally, the equivalence with
 the set of linear (Hamiltonian) sections of the projection
 $\delta\colon J^1E^*\to\Pi$) is proved by taking into account that
 every linear section of $\delta$ is associated with a connection $\nabla$,
 since this linear section defines a linear map from ${\rm V} *(\pi)$
 to $\Tan^*E$, and hence a projection $\Tan E\to {\rm V}(\pi)$
(that is, a connection).
 \qed

 \subsection{Hamiltonian system
  associated with a hyper-regular Lagrangian system}
 \protect\label{hslsjpi1}

 The procedure is analogous to that in Section \ref{hsahrs}
 (see also diagram (\ref{diag1})).
 Let $\ls$ be a hyper-regular Lagrangian system, then:

 \begin{definition}
 Let ${\rm h}_\mu\colon J^1\pi^*\to{\cal M}\pi$ be the section of $\mu$
 given by
 $$
{\rm h}_\mu:= \widetilde{{\cal F}\Lag}\circ{\cal F}\Lag^{-1}
 $$
 which is a diffeomorphism connecting $J^1\pi^*$ and
 $\widetilde{{\cal F}\Lag}(J^1E)$
 (observe that it is just the inverse of
 $\mu$ restricted to $\widetilde{{\cal F}\Lag}(J^1E)$).
 We define the Hamilton-Cartan forms
  $$
 \Theta_{{\rm h}_\mu}:={\rm h}_\mu^*\Theta
 \quad ; \quad
 \Omega_{{\rm h}_\mu}:={\rm h}_\mu^*\Omega
 $$
  \label{probiss}
 \end{definition}

 \begin{prop}
 The Hamilton-Cartan forms satisfy that
 $$
 {\cal F}\Lag^*\Theta_{{\rm h}_\mu}=\Theta_{\Lag}
 \quad ,\quad
 {\cal F}\Lag^*\Omega_{{\rm h}_\mu}=\Omega_{\Lag}
 $$
 Then $\hsljpi$ is the (unique) Hamiltonian system
 associated with the hyper-regular Lagrangian system $\ls$.
 \end{prop}
\proof
 We have the diagram
 $$
\begin{array}{ccccc}
\begin{picture}(15,52)(0,0)
\put(0,0){\mbox{$J^1E$}}
\end{picture}
&
\begin{picture}(65,52)(0,0)
 \put(20,27){\mbox{${\cal F}\Lag$}}
 \put(25,6){\mbox{$\widetilde{{\cal F}\Lag}$}}
 \put(0,7){\vector(2,1){65}}
 \put(0,4){\vector(1,0){65}}
\end{picture}
&
\begin{picture}(15,52)(0,0)
 \put(0,0){\mbox{${\cal M}\pi$}}
 \put(0,41){\mbox{$J^1\pi^*$}}
 \put(10,13){\vector(0,1){25}}
 \put(0,22){\mbox{$\mu$}}
\end{picture}
 &
 \begin{picture}(25,52)(0,0)
 \put(0,44){\vector(1,0){25}}
 \put(25,39){\vector(-1,-1){25}}
 \put(13,16){\mbox{${\rm h}_\mu$}}
 \put(1,50){\mbox{${\rm Id}_{J^1\pi^*}$}}
 \end{picture}
 &
 \begin{picture}(10,52)(0,0)
 \put(0,41){\mbox{$J^1\pi^*$}}
 \end{picture}
\end{array}
 $$
 Taking into account the commutativity of this diagram,
 and proposition \ref{propositio5}, we have
 $$
{\cal F}\Lag^*\Theta_{{\rm h}_\mu}=
{\cal F}\Lag^*{\rm h}_\mu^*\Theta=
\widetilde{{\cal F}\Lag}^*\Theta=\Theta_\Lag
 $$
and the same result follows for $\Omega_{{\rm h}_\delta}$.

 Using charts of natural coordinates in $J^1\pi^*$ and
 ${\cal M}\pi$, and the expression (\ref{restlt}) of the Legendre map,
 we obtain that the local Hamiltonian function $H_{{\rm h}_\mu}$
 representing this Hamiltonian section is
 \beq
 H_{{\rm h}_\mu}(x^\nu,y^A,p^\nu_A)=
 {\cal F}\Lag^{-1*}\left(v^A_\nu\derpar{\lag}{v^A_\nu}-\lag\right)=
 p^\nu_A{\cal F}\Lag^{-1^*}v_\nu^A- {\cal F}\Lag^{-1^*}\lag
 \label{otraH}
 \eeq
 and the local expressions of the corresponding Hamilton-Cartan
 forms are
 \beann
 \Theta_{{\rm h}_\mu} &=& p_A^\nu\d y^A\wedge \d^{m-1}x_\nu -
 (p^\nu_A{\cal F}\Lag^{-1^*}v_\nu^A-{\cal F}\Lag^{-1^*}\lag )\d^mx
 \\
 \Omega_{{\rm h}_\mu} &=& -\d p_A^\nu\wedge\d y^A\wedge\d^{m-1}x_\nu +
 \d (p^\nu_A{\cal F}\Lag^{-1^*}v_\nu^A-
 {\cal F}\Lag^{-1^*}\lag )\wedge\d^mx
 \eeann

 We can construct this Hamiltonian system using connections.
 Thus, if $\nabla$ is a connection in $\pi\colon E\to M$, and
 $h_\mu^\nabla$ is the linear Hamiltonian section of $\mu$
 associated with $\nabla$,
 following the same pattern as in Proposition \ref{delpro},
 we can prove:

 \begin{prop}
 \ben
 \item
 The $m$-form  ${\cal F}\Lag^*\Theta_{h_\mu^\nabla}-\Theta_{\Lag}$
 is $\bar\pi^1$-semibasic and
 $$
 {\cal F}\Lag^*\Theta_{h_\mu^\nabla}-\Theta_{\Lag}={\cal E}_\Lag^{\nabla}
 $$
 \item
 There exists a unique Hamiltonian density
 ${\cal H}^\nabla\in\df^m(J^1\pi^*)$ such that
 \beq
 {\cal F}\Lag^*{\cal H}^\nabla=
 {\cal F}\Lag^*\Theta_{h_\mu^\nabla}-\Theta_{\Lag}=\del
 \label{form2bis}
 \eeq
 Then there exists a function $H^\nabla\in\Cinfty (J^1\pi^*)$
 such that ${\cal H}^\nabla =H^\nabla(\bar\tau^{1*}\omega)$.

 ${\cal H}^\nabla$ and $H^\nabla$ are called the
 {\rm Hamiltonian density} and the {\sl Hamiltonian function}
 associated with the Lagrangian system,
 the connection $\nabla$ and $\omega$.
 \item
 The Hamilton-Cartan forms of definition \ref{probiss} split as
 \bea
 \Theta_{{\rm h}_\mu}&=&
 \Theta_{h_\mu^\nabla}-{\cal H}^\nabla=
 \Theta_{h_\mu^\nabla}-{\cal H}^\nabla
 \nonumber  \\
 \Omega_{{\rm h}_\mu}&=&
 -\d\Theta_{{\rm h}_\mu}=
 \Omega_{h_\mu^\nabla}+\d{\cal H}^\nabla
 \label{splithr}
 \eea
 \een
 \label{delprobis}
 \end{prop}

 We can obtain this Hamiltonian density
 using only Hamiltonian sections. In fact:

 \begin{prop}
 \ben
 \item
 Consider the Hamiltonian section
 ${\rm h}_\mu:=\widetilde{{\cal F}\Lag}\circ {\cal F}\Lag^{-1}$,
 and a connection $\nabla$. Then we have
 $$
 h_\mu^{\nabla *}\Theta-{\rm h}_\mu^*\Theta = {\cal H}^\nabla
 $$
 and hence the splitting (\ref{splithr}) holds.
 \een
 \end{prop}
 \proof
 We have the diagram
 $$
\begin{array}{ccccccc}
\begin{picture}(15,52)(0,0)
\put(0,0){\mbox{$J^1E$}}
\end{picture}
&
\begin{picture}(65,52)(0,0)
 \put(20,27){\mbox{${\cal F}\Lag$}}
 \put(25,6){\mbox{$\widetilde{{\cal F}\Lag}$}}
 \put(0,7){\vector(2,1){65}}
 \put(0,4){\vector(1,0){65}}
\end{picture}
&
\begin{picture}(15,52)(0,0)
 \put(0,0){\mbox{${\cal M}\pi$}}
 \put(0,41){\mbox{$J^1\pi^*$}}
 \put(10,13){\vector(0,1){25}}
 \put(0,22){\mbox{$\mu$}}
\end{picture}
 &
 \begin{picture}(25,52)(0,0)
 \put(0,44){\vector(1,0){25}}
 \put(25,39){\vector(-1,-1){25}}
 \put(13,16){\mbox{${\rm h}_\mu$}}
 \put(1,50){\mbox{${\rm Id}_{J^1\pi^*}$}}
 \end{picture}
 &
 \begin{picture}(10,52)(0,0)
 \put(7,13){\vector(0,1){25}}
 \put(-5,24){\mbox{$\mu$}}
 \put(-5,41){\mbox{$J^1\pi^*$}}
 \put(-5,0){\mbox{${\cal M}\pi$}}
 \end{picture}
 &
 \begin{picture}(25,52)(0,0)
 \put(0,44){\vector(1,0){25}}
 \put(25,39){\vector(-1,-1){25}}
 \put(15,16){\mbox{$h_\mu^\nabla$}}
 \put(1,50){\mbox{${\rm Id}_{J^1\pi^*}$}}
 \end{picture}
 &
 \begin{picture}(10,52)(0,0)
 \put(0,41){\mbox{$J^1\pi^*$}}
 \end{picture}
\end{array}
 $$
 The first item is a consequence of the third item of Proposition
 \ref{hrprop}.
 For the second item, taking into account definition
 \ref{def7} and (\ref{def19}), Proposition
 \ref{propositio5}, and (\ref{form2bis}), we have
 $$
  h_\mu^{\nabla*}\Theta-{\rm h}_\mu^*\Theta =
 \Theta_{h_\mu^\nabla}-
 ({\cal F}\Lag^{-1})^*\widetilde{{\cal F}\Lag}^*\Theta=
 \Theta_{h_\mu^\nabla}-({\cal F}\Lag^{-1})^*\Theta_\Lag=
 {\cal H}^\nabla
 $$
 Then the result for the Hamilton-Cartan forms follows immediately.
 \qed

 Of course, all the results stated in section \ref{vpfe}
 concerning to the variational principle and the characterization of
 critical sections are true, and the local Hamiltonian function
 $H_{{\rm h}_\mu}$ appearing in the Hamilton-De Donder-Weyl equations
 is  given by (\ref{otraH}).

 Finally, the equivalence between the Hamiltonian formalisms (in $\Pi$ and
 $J^1\pi^*$) associated with a hyper-regular
 Lagrangian system, and between the
 Lagrangian formalism and the Hamiltonian formalism in $J^1\pi^*$ is
 given by the following:

 \begin{teor}
 Let $\ls$ be a hyper-regular Lagrangian system. Then
 $$
 {\mit\Psi}^*\Theta_{{\rm h}_\delta}=\Theta_{{\rm h}_\mu}
 \quad ,\quad
 {\mit\Psi}^*\Omega_{{\rm h}_\delta}=\Omega_{{\rm h}_\mu}
 $$
 and hence its associated Hamiltonian systems $\hslpi$ and $\hsljpi$ are
 equivalent.
 \label{equivalencia1}
 \end{teor}
 \proof
 It is immediate.
 Observe also that the sections
 ${\rm h}_\mu$ and ${\rm h}_\delta$ are equivalent,
 by the commutativity of diagram (\ref{diag1}).
 \qed

 Observe that, as ${\cal F}\Lag$ is a diffeomorphism, we also have that
 ${\mit\Psi}^*{\mbox{\es H}}^\nabla={\cal H}^\nabla$.

 \subsection{Hamiltonian system
  associated with an almost-regular Lagrangian system}
 \protect\label{hslsjpi2}

 The procedure is analogous to that in Section \ref{hspiar}
 (see also diagram (\ref{diag2})).
 Now, $\ls$ is an almost-regular Lagrangian system,
 and the submanifold
 $\jmath_0\colon {\cal P}\hookrightarrow J^1\pi^*$,
 is a fiber bundle over $E$ (and $M$). The
 corresponding projections will be denoted by
 $\tau^1_0\colon {\cal P}\to E$ and $\bar\tau^1_0\colon{\cal P}\to M$,
 satisfying that $\tau^1\circ\jmath_0=\tau^1_0$ and
 $\bar\tau^1\circ\jmath_0=\bar\tau^1_0$.

 Taking into account Proposition \ref{kerfl}, and
 following the same pattern as in Propositions \ref{flproj1},
 we can prove that the Lagrangian forms
 $\Theta_{\Lag}$ and $\Omega_{\Lag}$ are
 ${\cal F}\Lag$-projectable, and then:

 \begin{definition}
 Given the diffeomorphism $\tilde{\rm h}_\mu=\tilde \mu^{-1}$,
 we define the Hamilton-Cartan forms
  $$
 \Theta^0_{\tilde{\rm h}_\mu}=\tilde{\rm h}_\mu^*\Theta
 \quad ; \quad
 \Omega^0_{\tilde{\rm h}_\mu}=\tilde{\rm h}_\mu^*\Omega
 $$
 \label{hsar0}
 \end{definition}

 \begin{prop}
 The Hamilton-Cartan forms satisfy that
 $$
 {\cal F}\Lag_0^*\Theta^0_{\tilde{\rm h}_\mu}=\Theta_{\Lag}
 \quad ,\quad
 {\cal F}\Lag_0^*\Omega^0_{\tilde{\rm h}_\mu}=\Omega_{\Lag}
 $$
 Then $\hsjpio$ is the (unique) Hamiltonian system
 associated with the almost-regular Lagrangian system $\ls$.
 \end{prop}
\proof
 In fact, taking into account the commutativity of diagram (\ref{diag2}),
 and proposition \ref{propositio5}, we have
 $$
{\cal F}\Lag_0^*\Theta_{\hat{\rm h}_\mu}^0=
{\cal F}\Lag_0^*(\tilde\jmath_0\circ\hat{\rm h}_\mu)^*\Theta=
(\tilde\jmath_0\circ\hat{\rm h}_\mu\circ{\cal F}\Lag_0)^*\Theta=
(\tilde\jmath_0\circ\widetilde{{\cal F}\Lag}_0)^*\Theta=
\widetilde{{\cal F}\Lag}^*\Theta=\Theta_\Lag
 $$
and the same result follows for $\Omega_{{\rm h}_\mu}^0$.
 \qed

 We can construct this Hamiltonian system
 using a connection. Thus, let $\nabla$ be a
 connection in $\pi\colon E\to M$, and
 $h_\mu^\nabla\colon J^1\pi^*\to{\cal M}\pi$
 the induced linear section of $\mu$.
 Let $\del\in\df^m(J^1E)$ be the density of Lagrangian energy
 associated with $\nabla$. Then,
 as in Proposition \ref{proalmost} we can prove:

 \begin{prop}
 \ben
 \item
 The $\bar\pi^1$-semibasic $m$-form
 ${\cal F}\Lag^*\Theta_{h_\mu^\nabla}-\Theta_{\Lag}$
 is ${\cal F}\Lag$-projectable and,
 if $\Theta^0_{h_\mu^\nabla}=\jmath^*_0\Theta_{h_\mu^\nabla}$, then
 $$
 \del= {\cal F}\Lag^*\Theta_{h_\mu^\nabla}-\Theta_{\Lag}=
 {\cal F}\Lag_0^*\Theta^0_{h_\mu^\nabla}-\Theta_{\Lag}
 $$
 \item
 There exists a unique $\bar\tau^1_0$-semibasic form
 ${\cal H}_0^\nabla\in\df^m({\cal P})$,
 such that
 ${\cal F}\Lag^*_0{\cal H}^\nabla_0=\del$.
 Then, there is a function $H^\nabla_0\in\Cinfty ({\cal P})$
 such that
 ${\cal H}^\nabla_0=H^\nabla_0(\bar\tau_0^{1*}\omega)$.
 Obviously we have that
 ${\cal F}\Lag^*_0H^\nabla_0={\rm E}^\nabla_\Lag$.

 ${\cal H}^\nabla_0$ and $H^\nabla_0$ are called
 the {\rm Hamiltonian density} and  the {\rm Hamiltonian function}
 associated with the Lagrangian system,
 the connection $\nabla $ and $\omega$.
 \item
 The Hamilton-Cartan forms of definition \ref{hsar0} split as
 \bea
 \Theta_{\tilde{\rm h}_\mu}^0&=&
 \jmath_0^*\Theta_{h_\mu^\nabla}-{\cal H}^\nabla_0 =
 \Theta_{h_\mu^\nabla}^0-{\cal H}^\nabla_0
 \nonumber
\\
 \Omega_{\tilde{\rm h}_\mu}^0&=&
 -\d\Theta_{\tilde{\rm h}_\mu}^0=\jmath_0^*\Omega_{h_\mu^\nabla}+
 \d{\cal H}^\nabla_0=\Omega_{h_\mu^\nabla}^0+\d{\cal H}^\nabla_0
 \label{formcalh0}
 \eea
 \een
 \label{proprel3}
 \end{prop}

 We can obtain this Hamiltonian system
 in the following equivalent way:

 \begin{prop}
 Consider the map $\tilde{\rm h}_\mu$, and a connection $\nabla$.
 Then $h^\nabla_\mu$ induces a map
 $\tilde h_\mu^\nabla\colon{\cal P}\to\tilde{\cal P}$,
 defined by the relation
 $\tilde\jmath_0\circ\tilde h_\mu^\nabla= h_\mu^\nabla\circ\jmath_0$.
 Therefore
 $$
 (\tilde\jmath_0\circ\tilde h^\nabla_0)^*\Theta-
 (\tilde\jmath_0\circ\tilde{\rm h}_\delta)^*\Theta=
 {\cal H}^\nabla_0
 $$
 and hence the splitting (\ref{formcalh0}) holds.
 \end{prop}
 \proof
 We have the diagram
 $$
\begin{array}{cccccc}
\begin{picture}(15,52)(0,0)
\put(0,0){\mbox{$J^1E$}}
\end{picture}
&
\begin{picture}(65,52)(0,0)
 \put(17,28){\mbox{${\cal F}\Lag_0$}}
 \put(24,7){\mbox{$\widetilde{{\cal F}\Lag_0}$}}
 \put(0,7){\vector(2,1){65}}
 \put(0,4){\vector(1,0){65}}
\end{picture}
&
\begin{picture}(90,52)(0,0)
 \put(5,0){\mbox{$\tilde{\cal P}$}}
 \put(5,42){\mbox{${\cal P}$}}
 \put(5,13){\vector(0,1){25}}
 \put(10,38){\vector(0,-1){25}}
 \put(-5,22){\mbox{$\tilde\mu$}}
 \put(12,22){\mbox{$\tilde{\rm h}_\mu$}}
 \put(46,35){\vector(-1,-1){25}}
 \put(40,15){\mbox{$\tilde h_\mu^\nabla$}}
 \put(50,39){\mbox{${\cal P}$}}

 \put(30,50){\vector(1,0){55}}
 \put(60,40){\vector(1,0){25}}
 \put(30,4){\vector(1,0){55}}
 \put(70,12){\mbox{$\tilde\jmath_0$}}
 \put(70,44){\mbox{$\jmath_0$}}
 \end{picture}
&
\begin{picture}(15,52)(0,0)
 \put(0,0){\mbox{${\cal M}\pi$}}
 \put(0,41){\mbox{$J^1\pi^*$}}
 \put(10,13){\vector(0,1){25}}
 \put(0,22){\mbox{$\mu$}}
\end{picture}
 &
 \begin{picture}(25,52)(0,0)
 \put(0,44){\vector(1,0){25}}
 \put(25,39){\vector(-1,-1){25}}
 \put(15,17){\mbox{${\rm h}^\nabla_\mu$}}
 \put(1,49){\mbox{${\rm Id}_{J^1\pi^*}$}}
 \end{picture}
&
 \begin{picture}(10,52)(0,0)
 \put(0,41){\mbox{$J^1\pi^*$}}
 \end{picture}
\end{array}
 $$
 The first part of the statement is a consequence of the fact that
 $\tilde \mu$ is a diffeomorphism.
 For the second part, taking into account the commutativity
 of this diagram, and bearing in mind the first item of Proposition
 \ref{proprel3} and Proposition \ref{propositio5}, we have
 \beann
 {\cal F}\Lag_0^*{\cal H}^\nabla_0 &=&
 {\cal F}\Lag_0^*[(\tilde\jmath_0\circ\tilde h^\nabla_0)^*\Theta-
 (\tilde\jmath_0\circ\tilde{\rm h}_\delta)^*\Theta]=
 {\cal F}\Lag_0^*\Theta_{h_\mu^\nabla}^0-
 {\cal F}\Lag_0^*\tilde{\rm h}_\mu^*\jmath_0^*\Theta
 \\ &=&
 {\cal F}\Lag_0^*\Theta_{h_\mu^\nabla}^0-
 \widetilde{{\cal F}\Lag}^*\Theta=
 \del+\Theta_\Lag-\Theta_\Lag=\del
 \eeann
 and the result follows as a consequence of the above Proposition.
 The last statement is immediate.
 \qed

 Of course, the result stated in (\ref{equics2})
 concerning to the variational principle and the characterization of
 critical sections holds in the same way.

 Finally, the equivalence between the Hamiltonian formalisms (in $\Pi$ and
 $J^1\pi^*$) associated with an almost-regular
 Lagrangian system, and between the
 Lagrangian formalism and the Hamiltonian formalism in $J^1\pi^*$ is
 given by the following:

 \begin{teor}
 Let $\ls$ be an almost-regular Lagrangian system. Then
 $$
 {\mit\Psi}_0^*\Theta^0_{\hat{\rm h}_\delta}=\Theta^0_{\tilde{\rm h}_\mu}
 \quad ,\quad
 {\mit\Psi}_0^*\Omega^0_{\hat{\rm h}_\delta}=\Omega^0_{\tilde{\rm h}_\mu}
 $$
 and hence its
 associated Hamiltonian systems $\hspio$ and $\hsjpio$ are equivalent.
 \end{teor}
 \proof
 First, we have the following relation
 $$
 \Theta_{h_\mu^\nabla}^0=\jmath^*_0\Theta_{h_\mu^\nabla}=
 \jmath^*_0{\mit\Psi}^*\Theta_{h_\delta^\nabla}=
 ({\mit\Psi}\circ\jmath_0)^*\Theta_{h_\delta^\nabla}=
 (\j_0\circ{\mit\Psi}_0)^*\Theta_{h_\delta^\nabla}=
 {\mit\Psi}_0^* \j^*_0\Theta_{h_\delta^\nabla}=
 {\mit\Psi}_0^*\Theta_{h_\delta^\nabla}^0
 $$
 Furthermore, as ${\rm F}\Lag_0={\mit\Psi}_0\circ{\cal F}\Lag_0$,
 we have that
 $$
 \del = {\cal F}\Lag_0^*{\cal H}_0^\nabla=
 {\rm F}\Lag_0^*\mbox{\es H}_0^\nabla=
 ({\mit\Psi}_0\circ{\cal F}\Lag_0)^*\mbox{\es H}_0^\nabla=
 {\cal F}\Lag_0^*{\mit\Psi}_0^*\mbox{\es H}_0^\nabla
 \quad \Longleftrightarrow \quad
 {\mit\Psi}_0^*\mbox{\es H}_0^\nabla={\cal H}_0^\nabla
 $$
 since ${\cal F}\Lag_0$ is a submersion.
 (In the same way ${\mit\Psi}_0^*{\rm H}_0^\nabla=H_0^\nabla$).
 Therefore,
 ${\mit\Psi}_0^*\Theta^0_{\hat{\rm h}_\delta}=\Theta^0_{\tilde{\rm h}_\mu}$,
 and hence
 ${\mit\Psi}_0^*\Omega^0_{\hat{\rm h}_\delta}=\Omega^0_{\tilde{\rm h}_\mu}$.

 Observe also that the section
 $\tilde{\rm h}_\mu$ of $\mu$ and the family
 of sections $\hat{\rm h}_\delta$ of $\delta$
 (given in Proposition \ref{once}) are equivalent,
 by the commutativity of diagram (\ref{diag2}).
 \qed

 \section{Examples}

 \subsection{Non-autonomous Mechanics}

 The jet bundle description of time-dependent mechanical systems
 (see, for instance, \cite{EMR-91} and \cite{GMS-98})
 takes $M=\Real$, and $E=\Real\times Q$, where $Q$ is a $N$-dimensional
 manifold (and thus $\pi\colon \Real\times Q\to\Real$ is a trivial bundle).
 Then $J^1E=\Real\times\Tan Q$. Natural adapted coordinates are denoted
 by $(t,q^i,v^i)$. Lagrangian densities are written $\Lag=\lag\d t$,
 where $\lag\in\Cinfty (\Real\times\Tan Q)$ is a
 {\sl time-dependent Lagrangian function}.

 Next we will identify the different multimomentum bundles.
 First observe that, as $\Tan M\simeq\Real\times\Real$, we obtain
 $$
 \Lambda^m\Tan^*M\equiv\Lambda^1\Tan^*M\simeq\Real\times\Real^*
 $$
 Therefore we have:

 \underline{Generalized multimomentum bundle}:
 Observe that
 \beann
 \pi^*\Tan M&=&\pi^*(\Real\times\Real)\simeq
 (\Real\times Q)\times_{\Real} (\Real\times\Real )\simeq
 \Real\times Q\times\Real \\
 \Tan^*E&=&\Tan^*(\Real\times Q)=\Tan^*\Real\times\Tan^*Q \\
 \pi^*\Lambda^m\Tan^*M&\simeq&\Real\times Q\times\Real^*\simeq
 Q\times\Tan^*\Real
 \eeann
 and hence
 \beann
 J^1E^*&:=&\pi^*\Tan M\times_E\Tan^*E\times_E\pi^*\Lambda^m\Tan^*M=
 (\Real\times Q\times\Real)\times_{(\Real\times Q)}
 (\Tan^*\Real\times\Tan^*Q)\times_{(\Real\times Q)}
 (Q\times\Tan^*\Real)
 \\ &\simeq&
 (\Real\times Q\times\Real)\times_{(\Real\times Q)}
 (\Tan^*\Real\times\Tan^*Q)\times_{(\Real\times Q)}
 ((\Real\times Q)\times\Real^*)\simeq
 \Tan^*\Real\times\Tan^*Q\simeq\Tan^*(\Real\times Q)
 \eeann
 Notice that $\dim\,J^1E^*=2N+2$.

 \underline{Reduced multimomentum bundle}:
 As ${\rm V}^*(\pi)=\Real\times\Tan^*Q$, we have that
 \beann
 \Pi&:=&\pi^*\Tan M\times_E{\rm V}^*(\pi)\times_E\pi^*\Lambda^m\Tan^*M=
 (\Real\times Q\times\Real)\times_{(\Real\times Q)}
 (\Real\times\Tan^*Q)\times_{(\Real\times Q)}
 (Q\times\Tan^*\Real)
 \\ &\simeq&
 (\Real\times Q\times\Real)\times_{(\Real\times Q)}
 (\Tan^*\Real\times\Tan^*Q)\times_{(\Real\times Q)}
 ((\Real\times Q)\times\Real^*)\simeq
 \Real\times\Tan^*Q
 \eeann
 and $\dim\,\Pi=2N+1$.

 \underline{Extended multimomentum bundle}:
 Now we have
 $$
 {\cal M}\pi:=\Lambda^m_1\Tan^*E\equiv\Lambda^1_1\Tan^*E=
 \Lambda^1_1\Tan^*E\simeq\Tan^*E\simeq\Tan^*(\Real\times Q)\simeq
 \Tan^*\Real\times\Tan^*Q
 $$
 with $\dim\,{\cal M}\pi=2N+2$.

 \underline{Restricted multimomentum bundle}:
 Observe that
 $\Lambda^m_0\Tan^*E\equiv\Lambda^1_0\Tan^*E=(\Real\times Q)\times\Real^*$
 and then
 $$
 J^1\pi^*:={\cal M}\pi/\Lambda^m_0\Tan^*E\simeq
 (\Tan^*\Real\times\Tan^*Q)/(\Real\times Q\times\Real^*)\simeq
 \Real\times\Tan^*Q
 $$
 with $\dim\,J^1\pi^*=2N+1$.

 {\bf Comments}:
 \bit
 \item
 It is interesting to point out that in the Hamiltonian
 formalism of non-autonomous mechanics,
 $M\simeq\Real\times\Tan^*Q$ and
 $J^1\pi^*\simeq\Real\times\Tan^*Q$ make the canonical diffeomorphism
 between the generalized and the extended multimomentum bundle evident.
 They correspond to the so-called
 {\sl extended momentum phase space} of the {\sl symplectic formulation}
 of time-dependent systems \cite{EMR-91}, \cite{Ku-84}, \cite{Ra-92}.
 As a consequence, the generalized and the (first) extended Legendre maps
 are really the same.
 \item
 The case of {\sl singular (almost-regular) time-dependent
 mechanical systems} has been extensively studied in this context
 in \cite{CLM-94}, \cite{GMS-99a}.
 \eit

 \subsection{Electromagnetic field (with fixed background)}

    In this case $M$, is space-time endowed with a semi-Riemannian
    metric $g$, $E=\Tan^*M$ is a vector bundle over $M$ and
    $\pi\colon \Tan^*M\to M$ denotes the natural projection.
    Sections of $\pi$ are the so-called {\sl electromagnetic
    potentials}.  Using the linear connection associated with the
    metric $g$, one assures that $J^1E\to \Tan^*M$ is a vector
    bundle and, since ${\rm V}E=\pi^*\Tan^*M$, we have
    $J^1E=\pi^*\Tan^*M\otimes_E \pi^*\Tan^*M$.

    Let $\phi\colon M\to \Tan^*M$ be a section of $\pi$. Then
    $j^1\phi\colon M\to\pi^*\Tan^*M\times\pi^*\Tan^*M$ is just
    $j^1\phi =\Tan\phi$
    (observe that $j^1\phi$ is a metric tensor on $M$).
    Now, considering $\bar y\in J^1E$,
    and $\phi\colon M\to\Tan^*M$ being a representative of $\bar y$;
    we have that the Lagrangian density is
    $$
    \Lag =\frac{1}{4}\|\d\phi\|^2\d V_g
    $$
    where $\|\ \|$ denotes
    the norm induced by the metric $g$ on the $2$-forms on $M$, and
    $\d V_g$ is the volume element associated with the metric $g$.
    Observe that $\d\phi$ is the skew-symmetric part of the matrix
    giving $\Tan\phi$ or, in other words, the skew-symmetric part of
    the metric $\Tan\phi$ on $M$.

    For simplifying calculations, we take
    $M=\Real^3$ and the metric is $-++$. Then
    $E=\Real^3\times\Real^3$ and
    $$
 J^1E=(\Real^3\times\Real^3)\times(\Real^3\times\Real^{3*})
    $$
    with $\dim\, J^1E=15$. Coordinates in $J^1E$ are usually denoted
    $(x^\nu,A^j,v^j_\nu )$, with $\nu,j=0,1,2$.
    The coordinates $(A^1,A^2)$ constitute the
    {\sl vector potential} and $A^0$ is the {\sl scalar potential}.
    Then, locally $\phi =\phi_\eta\d x^\eta$, and therefore
    \dst j^1\phi =\derpar{\phi_\eta}{x^\nu}\d x^\nu\otimes\d x^\eta\) .
    It is usual to write $\phi$ in the form $A=\delta_{j\nu}A^j\d x^\nu$ and
    $\d A=\delta_{j\eta}v^j_\nu\d x^\eta\wedge\d x^\nu$
    ($\delta_{j\eta}$ is the {\sl Kronecker's delta}).
    Then, in natural coordinates we have
    the following expression for the Lagrangian function
    $$
    \lag =\frac{1}{4}[(v^1_2-v^2_1)^2-(v^2_0-v^0_2)^2-(v^1_0-v^0_1)^2]
    $$
    Obviously, this is a singular Lagrangian, since its
    Hessian matrix
    $$
    \frac{\partial^2\lag}{\partial v^i_\nu\partial v^j_\eta}=
    \frac{1}{2}\left(\matrix{
    0 &  0   &  0   &  0   & 0 &  0   &  0   &  0   & 0 \cr
    0 & -1   &  0   &  1   & 0 &  0   &  0   &  0   & 0 \cr
    0 &  0   & -1   &  0   & 0 &  0   &  1   &  0   & 0 \cr
    0 &  1   &  0   & -1   & 0 &  0   &  0   &  0   & 0 \cr
    0 &  0   &  0   &  0   & 0 &  0   &  0   &  0   & 0 \cr
    0 &  0   &  0   &  0   & 0 &  1   &  0   & -1   & 0 \cr
    0 &  0   &  1   &  0   & 0 &  0   & -1   &  0   & 0 \cr
    0 &  0   &  0   &  0   & 0 & -1   &  0   &  1   & 0 \cr
    0 &  0   &  0   &  0   & 0 &  0   &  0   &  0   & 0 \cr
    }\right)
    $$
   is singular (its rank is equal to 3).

    Next we study the several Legendre maps associated with this
    system. As we know, all of them leave the coordinates
    $(x^\nu,A^j)$ invariant, thus we will write the
    relations only for the multimomentum coordinates.

    \underline{Generalized Legendre map}:
    The generalized multimomentum bundle is
 $$
 J^1E^*=(\Real^3\times\Real^{3*})\times
 (\Real^3\otimes(\Real^3\otimes\Real^{3*})\otimes\Lambda^3\Real^{3*})
 $$
   From (\ref{coorglt}) we have
    $$
    \begin{array}{ccccccccc}
    \widehat{{\rm F}\Lag}^*{\rm p}^0_0&=&0 & \qquad
    \widehat{{\rm F}\Lag}^*{\rm p}^0_1&=&-\frac{1}{2}(v^1_0-v^0_1) \qquad &
    \widehat{{\rm F}\Lag}^*{\rm p}^0_2&=&-\frac{1}{2}(v^2_0-v^0_2)
    \\
    \widehat{{\rm F}\Lag}^*{\rm p}^1_0&=&\frac{1}{2}(v^1_0-v^0_1) & \qquad
    \widehat{{\rm F}\Lag}^* {\rm p}^1_1&=&0 \qquad &
    \widehat{{\rm F}\Lag}^*{\rm p}^1_2&=&-\frac{1}{2}(v^1_2-v^2_1)
    \\
    \widehat{{\rm F}\Lag}^*{\rm p}^2_0&=&\frac{1}{2}(v^2_0-v^0_2) & \qquad
    \widehat{{\rm F}\Lag}^*{\rm p}^2_1&=&\frac{1}{2}(v^1_2-v^2_1) \qquad &
    \widehat{{\rm F}\Lag}^*{\rm p}^2_2&=&0
    \end{array}
    $$
    and for the additional multimomentum coordinates (${\rm p}^\nu_\eta$
    in (\ref{coorglt}))
$$
    \begin{array}{ccc}
    \widehat{{\rm F}\Lag}^*\hat{\rm p}^0_0 =
   \frac{1}{2}(v^1_0(v^1_0-v^0_1)+v^2_0(v^2_0-v^0_2)) & \quad &
    \widehat{{\rm F}\Lag}^*\hat{\rm p}^0_1=
   \frac{1}{2}(v^1_1(v^1_0-v^0_1)+v^2_1(v^2_0-v^0_2)) \\
    \widehat{{\rm F}\Lag}^*\hat{\rm p}^0_2=
   \frac{1}{2}(v^1_2(v^1_0-v^0_1)+v^2_2(v^2_0-v^0_2)) & \quad &
    \widehat{{\rm F}\Lag}^*\hat{\rm p}^1_0=
   -\frac{1}{2}(v^0_0(v^1_0-v^0_1)+v^2_0(v^1_2-v^2_1)) \\
    \widehat{{\rm F}\Lag}^*\hat{\rm p}^1_1=
   -\frac{1}{2}(v^0_1(v^1_0-v^0_1)+v^2_1(v^1_2-v^2_1)) & \quad &
    \widehat{{\rm F}\Lag}^*\hat{\rm p}^1_2=
   -\frac{1}{2}(v^0_2(v^1_0-v^0_1)+v^2_2(v^1_2-v^2_1)) \\
    \widehat{{\rm F}\Lag}^*\hat{\rm p}^2_0=
   -\frac{1}{2}(v^0_0(v^2_0-v^0_2)-v^1_0(v^1_2-v^2_1)) & \quad &
    \widehat{{\rm F}\Lag}^*\hat{\rm p}^2_1=
   -\frac{1}{2}(v^0_1(v^2_0-v^0_2)-v^1_1(v^1_2-v^2_1)) \\
    \widehat{{\rm F}\Lag}^*\hat{\rm p}^2_2=
   -\frac{1}{2}(v^0_2(v^2_0-v^0_2)-v^1_2(v^1_2-v^2_1)) & &
    \end{array}
$$
    We have the Hamiltonian constraints
    $$
    \begin{array}{ccc}
    \hat\xi^1\equiv{\rm p}^0_0=0 & \qquad
    \hat\xi^2\equiv{\rm p}^1_1=0 \qquad &
    \hat\xi^3\equiv{\rm p}^2_2=0 \\
    \hat\xi^4\equiv{\rm p}^0_1+{\rm p}^1_0=0 & \qquad
    \hat\xi^5\equiv{\rm p}^0_2+{\rm p}^2_0=0 \qquad &
    \hat\xi^6\equiv{\rm p}^1_2+{\rm p}^2_1=0
    \end{array}
    $$
    and the additional ones
 \beann
 \hat\xi^7&\equiv&
 {\rm p}^2_0{\rm p}^2_1(\hat{\rm p}^0_0+\hat{\rm p}^1_1
 -2(\hat{\rm p}^1_0)^2))
 +({\rm p}^2_0)^2\hat{\rm p}^1_0+({\rm p}^2_1)^2\hat{\rm p}^0_1+
 {\rm p}^1_0({\rm p}^2_1\hat{\rm p}^2_1-{\rm p}^2_0\hat{\rm p}^2_0)=0
 \\
 \hat\xi^8&\equiv&
 {\rm p}^1_0{\rm p}^2_1(\hat{\rm p}^0_0+\hat{\rm p}^2_2
 -2(\hat{\rm p}^2_0)^2))
 +({\rm p}^1_0)^2\hat{\rm p}^2_0+({\rm p}^2_1)^2\hat{\rm p}^0_2+
 {\rm p}^2_0({\rm p}^2_1\hat{\rm p}^1_2-{\rm p}^1_0\hat{\rm p}^1_0)=0
 \\
 \hat\xi^9&\equiv&
  {\rm p}^1_0{\rm p}^2_0(\hat{\rm p}^1_1-\hat{\rm p}^2_2
 -2(\hat{\rm p}^2_1)^2))
 +({\rm p}^2_0)^2\hat{\rm p}^1_2-({\rm p}^1_0)^2\hat{\rm p}^2_1+
 {\rm p}^2_1({\rm p}^2_0\hat{\rm p}^0_2-{\rm p}^1_0\hat{\rm p}^0_1)=0
\eeann
   which define locally the submanifold $\hat P$ in $J^1E^*$.

    Observe that $\dim\ \hat P=\dim\ J^1E$, as
 ${\rm rank}\ \widehat{{\rm F}\Lag}_*$
 is maximal. Then, taking into account the commutativity of diagram
 (\ref{diag2}), we can conclude that, for this system,
 the degeneracy is on the fibers of the projection $\delta$.

      \underline{Reduced Legendre map}:
 The reduced multimomentum bundle is
 $$
 \Pi=(\Real^3\times\Real^{3*})\times(\Real^3\otimes\Real^3\otimes
 \Lambda^3\Real^{3*})
 $$
 For the reduced Legendre map the results are the same as
 for the restricted Legendre map,
 but changing the multimomentum coordinates $p^j_\nu$ by ${\rm p}^j_\nu$.
 So, from (\ref{redlt}) we obtain
    $$
    \begin{array}{ccccccccc}
    {\rm F}\Lag^*{\rm p}^0_0&=&0 & \qquad
    {\rm F}\Lag^*{\rm p}^0_1&=&-\frac{1}{2}(v^1_0-v^0_1) \qquad &
    {\rm F}\Lag^*{\rm p}^0_2&=&-\frac{1}{2}(v^2_0-v^0_2)
    \\
    {\rm F}\Lag^*{\rm p}^1_0&=&\frac{1}{2}(v^1_0-v^0_1) & \qquad
    {\rm F}\Lag^* {\rm p}^1_1&=&0 \qquad &
    {\rm F}\Lag^*{\rm p}^1_2&=&-\frac{1}{2}(v^1_2-v^2_1)
    \\
    {\rm F}\Lag^*{\rm p}^2_0&=&\frac{1}{2}(v^2_0-v^0_2) & \qquad
    {\rm F}\Lag^*{\rm p}^2_2&=&\frac{1}{2}(v^1_2-v^2_1) \qquad &
    {\rm F}\Lag^*{\rm p}^2_2&=&0
    \end{array}
    $$
     (${\rm F}\Lag$ is a submersion onto its image,
    and hence the system is almost-regular).

    Now we have the same Hamiltonian constraints
    $$
    \begin{array}{ccc}
    \xi^1\equiv{\rm p}^0_0=0 & \qquad
    \xi^2\equiv{\rm p}^1_1=0 \qquad &
    \xi^3\equiv{\rm p}^2_2=0 \\
    \xi^4\equiv{\rm p}^0_1+{\rm p}^1_0=0 & \qquad
    \xi^5\equiv{\rm p}^0_2+{\rm p}^2_0=0 \qquad &
    \xi^6\equiv{\rm p}^1_2+{\rm p}^2_1=0
    \end{array}
    $$
    which define locally the submanifold $P$ in $\Pi$.
     Observe that, as these constraints are conditions of skew-symmetry,
     we have that
     $$
     {\rm P}={\rm F}\Lag(J^1E)=
    \pi^*({\cal A}(\Real^3)\otimes\Lambda^3(\Real^{3*}))
     $$
     where ${\cal A}(\Real^3)$ denotes the bundle whose sections are
     the $2$-contravariant skew-symmetric tensor fields on $\Real^3$.
 Note also that $\hat P$ is not diffeomorphic to $P$.

     \underline{First extended Legendre map}:
    The extended multimomentum bundle is
 $$
 {\cal M}\pi=(\Real^3\times\Real^{3*})\times
 \Lambda^3_1(\Real^3\times\Real^{3*})
 $$
    From (\ref{extlt}) we obtain that
    $$
    \begin{array}{ccccccccc}
    \widehat{{\cal F}\Lag}^*p^0_0&=&0 &  \qquad
    \widehat{{\cal F}\Lag}^*p^0_1&=&-\frac{1}{2}(v^1_0-v^0_1)  \qquad &
    \widehat{{\cal F}\Lag}^*p^0_2&=&-\frac{1}{2}(v^2_0-v^0_2)
    \\
    \widehat{{\cal F}\Lag}^*p^1_0&=&\frac{1}{2}(v^1_0-v^0_1) & \qquad
    \widehat{{\cal F}\Lag}^*p^1_1&=&0 \qquad &
    \widehat{{\cal F}\Lag}^*p^1_2&=&-\frac{1}{2}(v^1_2-v^2_1)
    \\
    \widehat{{\cal F}\Lag}^*p^2_0&=&\frac{1}{2}(v^2_0-v^0_2) & \qquad
    \widehat{{\cal F}\Lag}^*p^2_1&=&\frac{1}{2}(v^1_2-v^2_1) \qquad &
    \widehat{{\cal F}\Lag}^*p^2_2&=&0
  \end{array}
    $$
    and the additional relation
    $$
    \widehat{{\cal F}\Lag}^*p=
    -\frac{1}{2}[(v^1_2-v^2_1)^2-(v^2_0-v^0_2)^2-(v^1_0-v^0_1)^2]\equiv-2\lag
    $$
    The corresponding Hamiltonian constraints are
    $$
    \begin{array}{ccc}
    \hat\chi^1\equiv p^0_0=0 & \qquad
    \hat\chi^2\equiv p^1_1=0 \qquad &
    \hat\chi^3\equiv p^2_2=0
    \\
    \hat\chi^4\equiv p^0_1+p^1_0=0 & \qquad
    \hat\chi^5\equiv p^0_2+p^2_0=0 \qquad &
    \hat\chi^6\equiv p^1_2+p^2_1=0
    \end{array}
    $$
    and the additional one
    $$
    \hat\chi^7\equiv p+2(p^2_1)^2-2(p^2_0)^2-2(p^1_0)^2=0
    $$
    All of them define locally the submanifold
    $\hat{\cal P}$ in ${\cal M}\pi$.

  \underline{Second extended Legendre map}:
    From (\ref{extlt}) we obtain that
    $$
    \begin{array}{ccccccccc}
    \widetilde{{\cal F}\Lag}^*p^0_0&=&0 & \qquad
    \widetilde{{\cal F}\Lag}^*p^0_1&=&-\frac{1}{2}(v^1_0-v^0_1) \qquad &
    \widetilde{{\cal F}\Lag}^*p^0_2&=&-\frac{1}{2}(v^2_0-v^0_2)
    \\
    \widetilde{{\cal F}\Lag}^*p^1_0&=&\frac{1}{2}(v^1_0-v^0_1) & \qquad
    \widetilde{{\cal F}\Lag}^*p^1_1&=&0 \qquad &
    \widetilde{{\cal F}\Lag}^*p^1_2&=&-\frac{1}{2}(v^1_2-v^2_1)
    \\
    \widetilde{{\cal F}\Lag}^*p^2_0&=&\frac{1}{2}(v^2_0-v^0_2) & \qquad
    \widetilde{{\cal F}\Lag}^*p^2_1&=&\frac{1}{2}(v^1_2-v^2_1) \qquad &
    \widetilde{{\cal F}\Lag}^*p^2_2&=&0
  \end{array}
    $$
    and the additional relation
    $$
    \widetilde{{\cal F}\Lag}^*p=
    -\frac{1}{4}[(v^1_2-v^2_1)^2-(v^2_0-v^0_2)^2-(v^1_0-v^0_1)^2]\equiv-\lag
    $$
    and the Hamiltonian constraints are now
    $$
    \begin{array}{ccc}
    \tilde\chi^1\equiv p^0_0=0 & \qquad
    \tilde\chi^2\equiv p^1_1=0 \qquad &
    \tilde\chi^3\equiv p^2_2=0
    \\
    \tilde\chi^4\equiv p^0_1+p^1_0=0 & \qquad
    \tilde\chi^5\equiv p^0_2+p^2_0=0 \qquad &
    \tilde\chi^6\equiv p^1_2+p^2_1=0
    \end{array}
    $$
    and the additional one
    $$
    \tilde\chi^7\equiv p+(p^2_1)^2-(p^2_0)^2-(p^1_0)^2=0
     $$
     All of them define locally the submanifold
    $\tilde{\cal P}$ in ${\cal M}\pi$.
    Note that the last constraint identifies the extra coordinate
    $p$ with the Hamiltonian function which, for this system, is
    \dst H=-\frac{1}{4}((p^2_1)^2-(p^2_0)^2-(p^1_0)^2)\) .

     \underline{Restricted Legendre map}:
    The restricted multimomentum bundle is
 $$
 J^1\pi^*=(\Real^3\times\Real^{3*})\times
 [\Lambda^3_1(\Real^3\times\Real^{3*})/\Lambda^3_0(\Real^3\times\Real^{3*})]
 $$
    From (\ref{restlt}) we obtain
    $$
    \begin{array}{ccccccccc}
    {\cal F}\Lag^* p^0_0&=&0 & \qquad
    {\cal F}\Lag^* p^0_1&=&-\frac{1}{2}(v^1_0-v^0_1) \qquad &
    {\cal F}\Lag^* p^0_2&=&-\frac{1}{2}(v^2_0-v^0_2)
    \\
    {\cal F}\Lag^* p^1_0&=&\frac{1}{2}(v^1_0-v^0_1) &  \qquad
    {\cal F}\Lag^* p^1_1&=&0  \qquad &
    {\cal F}\Lag^* p^1_2&=&-\frac{1}{2}(v^1_2-v^2_1)
    \\
    {\cal F}\Lag^* p^2_0&=&\frac{1}{2}(v^2_0-v^0_2) &  \qquad
    {\cal F}\Lag^* p^2_1&=&\frac{1}{2}(v^1_2-v^2_1)  \qquad &
    {\cal F}\Lag^* p^2_2&=&0
    \end{array}
    $$
    Hence we obtain the following set of Hamiltonian constraints
    $$
    \begin{array}{ccc}
    \chi^1\equiv p^0_0=0 &  \qquad
    \chi^2\equiv p^1_1=0  \qquad &
    \chi^3\equiv p^2_2=0
    \\
    \chi^4\equiv p^0_1+p^1_0=0 &  \qquad
    \chi^5\equiv p^0_2+p^2_0=0  \qquad &
    \chi^6\equiv p^1_2+p^2_1=0
    \end{array}
    $$
    which define locally the submanifold ${\cal P}$ in $J^1\pi^*$.

   Observe that, for this example,
    $$
    {\rm rank}\ {\rm F}\Lag_*=
    {\rm rank}\ \widehat{{\cal F}\Lag}_*=
    {\rm rank}\ \widetilde{{\cal F}\Lag}_*=
    {\rm rank}\ {\cal F}\Lag_*
    $$
    and the submanifolds
    $P, {\cal P}, \hat{\cal P}$ and $\tilde{\cal P}$ are,
    in fact, diffeomorphic.

 \section{Conclusions}

 We have studied the Hamiltonian formalism for first-order
 Classical Field theories in the context of multisymplectic
 manifolds, taking different choices of multimomentum bundles
 as phase spaces, in particular the bundles $\Pi$ and $J^1\pi^*$.

 \bit
 \item
 First we have reviewed the construction of these and other
 auxiliary multimomentum bundles ($J^1E^*$ and ${\cal M}\pi$),
 as well as the definition of
 suitable Legendre maps when all these bundles are thought of,
 in a certain sense, as the dual bundles of a Lagrangian system $\ls$.
 The key result is the existence of a canonical diffeomorphism between
 $\Pi$ and $J^1\pi^*$.
 (See section \ref{mmblm}).
 \item
 In order to state the Hamiltonian formalism on $\Pi$ and $J^1\pi^*$,
 some additional geometric element is needed for obtaining
 the {\sl Hamilton-Cartan forms}
 from the canonical forms which $J^1E^*$ and ${\cal M}\pi$
 are endowed with.
 In particular, we can take sections
 of the projections $\delta\colon J^1E^*\to\Pi$ and
 $\mu\colon{\cal M}\pi\to J^1\pi^*$,
 (which are called {\sl Hamiltonian sections},
 and are the elements carrying the ``physical'' information
 in this construction), which allows us to pull-back
 the canonical forms from $J^1E^*$ and ${\cal M}\pi$ to
 $\Pi$ and $J^1\pi^*$ respectively. These are the Hamilton-Cartan forms
 which define the Hamiltonian system.
 Hamiltonian sections are associated with
 {\sl local Hamiltonian functions}, which
 appear explicitly in the local expression of the corresponding
 Hamilton-Cartan forms.
 (See sections \ref{dsc} and \ref{hsjpsect}).
 \item
 A relevant result is that different choices of
 Hamiltonian sections of $\delta$ may lead to the same  Hamilton-Cartan forms
 in $\Pi$, and this allows us to establish an equivalence
 relation in the set of sections of the projection $\delta$.
 Then, using the diffeomorphism ${\mit\Psi}$ between $\Pi$ and
 $J^1\pi^*$, it is proved that there is a one-to-one
 correspondence between sections of $\mu$
 and classes of equivalent sections of $\delta$.
 Therefore, the Hamilton-Cartan forms in $\Pi$ and $J^1\pi^*$
 are ${\mit\Psi}$-related and hence, Hamiltonian systems
 in $\Pi$ and $J^1\pi^*$ are equivalent.
 (See sections \ref{dsc} and \ref{hsjpsect}).
 Furthermore, another one-to-one correspondence
 exists between the set of connections in the bundle
 $E\to M$, and the set of linear sections of the respective projections
 $\delta$ and $\mu$.
 (See sections \ref{hshdc} and \ref{hsjpsect}).
 \item
 The difference between two Hamilton-Cartan $m$-forms defined by
 Hamiltonian sections is a semibasic $m$-form which is called a
 {\sl Hamiltonian density}. Hence the set of Hamilton-Cartan forms
 can be thought of as an affine space modelled on the
 module of Hamiltonian densities.
 As a particular case, 
 Given a connection, (classes of) Hamiltonian sections and
 Hamiltonian densities are in one-to-one correspondence.
 As a consequence of this fact,
 a Hamiltonian system can be also constructed starting from a Hamiltonian
 density and a connection.
 (See sections \ref{hshdc} and \ref{hsjpsect}).
 \item
 The field equations of the Hamiltonian formalism can be derived
 from the so-called {\sl Hamilton-Jacobi variational principle}.
 Different but equivalent ways of characterizing the critical
 sections by means of the Hamilton-Cartan forms are shown.
 In particular, in natural coordinates of the multimomentum
 bundles $\Pi$ or $J^1\pi^*$, these sections are obtained
 as solutions of a local system of first-order
 partial differential equations,
 which are known as the {\sl Hamilton-De Donder-Weyl equations}.
 Nevertheless, as the Hamiltonian function appearing in the local expression
 of the Hamilton-Cartan form is local, these equations are not covariant.
 Then, for obtaining a set of covariant equations,
 we must introduce a Hamiltonian density, and so a global
 Hamiltonian function.
 (See section \ref{vpfe}).
 \item
 The question of associating a
 Hamiltonian system to a Lagrangian one is also analyzed,
 both in the {\sl hyper-regular} and the {\sl almost-regular} cases.
 We can define this Hamiltonian system in three equivalent ways
 (see sections \ref{hsahrs}, \ref{hspiar}, \ref{hslsjpi1} and \ref{hslsjpi2}):
 \bit
 \item
 Using a natural Hamiltonian section,
 which is defined using the Legendre maps,
 for obtaining the Hamilton-Cartan forms. These forms
 are related to the Poincar\'e-Cartan forms
 of the Lagrangian formalism, through the Legendre map.
 \item
 Using a connection, the {\sl density of Lagrangian energy}
 of the Lagrangian formalism can be defined. Then, we construct
 a Hamiltonian density as the only semibasic $m$-form which is related
 to it by means of the suitable Legendre map.
 \item
 This last Hamiltonian density can be obtained from a connection
 and the above natural Hamiltonian section.
 In this case, the extended Legendre maps must also be used, and in particular,
 for the construction in the reduced multimomentum bundle $\Pi$,
 both extended Legendre maps are needed
 (in the hyper-regular and in the almost-regular cases).
 This fact would justify the introduction of two extended Legendre maps.
 \eit
 \item
 As an additional result, in the hyper-regular case, the
 equivalence between the Lagrangian and the Hamiltonian formalism
 is proved from a double point of view: showing the equivalence between
 the sections solution of the Lagrangian and Hamiltonian problems,
 and proving the equivalence of the Lagrangian and Hamiltonian
 variational principles.
 This equivalence is only partially proved in the almost-regular case.
 (See section \ref{vp}).
 \eit

 \appendix

 \section{Geometrical structures in first-order jet bundles}
 \protect\label{gsjb}

(See \cite{EMR-96}, \cite{Gc-73} and \cite{Sa-89}).

 Let $\pi\colon E\to M$ be a fiber bundle
 ($\dim\, M=m$, $\dim\, E=N+m$),
 and $\pi^1\colon J^1E\to E$ the
 1-jet bundle of local sections of $\pi$,
 which is also a differentiable bundle on $M$ with projection
 $\bar\pi^1 = \pi$. If $(x^\nu,y^A)$ (with $\nu = 1,\ldots,m$; $A= 1,\ldots,N$)
 is a local system of coordinates
 adapted to the bundle $\pi\colon E\to M$, then we
 denote by $(x^\nu,y^A,v^A_\nu)$ the
 local system of coordinates induced in $J^1E$.

Let $\phi\colon M\to E$ be a section of $\pi$, $x\in M$ and
$y=\phi(x)$. The {\sl vertical differential} of the section $\phi$
at the point $y\in E$ is the map $$
\begin{array}{ccccc}
\d^v_y\phi&\colon&\Tan_y E & \longrightarrow & {\rm V}_y(\pi) \\ &
& u & \longmapsto & u-\Tan_y(\phi\circ\pi)u
\end{array}
$$ Then, considering $\bar y\in J^1E$ with $\bar
y\stackrel{\pi^1}{\mapsto}y\stackrel{\pi}{\mapsto}x$ and $\bar
u\in\Tan_{\bar y}J^1E$. The {\sl structure canonical $1$-form} of
$J^1E$, denoted by $\theta$, is defined by $$ \theta (\bar y;\bar
u):=(\d^v_y \phi)(\Tan_{\bar y}\pi^1 (\bar u)) $$ where $\phi$ is
a representative of $\bar y$. Its expression in a natural local
system is \dst\theta =  ({\rm d}y^A - v^A_{\nu}{\rm d}x^{\nu})
\otimes \derpar{}{y^A}\) . $\theta$ is an element of
$\df^1(J^1E)\otimes_{J^1E}\Gamma (J^1E,\pi^{1*}{\rm V}(\pi))$;
then it can be thought of as a $\Cinfty (J^1E)$-linear map
$\theta\colon\Gamma (J^1E,\pi^{1*}{\rm V}(\pi))^*\longrightarrow\df^1(J^1E)$.

 Consider the canonical isomorphism
 ${\cal S}_{\bar y}\colon
 \Tan_{\bar\pi^1 (\bar y)}^*M\otimes{\rm V}_{\pi^1 (\bar y)}(\pi)
 \longrightarrow {\rm V}_{\bar y}(\pi^1 )$
 which consists in associating to an element
 $\alpha\otimes v\in \Tan_{\bar\pi^1(\bar y)}^*M
 \otimes{\rm V}_{\pi^1 (\bar y)}(\pi)$
 the directional derivative in $\bar y$ with respect to
 $\alpha\otimes v$. Taking into account that
 $\alpha\otimes v$ acts in $J^1_yE$ by
 translation, we have
 $$
 {\cal S}_{\bar y}(\alpha\otimes v ):=
 D_{\alpha\otimes v}(\bar y) \colon f\mapsto
 \lim_{t\to 0}\frac{f(\bar y+t(\alpha\otimes v))- f(\bar y)}{t}
 $$
 for $f\in\Cinfty (J^1_yE)$. Then we have the following isomorphism of
 $\Cinfty (J^1E)$-modules
 $$
 {\cal S}\colon \Gamma (J^1E,\bar\pi^{1*}\Tan^*M\otimes
 \pi^{1*}{\rm V}(\pi)) \longrightarrow \Gamma (J^1E,{\rm V}(\pi^1 ))
 $$
 which is called the {\sl vertical endomorphism} ${\cal S}$.
 ($\Gamma (A,B)$ denotes the set of sections of the projection $A\to B$).
 Notice that
 ${\cal S}\in\Gamma (J^1E,(\pi^{1*}{\rm V}(\pi))^*\otimes
 {\rm V}(\pi^1 )\otimes\bar\pi^{1*}\Tan M)$
 (where all the tensor products are on $\Cinfty (J^1E)$). Then,
 another {\sl vertical endomorphism} ${\cal V}$ arises from the
 natural contraction between the factor
 $\Gamma (J^1E,(\pi^{1*}{\rm V}(\pi))^*)$ of ${\cal S}$ and the factor
 $\Gamma(J^1E,\pi^{1*}{\rm V}(\pi))$ of $\theta$:
 $$
 {\cal V}= \inn ({\cal S})\theta\in\df^1(J^1E)\otimes
 \Gamma (J^1E,{\rm V}(\pi^1 )\otimes\bar\pi^{1*}\Tan M)
 $$
 so it is a morphism
 $$
 {\cal V}\colon \Gamma (J^1E,{\rm V}^*(\pi^1)\otimes\bar\pi^{1*}\Tan^*M)
 \longrightarrow \df^1 (J^1E)
 $$
 ${\cal S}$ can also be thought of as a morphism
 $$
 {\cal S}\colon \Gamma (J^1E,{\rm V}^*(\pi^1)\otimes\bar\pi^{1*}\Tan^*M)
 \longrightarrow \Gamma (J^1E,(\bar\pi^{1*}{\rm V}(\pi))^*)
 $$
 As every connection $\nabla$ on $\pi\colon E\to M$ gives an injection
 $\nabla^v\colon\Gamma (J^1E,(\bar\pi^{1*}{\rm V}(\pi))^*)
 \hookrightarrow\df^1(J^1E)$,
 then it makes sense to define
 $$
 {\cal S}^\nabla:=\nabla^v\circ{\cal S}\colon
 \Gamma (J^1E,{\rm V}^*(\pi^1)\otimes\bar\pi^{1*}\Tan^*M)
 \longrightarrow \df^1 (J^1E)
 $$
 As a consequence of the foregoing, the operation ${\cal S}^\nabla-{\cal V}$
 is meaningful.

 In a natural system of coordinates the local expressions
 of all these elements are
 \beann
 {\cal S}&=&\zeta^A\otimes\derpar{}{v^A_\nu}\otimes\derpar{}{x^\nu}
 \\
 {\cal V}&=&\left(\d y^A-v^A_\nu\d x^\nu\right)\otimes
 \derpar{}{v^A_\eta}\otimes\derpar{}{x^\eta}
 \\
{\cal S}^\nabla&=&\left(\d y^A-{\mit\Gamma}^A_\nu\d x^\nu\right)\otimes
 \derpar{}{v^A_\eta}\otimes\derpar{}{x^\eta}
 \eeann
 where $\{\zeta^A\}$ is the local basis of
 $\Gamma (J^1E,\pi^{1*}{\rm V}(\pi))^*$ which is dual of
 \dst\left\{\derpar{}{y^A}\right\}\) , and ${\mit\Gamma}^A_\nu$
 are the component functions of the connection $\nabla$.

\subsection*{Acknowledgments}

We are grateful for the financial support of the
CICYT TAP97-0969-C03-01 and the CICYT PB98-0821.
We wish to thank Mr. Jeff Palmer for his
assistance in preparing the English version of the manuscript.

\end{document}